\documentclass[useAMS,usenatbib]{mn2e}

\usepackage{graphics,graphicx}
\usepackage{subfigure}
\usepackage{color}
\usepackage{lscape}
\usepackage{deluxetable}

\begin{document}

\title{Finding Rare AGN: {\it XMM-Newton} and {\it Chandra} Observations of SDSS Stripe 82}

\author[Stephanie M. LaMassa et al.]{Stephanie M. LaMassa$^1$\thanks{E-mail:stephanie.lamassa@yale.edu}, 
C. Megan Urry$^1$, 
Nico Cappelluti$^{2,3}$, 
Francesca Civano$^{4,5}$, 
\newauthor
Piero Ranalli$^{6}$,
Eilat Glikman$^1$, 
Ezequiel Treister$^7$, 
Gordon Richards$^{8}$,
\newauthor
David Ballantyne$^{9}$,
Daniel Stern$^{10}$,
Andrea Comastri$^{2}$,
Carie Cardamone$^{11}$,
\newauthor
Kevin Schawinski$^{12}$
Hans B\"ohringer$^{13}$, 
Gayoung Chon$^{13}$,
Stephen S. Murray$^{14,4}$,
\newauthor
Paul Green$^{4}$,
Kirpal Nandra$^{13}$\\
$^1$Yale Center for Astronomy \& Astrophysics, Yale University, Physics Department, PO Box 208120, New Haven, CT, 06520-8120, USA;\\
$^2$INAF - Osservatorio Astronomico di Bologna, Via Ranzani 1, I-40127 Bologna, Italy;\\
$^3$University of Maryland Baltimore College, Center for Space Science \& Technology, Physics Department,\\ 1000 Hilltop Circle, Baltimore, MD 21250, USA;\\
$^4$Dartmouth College, Physics \& Astronomy Department, Wilder Lab, Hanover, NH 03755, USA;\\
$^5$Harvard-Smithsonian Center for Astrophysics, 60 Garden Street, Cambridge, MA 02138, USA;\\
$^6$National Observatory of Athens\\
$^7$Universidad de Concepci\'on, Casilla 160-c Concepci\'on, Chile;\\
$^{8}$Drexel University, Department of Physics, 3141 Chestnut Street, Philadelphia, PA 19104, USA;\\
$^{9}$Center for Relativistic Astrophysics, School of Physics, Georgia Institute of Technology, Atlanta, GA;\\
$^{10}$Jet Propulsion Laboratory, California Institute of Technology, 4800 Oak Grove Drive, Mail Stop 169-221, Pasadena, CA 91109, USA;\\
$^{11}$Brown University, The Harriet W. Sheridan Center for Teaching and Learning, Box 1912, 96 Waterman Street, Providence, RI 02912, USA;\\
$^{12}$Institute for Astronomy, Department of Physics, ETH Z\"urich, Wolfgang-Pauli-Strasse 16, CH-8093 Zurich, Switzerland;\\
$^{13}$Max-Planck-Institut F\"ur Extraterrestriche Physik, D-85748 Garching, Germany; \\
$^{14}$The Johns Hopkins University, Department of Physics \& Astronomy, 3400 N. Charles Street, Baltimore, MD 21218, USA
}  

\maketitle
\begin{abstract}
We have analyzed the {\it XMM-Newton} and {\it Chandra} data overlapping $\sim$16.5 deg$^2$ of Sloan Digital Sky Survey Stripe 82, including $\sim$4.6 deg$^2$ of proprietary {\it XMM-Newton} data that we present here. In total, 3362 unique X-ray sources are detected at high significance. We derive the {\it XMM-Newton} number counts and compare them with our previously reported {\it Chandra} Log$N$-Log$S$ relations and other X-ray surveys. The Stripe 82 X-ray source lists have been matched to multi-wavelength catalogs using a maximum likelihood estimator algorithm.  We discovered the highest redshift ($z=5.86$) quasar yet identified in an X-ray survey. We find 2.5 times more high luminosity (L$_x \geq 10^{45}$ erg s$^{-1}$) AGN than the smaller area {\it Chandra} and {\it XMM-Newton} survey of COSMOS and 1.3 times as many identified by XBo\"otes. Comparing the high luminosity AGN we have identified with those predicted by population synthesis models, our results suggest that this AGN population is a more important component of cosmic black hole growth than previously appreciated. Approximately a third of the X-ray sources not detected in the optical are identified in the infrared, making them candidates for the elusive population of obscured high luminosity AGN in the early universe.
\end{abstract}

\section{Introduction}
Supermassive black holes (SMBHs) that reside in galactic centers grow by accretion in a phase where they appear as Active Galactic Nuclei (AGN). To understand AGN demography and evolution, large samples over a range of redshifts and luminosities are necessary. Extragalactic surveys provide an ideal mechanism for locating large enough samples of growing black holes to study the ensemble statistically. Large area surveys have been undertaken in the optical via, e.g., the Sloan Digital Sky Survey \citep[SDSS,][]{dr9} and in the near-infrared (NIR) via the Wide-Field Infrared Survey Explorer \citep[{\it WISE},][]{wright} and the UKIRT Infrared Deep Sky Survey \citep[UKIDSS,][]{lawrence}, locating over 100,000 AGN in the optical and millions of AGN candidates in the infrared. 

However, optical selection is not ideal for studying high-luminosity, high-redshift AGN (quasars) that are heavily reddened or obscured. At redshifts greater than 0.5, diagnostic diagrams that use ratios of narrow emission lines to identify Type 2 (obscured) AGN \citep[e.g.,][]{bpt, kewley, kauff} become inefficient as H$\alpha$ is shifted out of the optical. Such Type 2 AGN can be found using alternate rest-frame optical diagnostics, using, e.g., ratios of narrow emission lines versus $g - z$ color \citep[TBT,][]{trouille} and versus stellar mass \citep[MEx,][]{juneau}, probing out to distances $z<1.4$ and $z<1$, respectively. Narrow rest-frame UV emission lines also allow identification of SMBH accretion at $z>0.5$. Alternatively, obscured AGN candidates can be followed up with ground based infrared spectroscopy to detect redshifted H$\alpha$ and [NII]$\lambda$6584. However, the \citet{kewley} and \citet{kauff} boundaries between star-forming galaxies, composites and Sy2s are only calibrated at low redshifts. As galaxies beyond $z>0.5$ have lower metallicities, it is unclear whether these dividing lines can unambiguously identify signatures of SMBH accretion. 

The reliability of infrared color selection varies with the depth of the data, with {\it Spitzer} IRAC color cuts \citep{stern1,lacy} and {\it WISE} color cuts \citep{stern,assef} being most applicable at shallow depths. At fainter fluxes, contamination from normal galaxies can become appreciable \citep{cardamone,donley,mendez}. The revised IRAC color selection from \citet{donley} is more reliable for deeper data, yet at X-ray luminosities exceeding 10$^{44}$ erg s$^{-1}$, 25\% (32\%) of the {\it XMM-Newton}- ({\it Chandra}-) selected AGN are not recovered with this MIR identification method. 

X-rays provide an alternate way to search for AGN, complementing the optical and MIR identification techniques to provide a comprehensive view of black hole growth over cosmic time, because X-rays can pierce through large amounts of dust and gas. Their emission is visible out to cosmological distances as long as it is not attenuated by Compton-thick (N$_H \geq 10^{24}$ cm$^{-2}$) obscuration. Normal star formation processes rarely exceed an X-ray luminosity above 10$^{42}$ erg s$^{-1}$ \citep[e.g.,][]{persic,bh}, whereas AGN luminosities extend to $\sim10^{46}$ erg s$^{-1}$, making X-ray selection an efficient means for locating AGN at all redshifts. Indeed, X-ray surveys such as the {\it Chandra} Deep Fields North \citep{cdfn} and South \citep{giacconi,cdfs}, Extended {\it Chandra} Deep Field South \citep{lehmer, virani}, {\it XMM-Newton} survey of the {\it Chandra} Deep Field South \citep{comastri,ranalli}, {\it XMM-Newton} and {\it Chandra} surveys of COSMOS \citep{cap,cap09,C-Cosmos,brusa3,civano_mle}, XBo\"otes \citep{murray,kenter}, the {\it XMM-Newton} survey of the Lockman Hole \citep{brunner}, {\it Chandra} observations of All-Wavelength Extended Groth Strip International Survey \citep[AEGIS,][]{aegis,aegis2}, XDEEP2 \citep{deep2}, the {\it XMM-Newton} Serendipitous Survey \citep{Mateos} and the {\it Chandra} multi-wavelength campaign \citep[ChaMP,][]{champ}, have identified thousands of AGN, contributing significantly to our knowledge of AGN demography and galaxy and SMBH co-evolution.

However, most of these X-ray surveys cover small ($<$1 deg$^2$) to moderate (3-5 deg$^2$) areas, sacrificing area for depth to uncover the faintest X-ray objects. The {\it XMM}-COSMOS \citep{cap,cap09,brusa3}, {\it Chandra} COSMOS \citep{C-Cosmos,civano_mle} and ongoing {\it Chandra} COSMOS Legacy Project (PI: Civano) strikes a good balance of moderate area at moderate depth to populate a large portion of the L$_x-z$ plane. But sources that are rare, like high luminosity and/or high redshift AGN, are under-represented in these small to moderate area X-ray samples as a larger volume of the Universe must be probed to locate them. 

Considerable follow-up (optical/near-infrared imaging and spectroscopy) is needed to identify X-ray sources, and multi-wavelength data are needed to classify these objects. Since spectroscopic campaigns and multi-wavelength follow-up are time intensive, the output from wide area surveys such as XBo\"otes \citep[$\sim$9 deg$^2$,][]{kenter,kochanek}, ChaMP \citep[$\sim$33 deg$^2$,][]{champ,trichas} and {\it XMM}-LSS \citep[$\sim$11 deg$^2$, the first part of the expanded {\it XMM}-XXL 50 deg$^2$ survey,][]{lss1,lss2}, has taken many years to achieve. The high-redshift X-ray-selected luminosity AGN population therefore remains poorly explored, prohibiting a comprehensive view of black hole growth.

\begin{figure}
{\includegraphics[scale=0.35,angle=90]{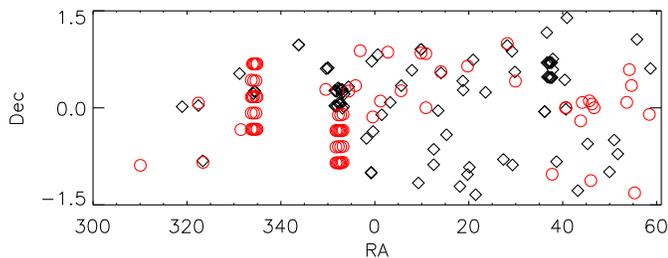}}
\caption[]{\label{pointings} X-ray observations overlapping Stripe 82 used in this analysis, with {\it Chandra} observations shown as black diamonds and {\it XMM-Newton} pointings depicted as red circles. The dense {\it Chandra} pointings are part of the XDEEP2 survey \citep{deep2} while the dense {\it XMM-Newton} groupings represent the positions of the proprietary mosaicked observations we were awarded in AO 10.}
\end{figure}

To address this gap, we have begun a wide area X-ray survey in a region that already has a rich investment in multi-wavelength data and a high level of optical spectroscopic completeness ($>$ 400 objects deg$^2$): the SDSS Stripe 82 region, which spans 300 deg$^2$ along the celestial equator (-60$^{\circ}$ $<$ R.A. $<$ 60$^{\circ}$, -1.25$^{\circ}$ $<$ Dec $<$ 1.25$^{\circ}$). The current non-overlapping X-ray coverage in Stripe 82 from archival {\it Chandra} and archival and proprietary {\it XMM-Newton} observations is $\sim$16.5 deg$^2$. The distribution of these pointings across Stripe 82 is shown in Figure \ref{pointings}. As we are endeavoring to increase the survey area to $\sim$100 deg$^2$, we dub the present survey `Stripe 82X Pilot.' Here we follow-up on the work presented in \citet{me} where we focused on just the {\it Chandra} overlap with Stripe 82, by adding in $\sim$10.5 deg$^2$ of {\it XMM-Newton} observations, 4.6 deg$^2$ of which were obtained by us as part of an approved AO10 proposal (PI: Urry), with the observations performed in `mosaic' mode. We then match both catalogs to large optical \citep[SDSS,][]{dr9}, NIR \citep[UKIDSS and {\it WISE},][]{lawrence,wright}, ultraviolet \citep[{\it GALEX},][]{morrissey} and radio datasets \citep[FIRST,][]{first} in this region. Observations covering Stripe 82 with {\it Spitzer} (P.I. Richards) and analysis of Herchel observation overlapping $\sim$55 deg$^2$ of the region (P.I. Viero) are on-going.

In Section 2, we discuss the reduction and analysis of the archival and proprietary mosaicked {\it XMM-Newton} data in Stripe 82. We use these data to calculate area-flux curves and in Section 3 present the Log$N$-Log$S$ relations, which we compare to the {\it Chandra} Stripe 82 number counts \citep{me} and those from other X-ray surveys. We then describe in Section 4 the matching of the {\it XMM-Newton} and {\it Chandra} X-ray source lists with multi-wavelength catalogs, producing multi-wavelength source lists. In Section 5, we describe the general characteristics of the Stripe 82X sample so far. In particular, we highlight the interesting science gaps our data are primed to fill: uncovering the population of rare high-luminosity AGN at high redshift and identifying candidates for high-luminosity obscured AGN at $z>1$. We have adopted a cosmology of H$_{0}$ = 70 km s$^{-1}$ Mpc$^{1}$, $\Omega_M$ = 0.27 and $\Lambda$=0.73 throughout the paper.

\section{{\it XMM-Newton} Data Reduction}

\subsection{Archival Observations}
Fifty-seven {\it XMM-Newton} EPIC non-calibration observations overlap Stripe 82. Of these, twenty-four were removed due to flaring, substantial pile-up and read-out streaks, small-window mode set-up, or extended emission spanning the majority of the detector, all of which complicate serendipitous detections of point sources in the field. We were left with 33 archival observations well suited for our analysis, listed in Table \ref{obs_summary} and shown as red circles in Figure \ref{pointings}; for 3 of these we dropped the PN detector due to significant pile-up which did not affect the MOS detectors as seriously.

The raw observational data files (ODFs) were processed with {\it XMM-Newton} Standard Analysis System (SAS) version 11. SAS tasks {\em emchain} and {\em epchain} were run to generate MOS1 and MOS2 event files as well as PN and PN out-of-Time (OoT) event files. OoT events result from photons detected during  CCD readout, when photons are recorded at random position along the readout column in the $y$ direction. The subsequent energy correction for these OoT events will then be incorrect. The fraction of OoT events is highest for the PN detector in full frame mode, affecting $\sim$6.3\% of observing time. By generating simulated OoT event files, the PN images can be statistically corrected for this effect.

Good time intervals (GTIs) were applied to the data by searching for flaring in the high energy background (10-12 keV for MOS, 12-14 keV for PN and PN OoT), removing intervals where the count rate was $\geq$3$\sigma$ above the average. Low energy flares were removed from this filtered event list by removing intervals where the count rate was $\geq$3$\sigma$ above the average in the 0.3-10 keV range. In both the high energy and low energy cleaning, GTIs were extracted from single events (i.e., PATTERN = 0).

MOS images were extracted from all valid events (PATTERN 0 to 12) whereas the PN and PN OoT images were extracted from the single and double events only (PATTERN 0 to 4). To avoid emission line features from the detector background (i.e., Al K$\alpha$ at 1.48 keV), the energy range 1.45 to 1.54 keV was excluded when extracting images from both the MOS and PN detectors. The PN background also has strong emission from Cu at $\sim$7.4 and $\sim$8.0 keV, so the 7.2-7.6 keV and 7.8-8.2 keV ranges were also excluded when extracting images from the PN detector. The PN OoT images were scaled by 0.063 to account for the loss of observing time due to photon detection during CCD readout, and were then subtracted from the PN images. Finally, MOS and PN images were extracted in the standard 0.5-2 keV, 2-10 keV and 0.5-10 keV ranges and were added among the detectors in each energy band.\footnote{We note that in some observations, only one or two detectors had data. See Table \ref{obs_summary}.} 

Exposure maps were generated using the SAS task {\em eexpmap} for each detector and energy range. Since vignetting, decrease in effective area with off-axis distance, increases as a function of energy, we created spectrally weighted exposure maps, i.e., the mean energy at which the maps were calculated was found assuming a spectral model where, consistent with previous {\it XMM-Newton} surveys \citep[e.g.][]{cap}, $\Gamma$=2.0 in the soft band and $\Gamma$=1.7 in the hard and full bands, since the specral slope of the soft band in AGN tends to be steeper than the hard band. The same spectral model was used to derive energy conversion factors (ECFs) to transform count rates to physical flux units, where the ECF depends on the filter for the observation and was calculated via PIMMS\footnote{http://heasarc.nasa.gov/Tools/w3pimms.html} (see Table \ref{ecf} for a summary). The exposure maps were added among the three detectors for each observation, normalized by these ECFs.\footnote{In observations where only 2 detectors were active instead of 3, the normalization was adjusted accordingly. No normalization was necessary for observations with only 1 detector.}

Two regions in Stripe 82 had multiple X-ray observations (ObsIDs 0056020301, 0312190401 and 0111200101, 0111200201). In order to detect sources from these overlapping observations simultaneously, the events files were mapped to a common set of WCS coordinates using SAS task {\em attcalc} to update the  `RA\_NOM' and `DEC\_NOM' header keywords. The subsequent data products (e.g., images, exposure maps, background maps, detector masks) then share common coordinates. Before running the source detection in `raster' mode (see Section 2.4), the header keywords `EXP\_ID' and `INSTRUME' for these files were updated to common values.

\clearpage
\begin{table}
\caption{\label{obs_summary}Archival {\it XMM-Newton} Observations in SDSS Stripe 82}
\begin{tabular}{lrrrr}
\hline
Obs. ID & R.A. & Dec & Detectors & Exp time \\
& & & &(ks)\\
\hline

0036540101 &  54.64 &    0.34 & MOS1,MOS2,PN & 21.77 \\
0041170101 &  45.68 &    0.11 & MOS1,MOS2,PN & 50.04 \\
0042341301$^{2}$ & 354.44 &    0.26 & MOS1,MOS2,PN & 13.36 \\
0056020301$^{2,3}$ &  44.16 &    0.08 & MOS1,MOS2,PN & 23.37 \\
0066950301$^{2}$ & 349.54 &    0.28 & MOS1,MOS2 & 11.45 \\
0084230401$^{2}$ &  28.20 &    0.99 & PN & 23.74 \\
0090070201$^{2}$ &  10.85 &    0.84 & MOS1,MOS2,PN & 20.53 \\
0093030201$^{2}$ & 322.43 &    0.07 & MOS1,MOS2,PN & 57.43 \\
0101640201$^{2}$ &  29.94 &    0.41 & MOS1,MOS2,PN & 10.54 \\
0111180201$^{2}$ & 310.06 &   -0.89 & MOS1,MOS2,PN & 16.31 \\
0111200101$^{1,2,3}$ &  40.65 &    0.00 & MOS1,MOS2 & 38.39 \\
0111200201$^{1,2,3}$ &  40.65 &    0.00 & MOS1,MOS2 & 37.99 \\
0116710901$^{2}$ &  54.20 &    0.59 & MOS1,MOS2 &  7.64 \\
0134920901 &  58.45 &   -0.10 & MOS1,MOS2,PN & 18.69 \\
0142610101$^{2}$ &  46.69 &    0.00 & PN & 65.89 \\
0147580401 & 356.88 &    0.88 & MOS1,MOS2,PN & 15.12 \\
0200430101 &  55.32 &   -1.32 & MOS1,MOS2,PN & 11.46 \\
0200480401 &  37.76 &   -1.03 & MOS1,MOS2,PN & 16.07 \\
0203160201$^{1,2}$ &  46.22 &    0.06 & MOS1,MOS2 & 15.08 \\
0203690101 &   9.83 &    0.85 & MOS1,MOS2,PN & 47.31 \\
0211280101$^{2}$ & 355.89 &    0.34 & MOS1,MOS2,PN & 40.68 \\
0303110401 &  14.07 &    0.56 & MOS1,MOS2,PN & 11.09 \\
0303110801 & 359.55 &   -0.14 & MOS1,MOS2,PN &  9.63 \\
0303562201 &  10.88 &    0.00 & MOS1,MOS2,PN &  6.57 \\
0304801201 & 323.39 &   -0.84 & MOS1,MOS2,PN & 13.27 \\
0305751001 &   1.20 &    0.11 & MOS1,MOS2,PN & 15.07 \\
0307000701 &  45.97 &   -1.12 & MOS1,MOS2,PN & 15.84 \\
0312190401$^{3}$ &  43.82 &   -0.20 & MOS1,MOS2,PN & 11.63 \\
0400570301 &  19.75 &    0.65 & MOS1,MOS2,PN & 25.94 \\
0401180101 & 331.47 &   -0.34 & MOS1,MOS2,PN & 40.13 \\
0402320201 &  53.64 &    0.09 & MOS1,MOS2,PN & 10.51 \\
0403760301 &   2.76 &    0.86 & MOS1,MOS2,PN & 25.46 \\
0407030101$^{2}$ &   5.58 &    0.26 & MOS1,MOS2,PN & 27.15 \\

\hline
\multicolumn{5}{l}{$^1$PN detector removed from analysis due to significant pile-up.}\\
\multicolumn{5}{l}{$^2$Detector mask manually updated to screen out regions of }\\
\multicolumn{5}{l}{pile-up and extended emission.}\\
\multicolumn{5}{l}{$^3$Overlapping observations that were run simultaneously through}\\
\multicolumn{5}{l}{source detection software: 0056020301 and  0312190401 grouped together;}\\
\multicolumn{5}{l}{0111200101 and 0111200201 grouped together.}
\end{tabular}
\end{table}

\begin{table}
\caption{\label{ecf} ECFs$^{1}$ for Each Detector and Filter$^2$}
\begin{tabular}{lllllll}
\hline
Band & PN & PN & PN & MOS & MOS & MOS \\
     & Thin & Medium & Thick & Thin & Medium & Thick \\
\hline

Soft (0.5-2 keV)   & 7.45 & 7.36 & 5.91 & 2.00 & 1.87 & 1.67 \\
Hard (2-10 keV)  & 1.22 & 1.24 & 1.19 & 0.45 & 0.42 & 0.43 \\
Full (0.5-10 keV) & 3.26 & 3.25 & 2.75 & 0.97 & 0.91 & 0.85 \\

\hline
\multicolumn{7}{l}{$^1$Energy conversion factors in units of counts s$^{-1}$/10$^{-11}$ erg cm$^{-2}$ s$^{-1}$.}\\
\multicolumn{7}{l}{$^2$Assuming a spectral model where N$_H=3\times 10^{20}$ cm$^{-2}$ and $\Gamma$=2.0}\\
\multicolumn{7}{l}{for the soft band and $\Gamma$=1.7 for the hard and full bands. We note that}\\
\multicolumn{7}{l}{for the PN detector, ECFs were adjusted to account for masking out}\\
\multicolumn{7}{l}{energy ranges corresponding to background emission lines, as described}\\
\multicolumn{7}{l}{in the text. For source detection, ECFs were summed among all}\\
\multicolumn{7}{l}{detectors turned on during the observation.}
\end{tabular}
\end{table}

\clearpage

\subsection{Proprietary Observations}
We were awarded 2 {\it XMM-Newton} mosaicked pointings in AO 10 (PI: C. Megan Urry, ObsIDs: 0673000101 (`Stripe 82 XMM field 1'), 0673002301 (`Stripe 82 XMM field 2')), covering $\sim$4.6 deg$^2$. With this observing strategy, each pointing has $\sim$4.56 ks of exposure time and is separated with 15$^{\prime}$ spacing. The exposure time in the regions with greatest overlap reaches a depth of $\sim$12 ks. The {\it XMM-Newton} mosaic procedure enables a relatively large region to be surveyed, in this case $\sim$2.5 deg$^2$ per mosaic, while minimizing overhead as after the first pointing, the EPIC offset tables do not need to be calculated (PN) and uploaded (MOS). Each mosaic was made up of 22 individual, overlapping pointings, for a total observing time of 240 ks between both mosaics.

We split the events files for the mosaicked observations into individual pseudo-exposures using the SAS task {\em emosaic\_prep}. Each pseudo-exposure is then reduced in the same way as the archival pointings, producing cleaned events files, spectrally weighted exposure maps and appropriately modeled background maps (see below). As with overlapping archival observations,  `RA\_NOM', `DEC\_NOM', `EXP\_ID' and `INSTRUME' were updated to common values, but `RA\_PNT' and `DEC\_PNT' also had to be set manually to reflect the center coordinates of each pointing for the point spread function (PSF) to be calculated correctly during source detection. One of the pointings from ObsID 0673002301 (pseudo-exposure field 22) was afflicted by flaring and consequently not used in the source detection. In total, approximately 4.6 deg$^2$ of Stripe 82 were covered in these observations. 



\subsection{Background Modeling}

Following \citet{cap}, we used the following algorithm to model the background. First we created detection masks for each detector in each energy band for each observation and then ran the SAS task {\em eboxdetect} with a low detection probability ($likemin$ = 4) to generate a preliminary list of detected sources. The positions of these sources were then masked out when generating the background maps. Regions of significant extended emission (radius $>$1$^{\prime}$), piled-up sources and read-out streaks were also masked out manually.

As noted by \citet{cap}, the background has two components: unresolved X-ray emission which comprises the cosmic X-ray background (CXB) and local particle and detector background. The former background is subject to vignetting while the latter is not. The residual area (i.e., regions where no sources are detected) was split into two parts based on the median of the effective exposure. Regions above the median, with low vignetting, are dominated by the CXB whereas the detector background becomes more important below the median effective exposure. We set up templates to account for these two components of the background:

\begin{equation} AM_{1,v} + BM_{1,unv} = C_1 \end{equation}
\begin{equation} AM_{2,v} + BM_{2,unv} = C_2, \end{equation}

\noindent where $M_{1,v}$ and $M_{2,v}$ are the vignetted exposure maps for the areas above and below the median effective exposure time, respectively; $M_{1,unv}$ and $M_{2,unv}$ are the unvignetted template exposure maps, and $C_{1}$ and $C_{2}$ are the background counts. We solve this system of linear equations for the normalizations $A$ and $B$. The vignetted and unvignetted exposure maps are normalized by $A$ and $B$ respectively and then added to obtain the background map for each detector and observation. The background maps among the multiple detectors were added, giving one background map per observation.

\subsection{\label{src_detect} Source Detection}

We ran the source detection algorithm using the combined images, exposure maps and background maps generated as described above. We created detector masks on the combined images using the SAS task {\em emask}. For 15 observations, we manually updated these masks to screen out regions of extended emission and piled-up sources and read out streaks, as noted in Section 2.3 (see Table \ref{obs_summary}). A preliminary list of sources was generated with the SAS task {\em eboxdetect}, which is a sliding box detection algorithm run in `map' mode, where source counts are detected in a 5$\times$5 pixel box with a low probability threshold ($likemin = 4$). The source list generated by {\em eboxdetect} is used as an input for the SAS task {\em emldetect} which performs a maximum likelihood point PSF fit to the source count distribution, using a likelihood threshold ($det\_ml$) of 6, where $det\_ml = -ln P_{\rm random}$, with $P_{\rm random}$ being the Poisson probability that a detection is due to random fluctuations. We ran {\em emldetect} with the option to fit extended sources, where the PSF is convolved with a $\beta$ model profile. All extended sources (i.e., $ext$ flag $>$0 in the {\em emldetect} outputted source list) are omitted from further analysis in this paper. 

For overlapping archival observations, {\em eboxdetect} and {\em emldetect} were run in `raster' mode, i.e., these tasks were run on an input list of images, exposure maps, detector masks and background maps, which as noted above were remapped to a common WCS grid. The source detection algorithm was run separately for the soft, hard and broad bands for the overlapping observations but simultaneously for the non-overlapping pointings; memory constraints precluded running {\em eboxdetect} and {\em emldetect} simultaneously for overlapping observations in multiple energy bands. The ECFs reported in Table \ref{ecf} are summed among the detectors turned on for each observation and given as input in the source detection algorithm, converting count rates into physical flux units.

The 22 pointings for each mosaicked observation could not be fit simultaneously for source detection due to computational memory constraints. Instead, each group of mosaicked pointings was split into sub-groups so that source detection was run on two adjacent `rows' in R.A. to accommodate overlapping pointings. Other than the pointings on the Eastern and Western edges of the mosaic, each R.A. row was included in two source detection runs to account for overlap and ensure the deepest possible exposures. Similar to the overlapping archival observations, the source detection was run separately for the soft, hard and full bands. From the source lists, we then generated a list of individual sources and searched for the inevitable duplicate identifications of the same source, since portions of every field were in more than one source detection fitting run. Similar to the algorithm used for the Serendipitous {\it XMM-Newton} Source Catalog to identify duplicates \citep{Watson}, if the distance between any two sources is less than $d_{\rm cutoff}$ (where $d_{\rm cutoff}$ = min(0.9$\times d_{nn,1}$,0.9$\times d_{nn,2}$,15$^{\prime\prime}$,3$\times$($\sqrt{ra\_dec\_err_{1}^2+sys\_err^2} +\sqrt{ra\_dec\_err_{2}^2+sys\_err^2}$), where $d_{nn}$ is the distance between a source and its nearest neighbor in that pointing, $ra\_dec\_err$ is the positional X-ray error returned by {\em emldetect}, and $sys\_err$ is the systematic positional error (taken to be 1$^{\prime\prime}$), we consider the sources to be the same. We then chose the source with the higher $det\_ml$ as the detection from which to derive the position, positional error, flux and flux error. We chose a maximum search radius of 15$^{\prime\prime}$ based in part from the results of the simulations and matching the input simulated list to the detected source list, with this threshold maximizing identification of counterparts while minimizing spurious associations.

To merge the separate soft, hard and full band source lists into one single source list for the archival overlapping and mosaicked observations, we identified duplicate sources using the method described above. The positions among (or between, for cases where a match was found in 2 rather than 3 bands) the bands were averaged and the positional errors were added in quadrature. In our final point source list, we remove extended objects (i.e., where $ext > 0$ as reported by $emldetect$) and only include the objects where $det\_ml \ge$15 (5$\sigma$ significance) in at least one of the energy bands, to reduce spurious identifications and assure our catalog contains reliable X-ray detections \citep[see][]{Mateos, Loaring}. As summarized in Table \ref{src_num}, we detected 2358 X-ray sources, of which 1607 were found in archival observations and 751 were discovered in our proprietary program. Of this total number, 182 were detected only in the full band, 261 were identified solely in the soft band and 18 in just the hard band.

\begin{table}
\caption{\label{src_num} Number of Detected {\it XMM-Newton} Sources$^{1}$}
\begin{tabular}{lrrr}
\hline
Band & Archival & Proprietary & Total \\
\hline

Soft (0.5-2 keV) & 1438 & 635 & 2073 \\
Hard (2-10 keV) &  432 & 175 &  607 \\
Full (0.5-10 keV)  & 1411 & 668 & 2079 \\
Total  & 1607 & 751 & 2358 \\

\hline
\multicolumn{4}{l}{$^1$The numbers for the individual bands refer to the}\\
\multicolumn{4}{l}{sources detected at $det\_ml \ge$ 15 in that band, while}\\
\multicolumn{4}{l}{the total band numbers indicate the sources detected}\\ 
\multicolumn{4}{l}{at $det\_ml \ge$ 15 in any given band.}
\end{tabular}
\end{table}

\subsection{Monte Carlo Simulations: Source Detection Reliability \& Survey Coverage}

To assess the source detection efficiency and the survey area as a function of limiting flux, we have performed detailed Monte Carlo simulations. First, we generated a list of random fluxes following a published Log$N$-Log$S$ distribution for each observation, using the fits to the XMM-COSMOS soft and hard bands number counts \citep{cap09} and the fit to the ChaMP full band number counts \citep{champ}. These simulated sources are placed in random positions across the detector. Using part of the simulator written for the {\it XMM-Newton} survey of the CDFS by \citet{ranalli}\footnote{https://github.com/piero-ranalli/cdfs-sim}, each input source list is convolved with the {\it XMM-Newton} PSF, generating simulated event lists for all detectors turned on during each observation. Similar to the procedure for the real data, images are extracted from these simulated events files and added among the detectors. The background map for each observation is added to the combined simulated image and then Poisson noise is added to the combined source image and background map to replicate real observations. The source detection on these simulated images is then executed in the same manner as the real data. We simulated 20 images per pointing, providing us with an adequate number of input and detected sources to gauge source detection reliability and assess survey sensitivity.

To estimate the fraction of spurious and confused sources, we compare the sources detected significantly from the simulations ($det\_ml \geq 15$) with the input source list. We consider a detected source within 15$^{\prime\prime}$ of an input source as a match. Any detected object lacking an input counterpart is deemed spurious. The fraction of spurious sources is 0.49\%, 0.37\%, and 0.20\% in the soft, hard and full bands, respectively. Following the prescription of \citet{cap}, a source is considered confused if $S_{\rm out}/(S_{\rm in} + 3\sigma_{\rm out}) > 1.5$, where $S_{\rm out}$ and $S_{\rm in}$ are the output and input fluxes of the counterparts and $\sigma_{out}$ is the error on the detected flux. We estimate our fraction of confused sources in the soft, hard, and full bands as 0.34\%, 0.23\%, and 0.34\%, respectively.

\begin{figure}
\subfigure[]{\includegraphics[scale=0.40,angle=90]{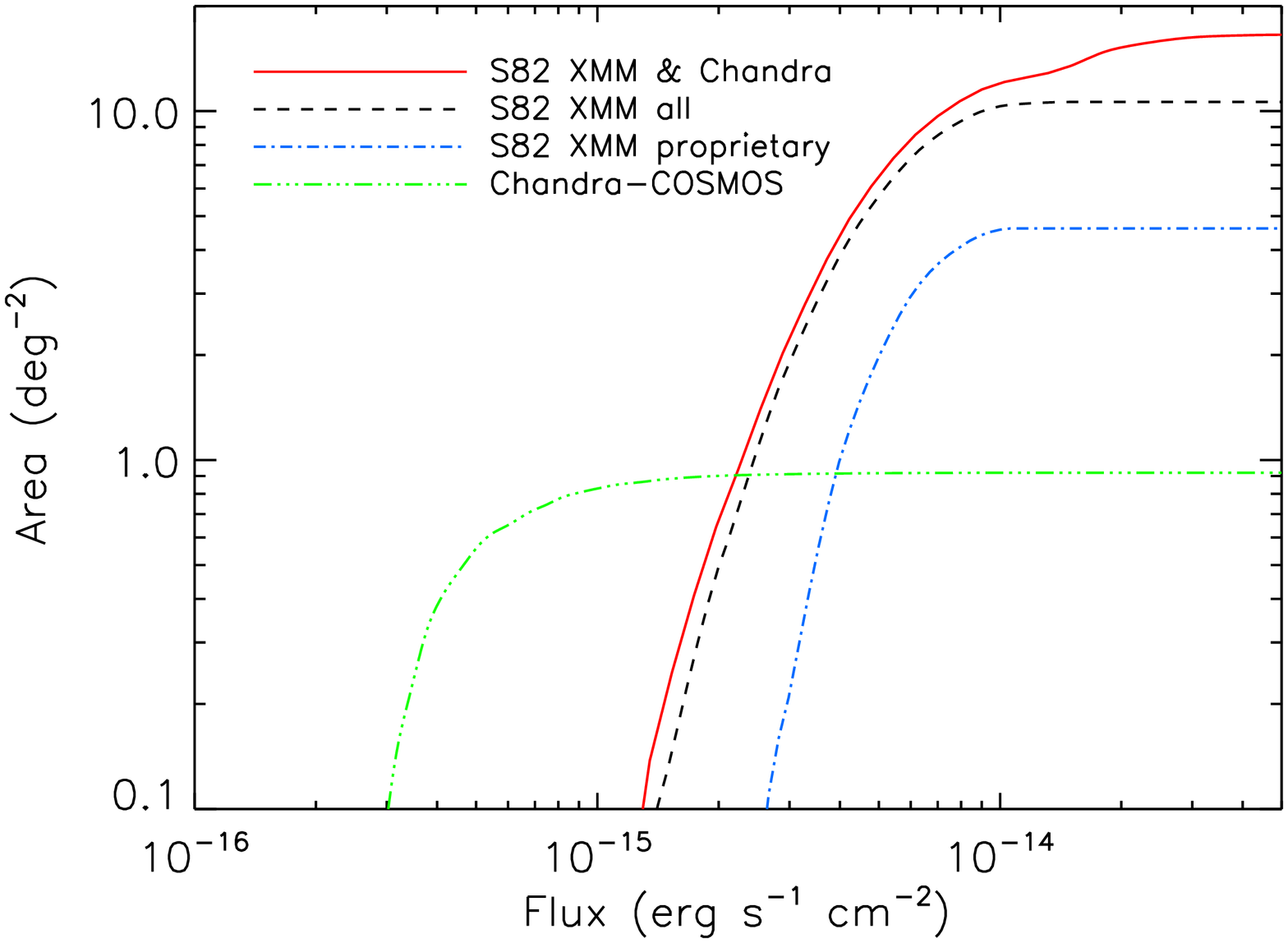}}
\subfigure[]{\includegraphics[scale=0.40,angle=90]{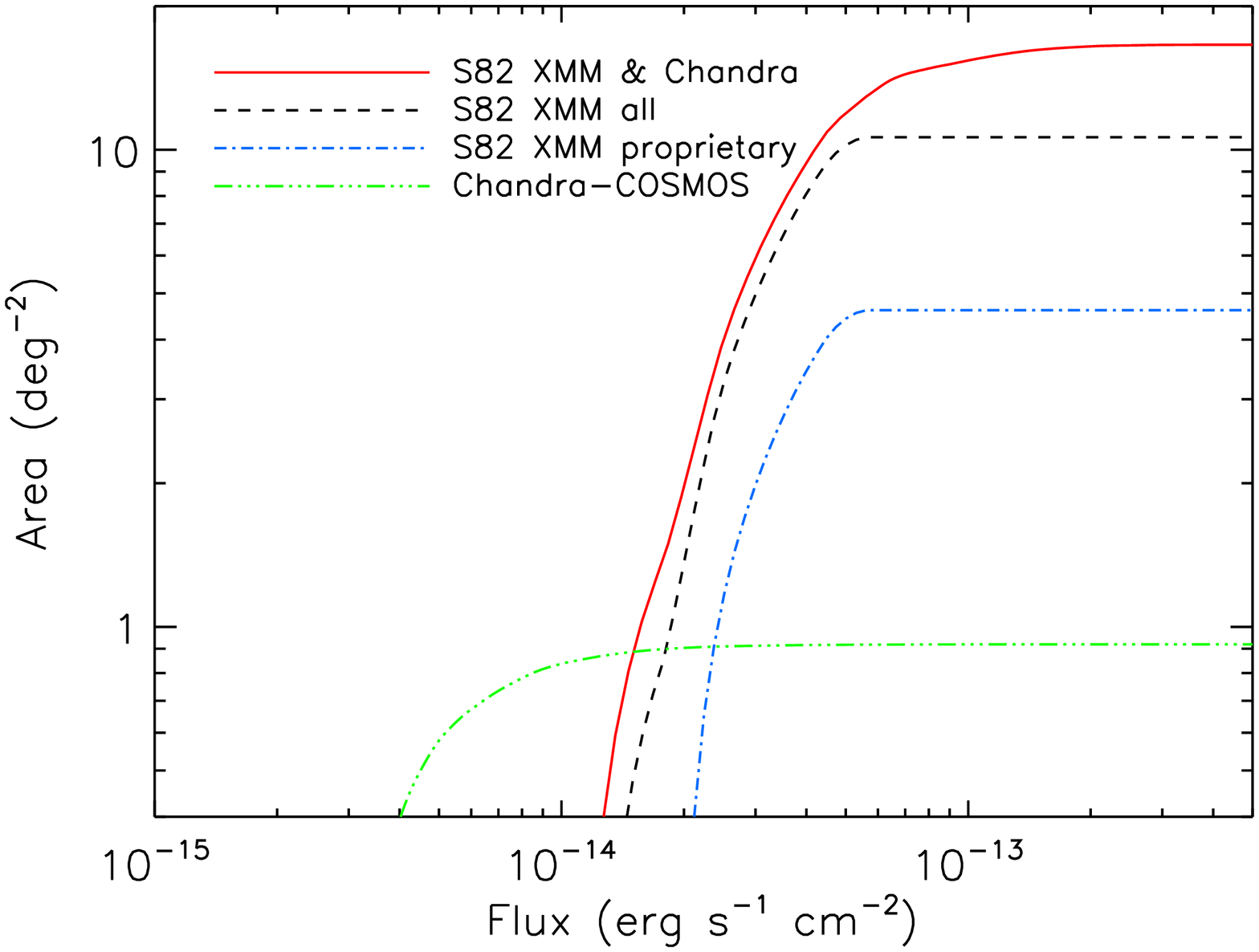}}
\subfigure[]{\includegraphics[scale=0.40,angle=90]{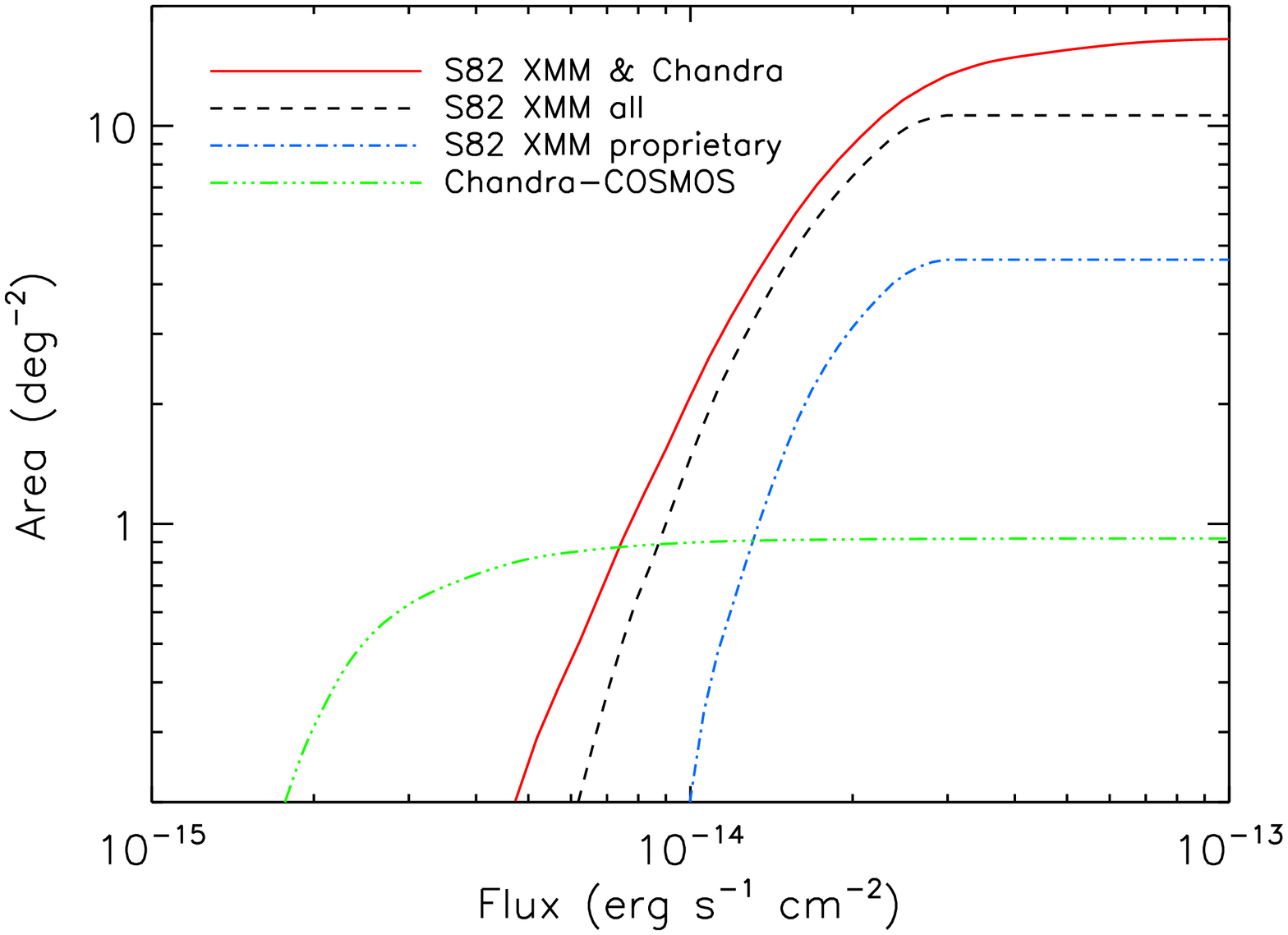}}
\caption[]{\label{area_flux}Area-flux curves for the Stripe 82 X-ray coverage and {\it Chandra}-COSMOS \citep{C-Cosmos} for comparison in the (a) soft, (b) hard, and (c) full bands. From the {\it XMM-Newton} proprietary + archival area-flux curves, we produced the Log$N$-Log$S$ relationships in Figure \ref{logn_logs}.}
\end{figure}

From these simulations, we also accurately gauge our survey sensitivity by determining the distribution of fluxes for both input and significantly detected sources. The ratio of these distributions as a function of flux provides us with the area-flux curves shown in Figure \ref{area_flux}, where we show the area-flux curves separately for the {\it XMM-Newton} proprietary data ($\sim$4.6 deg$^2$), proprietary and archival {\it XMM-Newton} data ($\sim$10.5 deg$^2$), {\it XMM-Newton} and {\it Chandra} coverage ($\sim$16.5 deg$^2$), and {\it Chandra}-COSMOS \citep[$\sim$0.9 deg$^2$][]{C-Cosmos} for comparison; we note that the fluxes in the {\it Chandra} hard (2-7 keV) and full (0.5-7 keV) bands were converted to 2-10 keV and 0.5-10 keV ranges using the assumed spectral models of $\Gamma$=1.7 for Stripe 82 and $\Gamma$=1.4 for {\it Chandra}-COSMOS. We reach down to approximate flux limits (at $\sim$0.1 deg$^2$ of coverage) of 1.4$\times10^{-15}$ erg s$^{-1}$ cm$^{-2}$, 1.2$\times10^{-14}$ erg s$^{-1}$ cm$^{-2}$ and 5.6$\times10^{-15}$ erg s$^{-1}$ cm$^{-2}$ with half-survey area at 4.7$\times10^{-15}$ erg s$^{-1}$ cm$^{-2}$, 3.1$\times10^{-14}$ erg s$^{-1}$ cm$^{-2}$ and 1.6$\times10^{-14}$ erg s$^{-1}$ cm$^{-2}$ in the soft, hard and full bands, respectively. From these curves, we then generate the number counts below.

\section{log$N$ - log$S$}
We present the number density of point sources as a function of flux, i.e., the log$N$ - log$S$ relation. In integral form, the cumulative source distribution is represented by:

\begin{equation} N(>S) = \sum_{i=1}^{N_s}  \frac{1}{\Omega_i}, \end{equation}
where N($>$S) is the number of sources with a flux greater than $S$ and $\Omega_{i}$ is the limiting sky coverage associated with the $i$th source. The associated error is the variance:

\begin{equation} \sigma^2 = \sum_{i=1}^{N_s}  (\frac{1}{\Omega_i})^2. \end{equation}

To avoid biasing our Log$N$-Log$S$ relations by the inclusion of targeted sources, we removed the closest object located within 30$^{\prime\prime}$ of the target R.A. and Dec, taken from $RA\_OBJ$ and $Dec\_OBJ$ in the FITS header. Of the 33 archival pointings, 18 had objects within 30$^{\prime\prime}$ of the nominal target positions. Three of these were not detected at a significant level (i.e., $det\_ml \geq 15$) in any given band and were not in our final source list. Thus, only 15 sources were excluded when generating the Log$N$-Log$S$. Of the remaining 15 archival pointings, 12 had central regions masked out due to extended emission or pile-up (presumably from the targeted source) while the other 3 had no sources detected within 30$^{\prime\prime}$ of the targeted position.

\begin{figure}
\subfigure[]{\includegraphics[scale=0.40,angle=90]{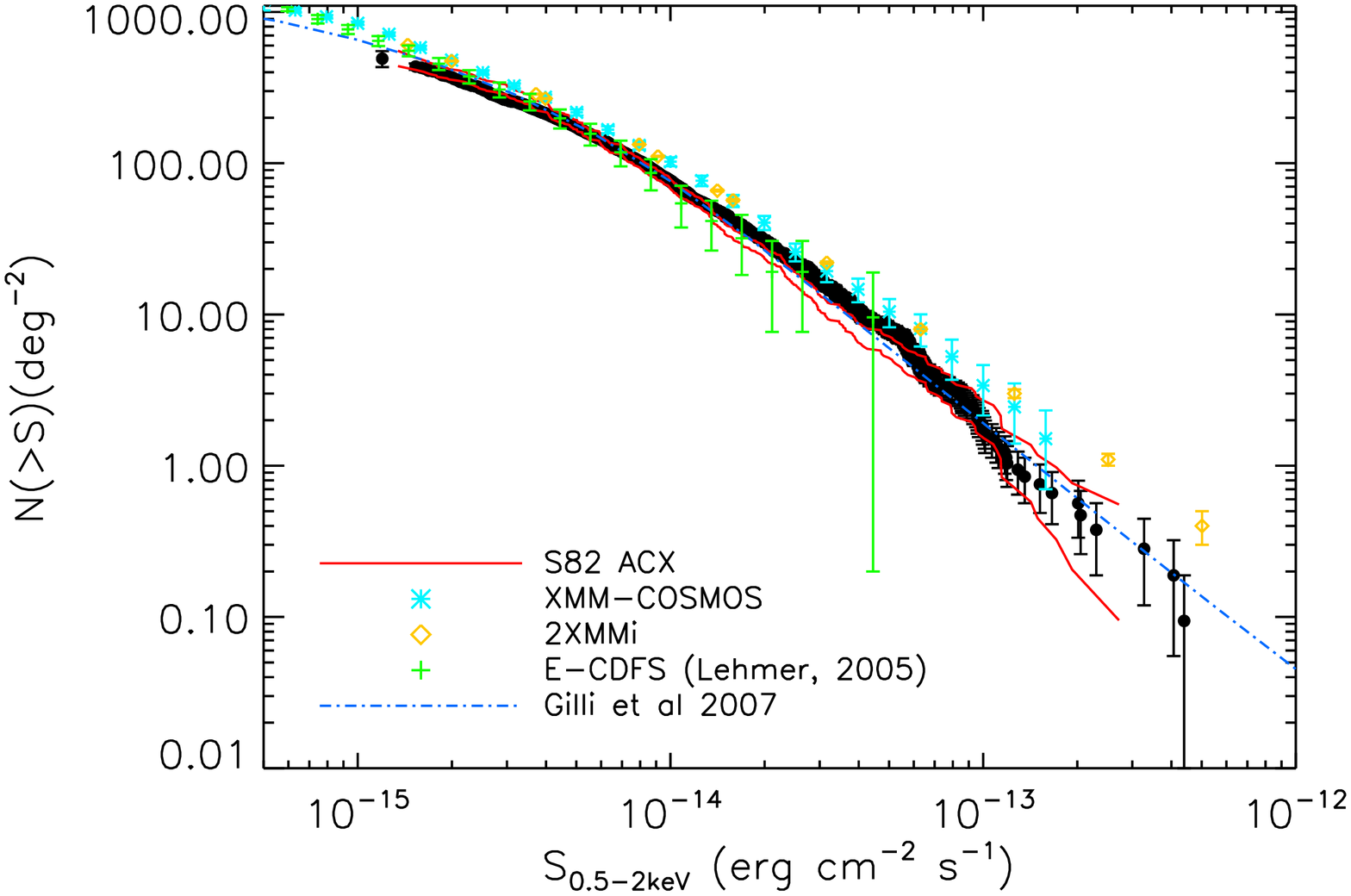}}
\subfigure[]{\includegraphics[scale=0.40,angle=90]{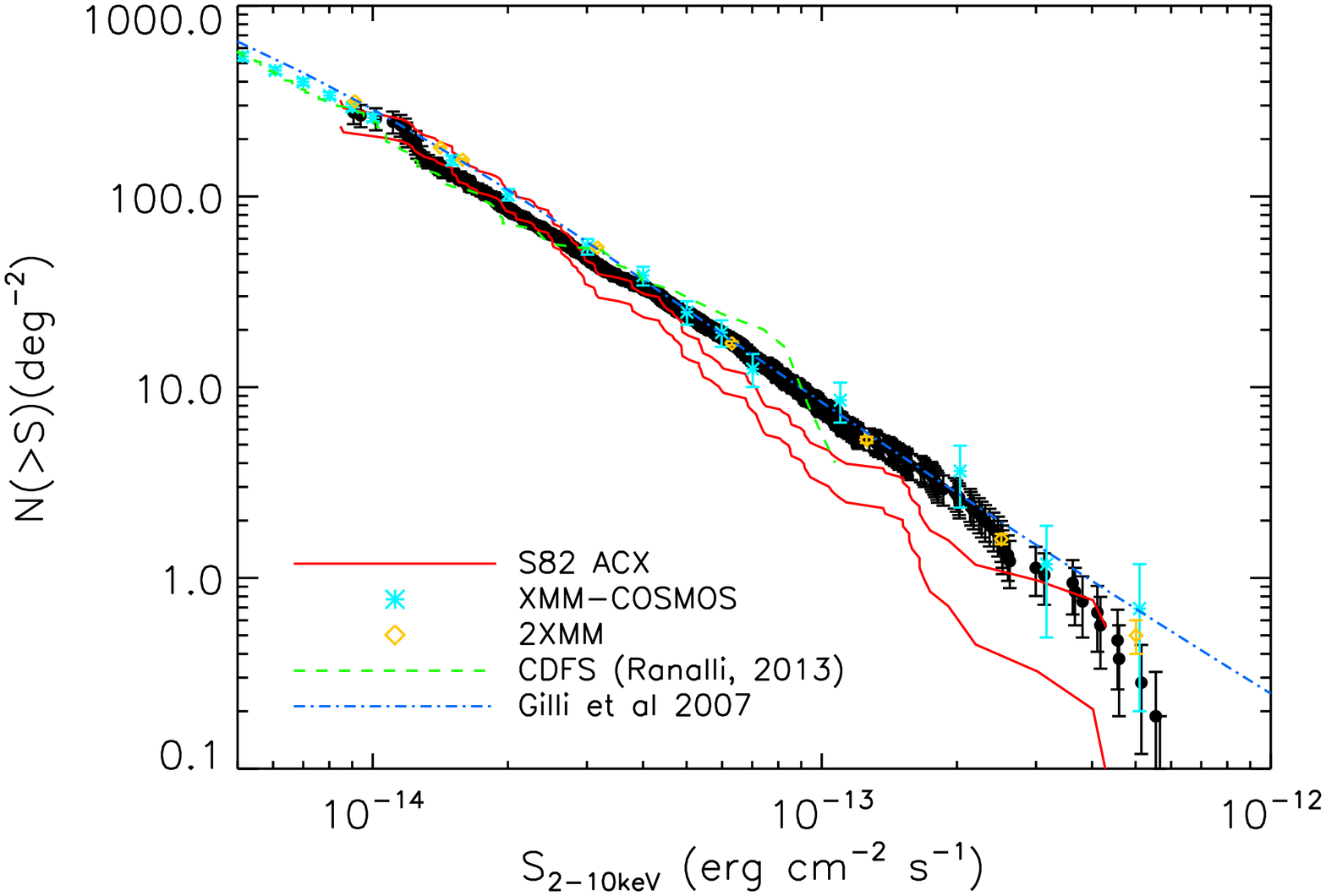}}
\subfigure[]{\includegraphics[scale=0.40,angle=90]{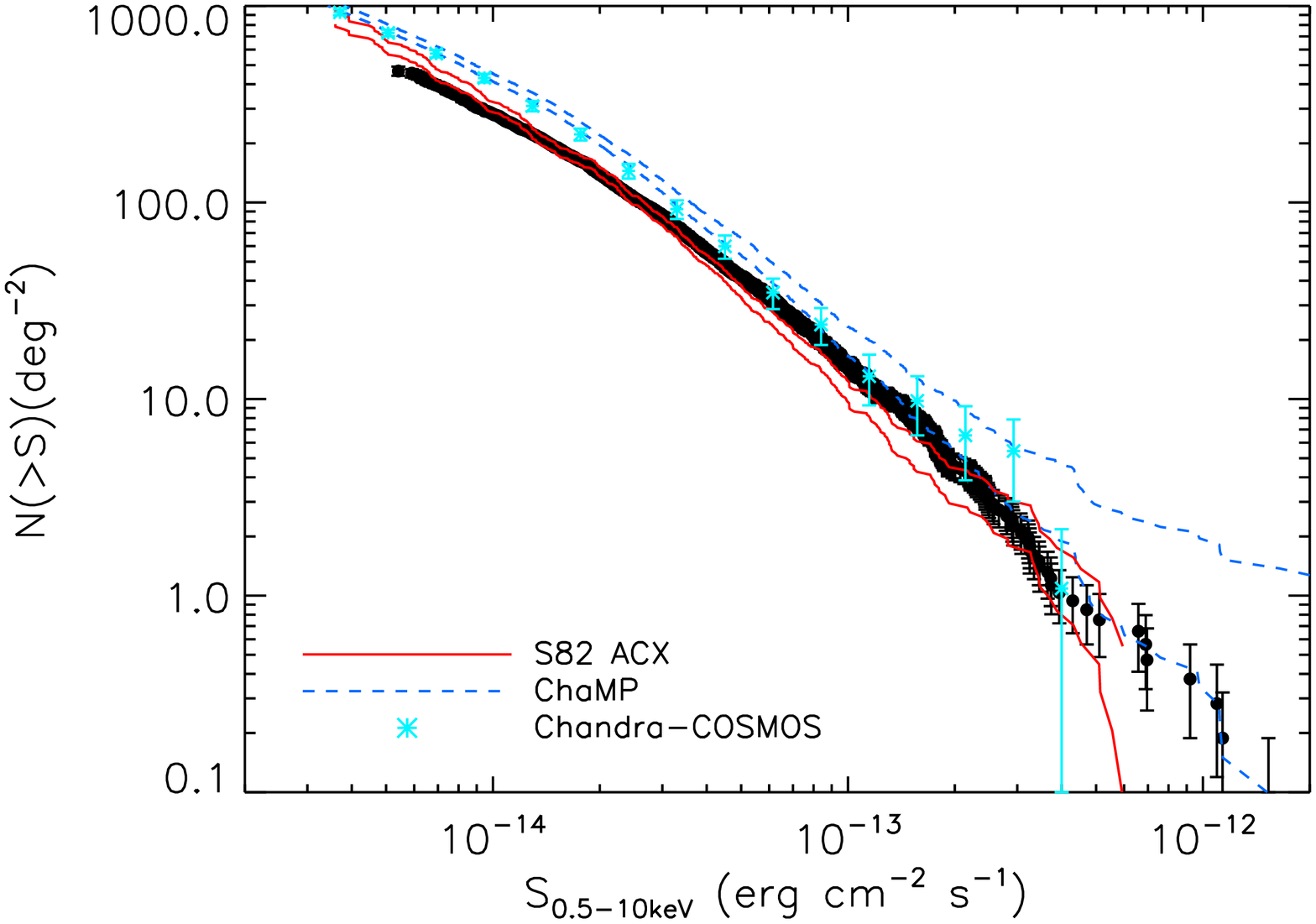}}
\caption[]{\label{logn_logs} Stripe 82 {\it XMM-Newton} number counts (filled black circles) in the (a) soft, (b) hard and (c) full bands, with the 1$\sigma$ confidence interval from Stripe 82 Archival {\it Chandra} (S82 ACX) overplotted in red. Comparison X-ray surveys are overplotted, ranging from small to moderate to wide area. There is general agreement; see text for discussion.}
\end{figure}

The number counts in the soft, hard and full bands are shown in Figure \ref{logn_logs}. We have also overplotted the upper and lower bounds of the {\it Chandra} Log$N$-Log$S$ from Stripe 82 (S82 ACX) for comparison, where we have re-calculated the source fluxes and survey sensitivity from \citet{me} using the same spectral model applied to the {\it XMM-Newton} data. We note that 12 {\it Chandra} non-cluster pointings used for generation of the Log$N$-Log$S$ presented in \citet{me} at least partially overlap the {\it XMM-Newton} observations, $\sim$1.2 deg$^2$. Since the hard and full bands are defined in S82 ACX up to 7 keV, the {\it Chandra} fluxes have been adjusted assuming a powerlaw model of $\Gamma$=1.7 to convert to the energy ranges used in our {\it XMM-Newton} analysis (i.e., the {\it Chandra} fluxes have been multiplied by factors of 1.36 and 1.2 for the hard and full bands, respectively). The {\it XMM-Newton} and S82 ACX number counts are largely consistent, with slight discrepancies apparent at moderate fluxes in the hard band ($\sim 5\times10^{-14}$ erg cm$^{-2}$ s$^{-1} <$ S$_{\rm 2-10keV} < 2\times10^{-13}$ erg cm$^{-2}$ s$^{-1}$) and at the low {\it XMM-Newton} flux limit in the full band ($< 10^{-14}$ erg cm$^{-2}$ s$^{-1}$). However, as noted in \citet{me}, short exposure times in {\it Chandra} observations, which constitute the majority of Stripe 82 ACX, has an effect on the Log$N$-Log$S$ normalization in the hard band, making the offset between {\it XMM-Newton} and {\it Chandra} in this energy range unsurprising. 

\begin{figure}
\subfigure[]{\includegraphics[scale=0.40,angle=90]{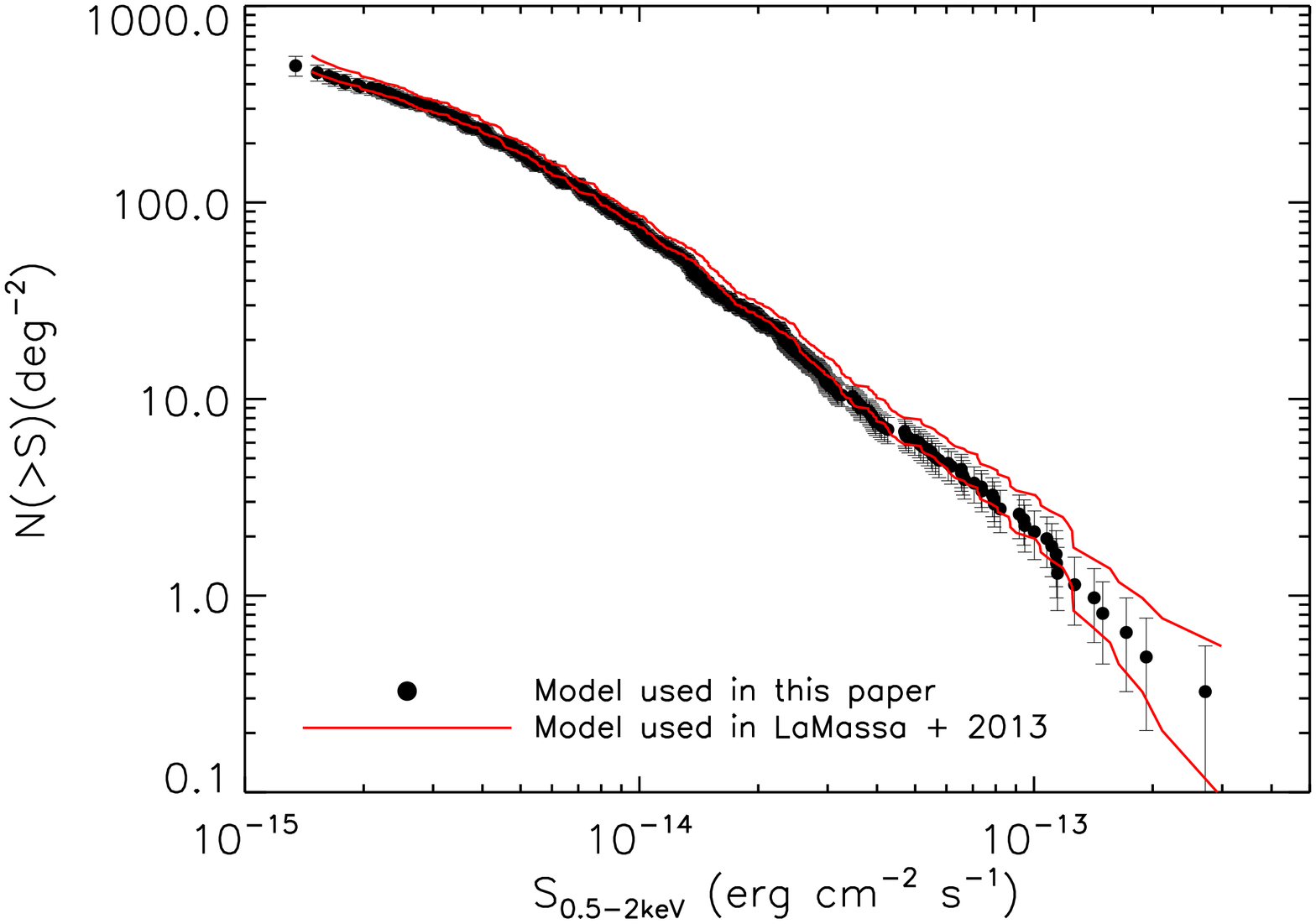}}
\subfigure[]{\includegraphics[scale=0.40,angle=90]{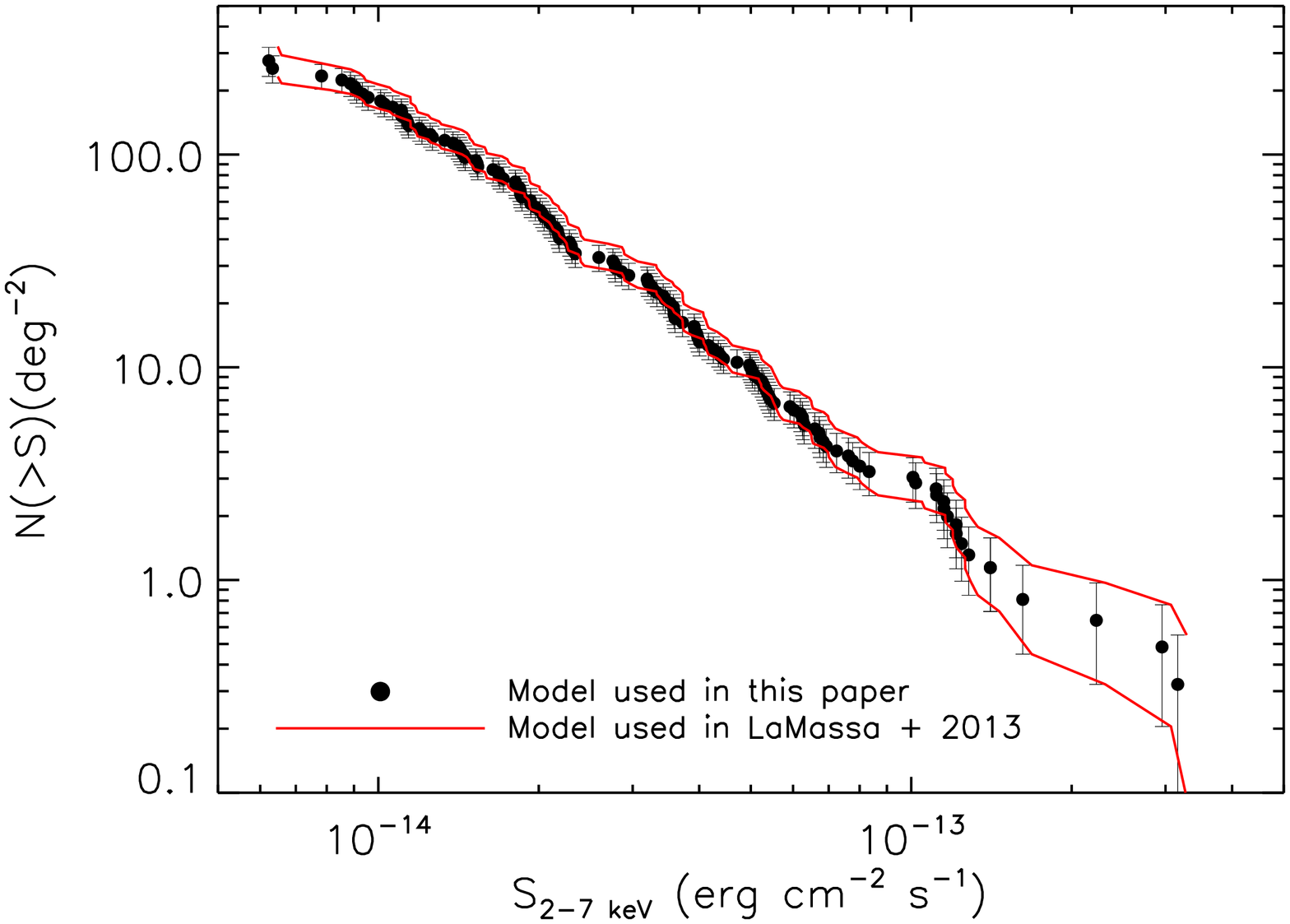}}
\subfigure[]{\includegraphics[scale=0.40,angle=90]{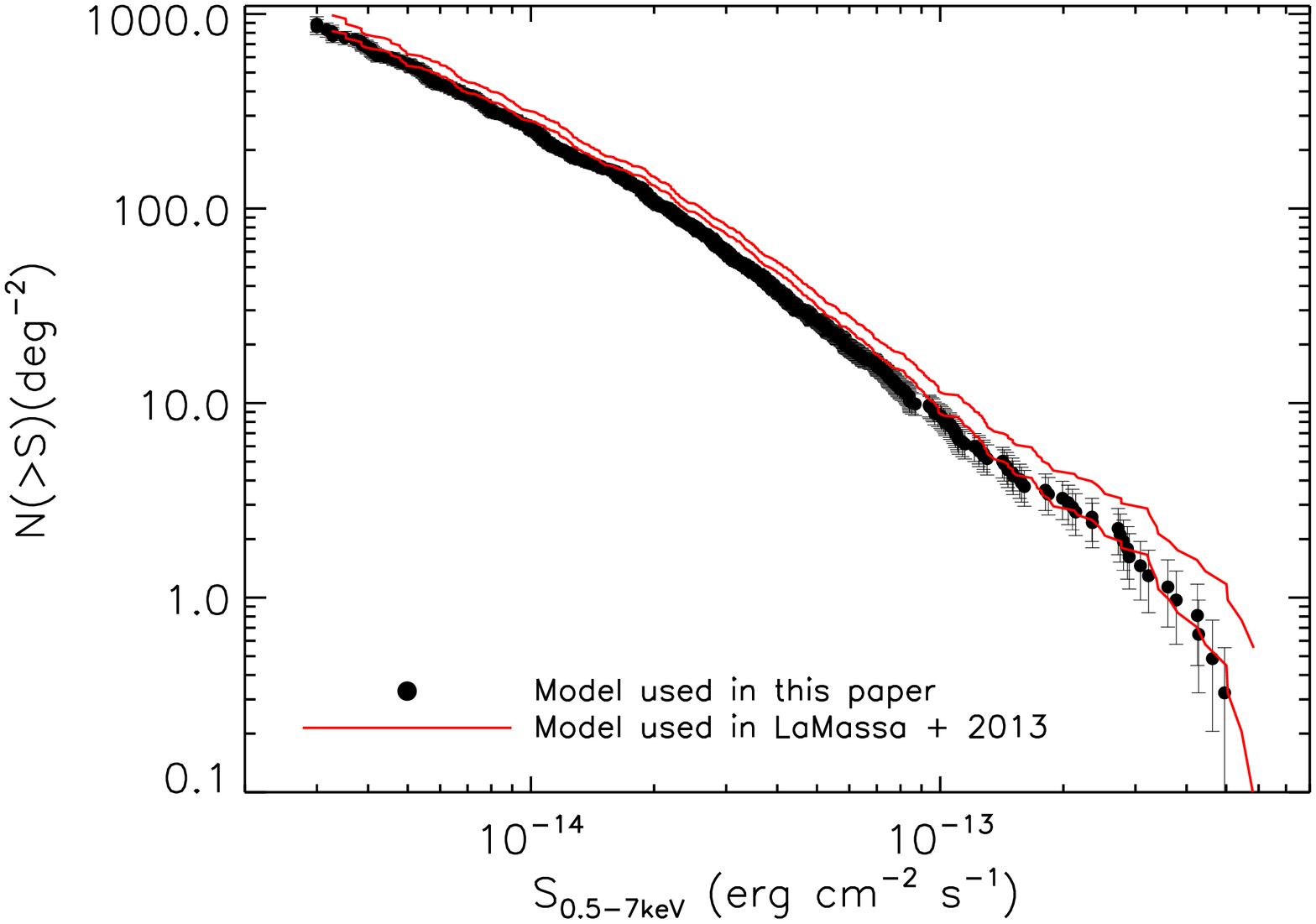}}
\caption[]{\label{comp_logns} Comparison of the {\it Chandra} number counts in the (a) soft, (b) hard and (c) full bands using the assumed spectral model from \citet{me}, $\Gamma=1.4$ in all bands (1$\sigma$ confidence interval shown by red lines) and the spectral model adopted for the {\it XMM-Newton} data, $\Gamma$=2 in the soft band and $\Gamma=1.7$ in the hard and full bands (black filled circles). The spectral model assumed for this paper shifts the number counts normalization to lower values with respect to the \citet{me} results, with the most significant offset (i.e., where the error ranges do not overlap) in the full band, with a median discrepancy of $\sim$10\%.}
\end{figure}

In Figure \ref{comp_logns}, we compare the Stripe 82 ACX Log$N$-Log$S$ using the spectral model from \citet{me} and the one used here. In \citet{me}, we adopted a spectral model used in {\it Chandra} surveys to which we compared our results while here we used a spectral model consistent with previous {\it XMM-Newton} surveys, such as {\it XMM}-COSMOS \citep{cap}. The difference in the hard band number counts is slight with this change of assumed spectral model, but shifts the normalization to lower values in the soft and especially the full band where the median offset between the 1$\sigma$ error bars in the discrepant ranges is $\sim$10\%. 

We also compare our Log$N$-Log$S$ relationships with those from previous X-ray surveys, spanning from wide \citep[2XMMi, 132 deg$^2$;][]{Mateos} to moderate \citep[XMM-COSMOS, 2 deg$^2$;][]{cap, cap09} to small areas \citep[E-CDFS, 0.3 deg$^2$, {\it XMM}-CDFS, $\sim$0.25 deg$^2$;][]{lehmer, ranalli}. Where possible, we aim to compare our data with other {\it XMM-Newton} surveys. However, the {\it XMM-Newton} survey in the CDFS \citep{ranalli} only produced the Log$N$-Log$S$ in the hard band, so we use the {\it Chandra} E-CDFS survey \citep{lehmer} for comparison in the soft band. No previous {\it XMM-Newton} survey has produced a full band Log$N$-Log$S$, so we compare our Stripe 82 number counts with the small area {\it Chandra}-COSMOS \citep[C-COSMOS, 0.9 deg$^2$;][]{C-Cosmos} and wide area ChaMP surveys \citep[9.6 deg$^2$;][]{champ}. As ChaMP defines the full band to be 0.5-8 keV, their fluxes were adjusted to match our 0.5-10 keV range using their adopted spectral model (i.e., multiplied by a factor of 1.18). We note that the spectral shapes over a broad band are not well constrained, making the energy conversion factors in this range approximate and comparisons with number counts using other model assumptions difficult to quantify; these comparisons are for illustrative purposes. The model predictions from \citet{Gilli} have also been overplotted in the soft and hard bands.

The Stripe 82 {\it XMM-Newton} number counts are consistent with previous {\it XMM-Newton} surveys in the hard band. The 2XMM Log$N$-Log$S$ from \citet{Mateos} is systematically higher than our data in the soft band. However, they note that their 0.5-2 keV number counts are higher than several other X-ray surveys, which they attribute to the inclusion of moderately extended sources in their catalog. Similar to other surveys, we include only point sources, making this soft band discrepancy with \citet{Mateos} not surprising. Stripe 82 {\it XMM-Newton} is fully consistent with E-CDFS \citep{lehmer} and the model predictions from \citet{Gilli} in this energy range. Though the normalization for the full band Stripe 82 {\it XMM-Newton} Log$N$-Log$S$ seems low compared to ChaMP and C-COSMOS, this is likely due to differences in spectral models to convert from count rate to fluxes: ChaMP and C-COSMOS adopt a powerlaw model with $\Gamma=1.4$ whereas we use $\Gamma=1.7$. As shown in Figure \ref{comp_logns} (c), the difference between these two spectral models shifts the full band number counts in the right sense to account for the observed disagreement between Stripe 82 {\it XMM-Newton} and C-COSMOS and ChaMP. We also note that the ChaMP number counts seem to be somewhat higher than other {\it Chandra} surveys \citep{me} while C-COSMOS shows better agreement with our calculations.

As we show below, these X-ray objects do preferentially sample the high luminosity AGN population and include candidates for interesting rare objects: reddened quasars and high luminosity AGN at high redshift. In a future paper, we will quantify the evolution of these sources by generating the quasar luminosity function, beginning with the Log$N$-Log$S$ relations presented here.

\section{Multi-wavelength Source Matching via Maximum Likelihood Estimator}
The Stripe 82 X-ray source lists represent the {\it XMM-Newton} objects found above and the {\it Chandra} sources detected at $\geq 4.5\sigma$ level from all pointings overlapping the Stripe 82 area. In \citet{me}, we presented only those observations that did not target galaxy clusters, covering an area of $\sim$6.2 deg$^2$, garnering 709 objects. Inclusion of the previously omitted {\it Chandra} pointings adds an additional 1.2 deg$^2$ to produce a total of 1146 X-ray sources. About 1.5 deg$^2$ of the full 7.4 deg$^2$ of {\it Chandra} coverage in Stripe 82 overlaps the {\it XMM-Newton} pointings. Using the method described above to find duplicate observations of the same X-ray object, we cross-matched the {\it XMM-Newton} and {\it Chandra} source lists, finding 3362 unique objects over $\sim$16.5 deg$^2$ of non-overlapping area.

To assign multi-wavelength counterparts to the Stripe 82 X-ray sources, we employed a maximum likelihood estimator (MLE) algorithm which takes into account the distance between potential matches and the brightness of the ancillary counterpart \citep{mle}. The ancillary source at the closest distance to the X-ray object, as found using the nearest neighbor method, may not be the true match, but may instead be a spurious association due to random chance. As there are many more faint than bright objects, an association between a bright source and an X-ray target is more likely to represent a true counterpart than a match to a faint source. The MLE technique codifies this statistically, assigning reliability values to each potential match and has been successfully implemented in multi-wavelength catalog matching in previous X-ray surveys \citep[e.g.,][]{brusa1,brusa2,cardamone,luo,brusa3,civano_mle}.

All the objects within a search radius ($r_{\rm search}$) around each X-ray target are assigned a likelihood ratio ($LR$), which is the probability that the correct counterpart is found within $r_{\rm search}$ divided by the probability of finding an unassociated object by chance: 

\begin{equation} LR = \frac{q(m)f(r)}{n(m)}, \end{equation}

\noindent where $q(m)$ is the expected normalized magnitude distribution of ancillary counterparts, $f(r)$ is the probability distribution of the positional errors (which is assumed to be a two-dimensional Gaussian, where $\sigma$ is derived by adding the X-ray and ancillary positional errors in quadrature), and $n(m)$ is the magnitude distribution of background sources. For the positional {\it Chandra} uncertainty, we added the major and minor axes of the 95\% confidence level error ellipse, $err\_ellipse\_r0$ and $err\_ellipse\_r1$, in quadrature, while {\it XMM-Newton} positional errors are from the {\it emldetect} source detection script added in quadrature to a 1$^{\prime\prime}$ systematic error.\footnote{This systematic uncertainty takes into account that we used the coordinates as reported from {\it emldetect} as our attempt to use {\it eposcorr} to correct systematic astrometric offsets was unsuccessful, introducing different systematic offsets. The 1$^{\prime\prime}$ systematic error used here is consistent with the {\it XMM-Newton} Serendipitious Catalog procedure for estimating positional uncertainty for sources lacking independent astrometric corrections \citep{Watson}.}  As noted below, for ancillary catalogs where a positional error is not quoted, we adopted a uniform, survey dependent, positional uncertainty. Since {\it Chandra} has higher resolution and a smaller on-axis PSF than {\it XMM-Newton}, we chose different radii to search for ancillary counterparts for each catalog. For {\it Chandra} objects, $r_{\rm search}$ = 5$^{\prime\prime}$ \citep{civano_mle} while for {\it XMM-Newton} sources, $r_{\rm search}$ = 7$^{\prime\prime}$ \citep{brusa3}; the positional errors for 88\% of {\it Chandra} and 99.9\% of {\it XMM-Newton} sources are below the adopted search radii.

To determine the background distribution $n(m)$, we isolate the ancillary sources within an annulus around each X-ray source, with inner and outer radii of 7$^{\prime\prime}$ and 30$^{\prime\prime}$ for {\it Chandra} and 10$^{\prime\prime}$ and 45$^{\prime\prime}$ for {\it XMM-Newton}. The inner radius is chosen to avoid the inclusion of real counterparts and the outer radius is picked to ensure a large number of sources to estimate the background while minimizing overlap with other X-ray sources. Within these annular regions, there were 53 {\it Chandra} pairs and 7 {\it Chandra} triples ($\sim11\%$ of the sample) and 49 {\it XMM-Newton} pairs and 3 {\it XMM-Newton} triples ($\sim4\%$ of the sample), i.e., only a small fraction of the background histogram has duplicate objects.

We then calculate $q^\prime(m)$ by first finding the magnitude distribution of ancillary objects within $r_{\rm search}$ of each X-ray source and dividing by the area to obtain the source density magnitude distribution. Similarly, we divide $n(m)$ by the search area, and take the difference between the former and latter which gives us the expected source density magnitude distribution. Finally, multiplying this distribution by the search area gives us $q^\prime(m)$. We then normalize $q^\prime(m)$ to $Q$, the ratio of the number of X-ray sources with counterparts found within $r_{\rm search}$ to the total number of X-ray sources, producing $q(m)$ \citep[see][]{civano_mle}.

From $LR$, we then calculate a reliability value for each source:

\begin{equation} R = \frac{LR}{\Sigma_i (LR)_i + (1 - Q)}, \end{equation}

\noindent where the sum over $LR$ is for each possible counterpart found within $r_{\rm search}$ around an individual X-ray source. We use $R$ to discriminate between true counterparts and spurious associations. Since $R$ depends on the source density and magnitude distribution of the ancillary sources, the critical $R$ value ($R_{\rm crit}$) we adopt to accept a match as `real' differs among catalogs and strikes a fine balance between missing true counterparts and adding contamination from chance proximity to an unrelated source. To calibrate $R_{\rm crit}$, we shifted the positions of the X-ray sources by random amounts, with offsets ranging from $\sim$21$^{\prime\prime}$ to $\sim$35$^{\prime\prime}$, and re-ran the matching code. Any matches found should be due to random chance. We then plotted the distribution of reliability values for these spurious associations to estimate the contamination above $R_{\rm crit}$; full details regarding the estimate of false matches are given in Appendix A. We impose a lower limit on $R_{\rm crit}$ of 0.5, even in the cases where the reliability values for the shifted X-ray positions are consistent with zero. If there were multiple counterparts per X-ray source, or multiple X-ray sources per counterpart, the match with the highest reliability was favored. 

\begin{figure}
\center
\subfigure[]{\includegraphics[scale=0.2,angle=90]{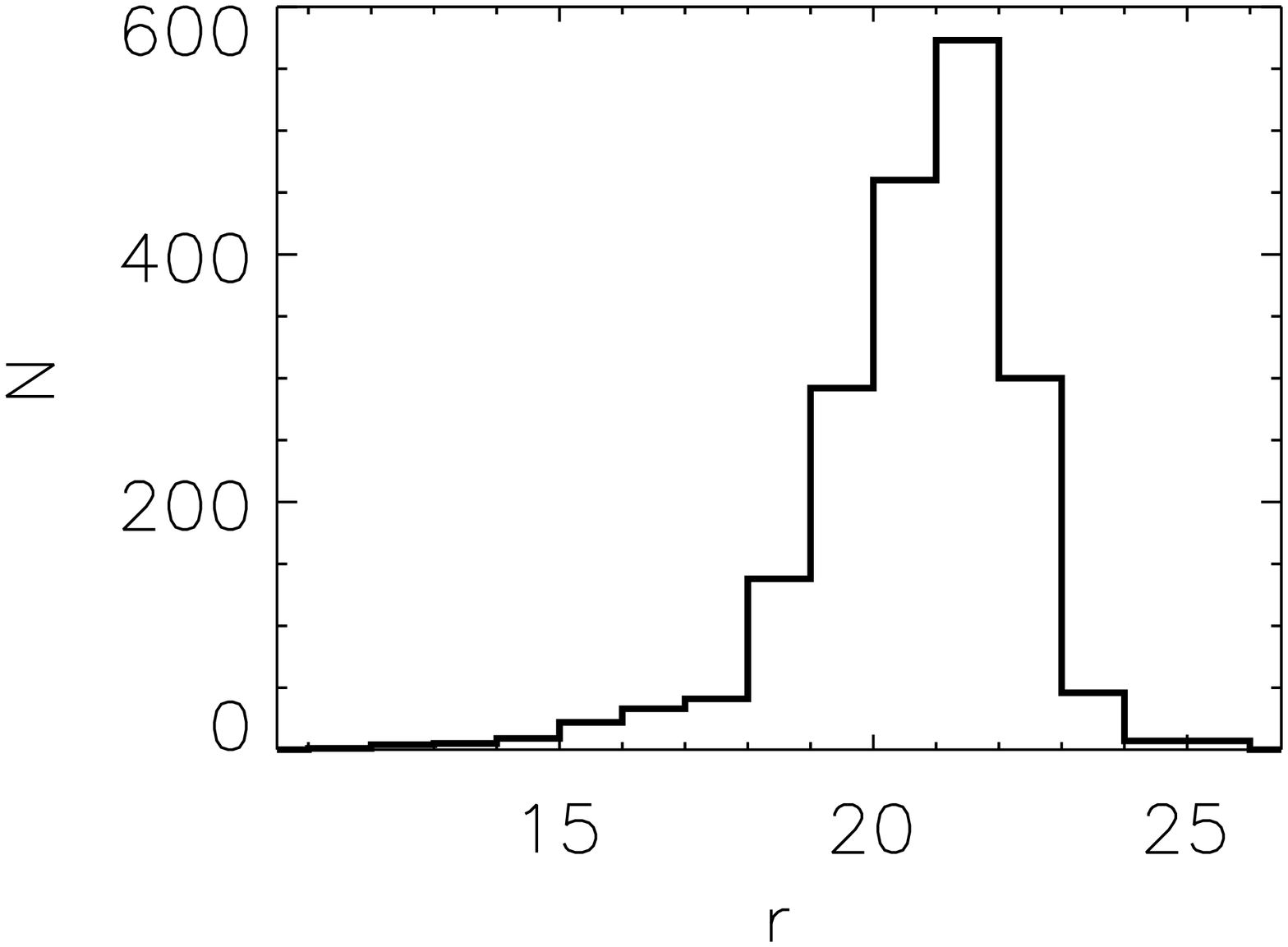}}~
\subfigure[]{\includegraphics[scale=0.2,angle=90]{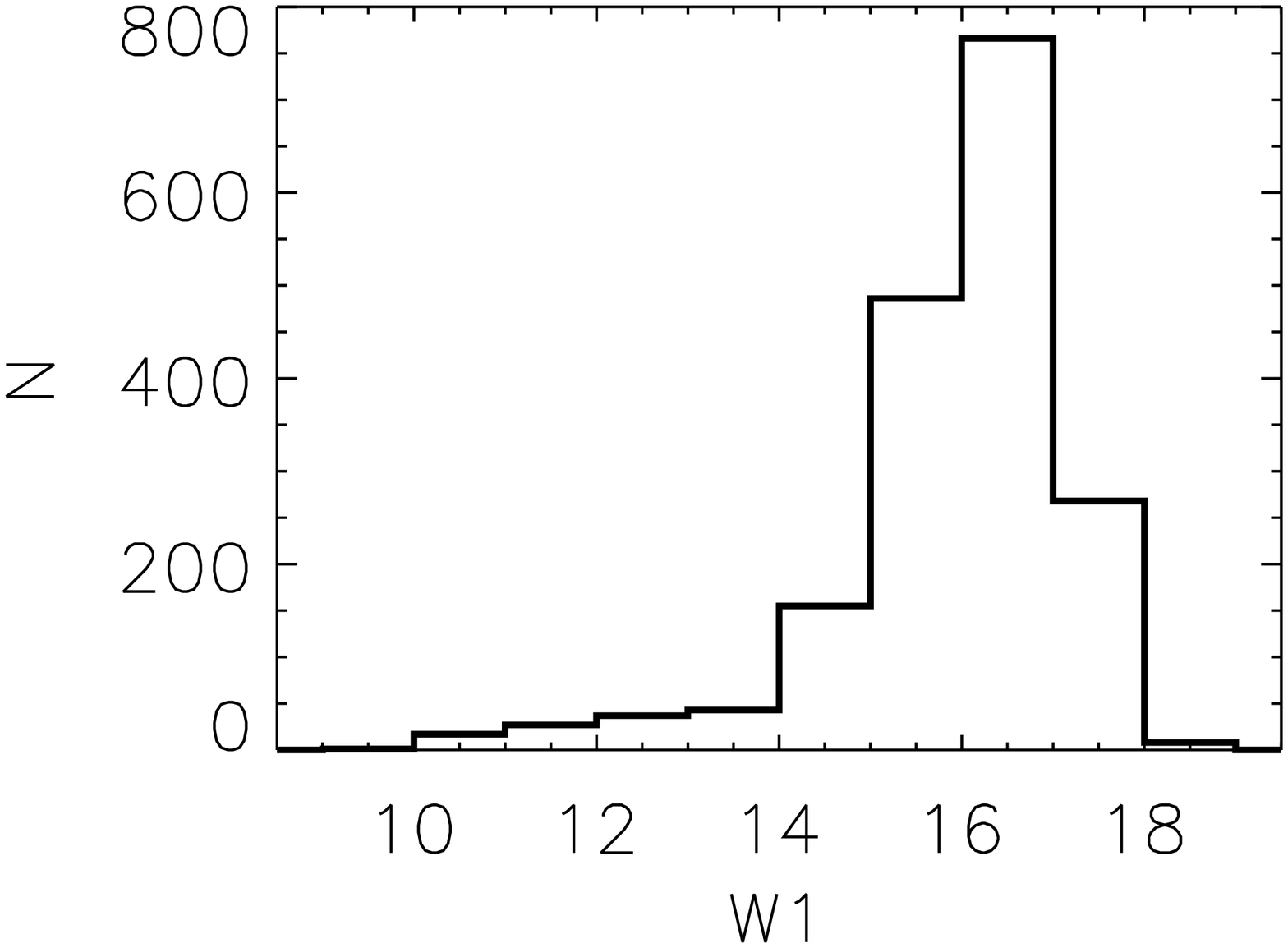}}
\subfigure[]{\includegraphics[scale=0.2,angle=90]{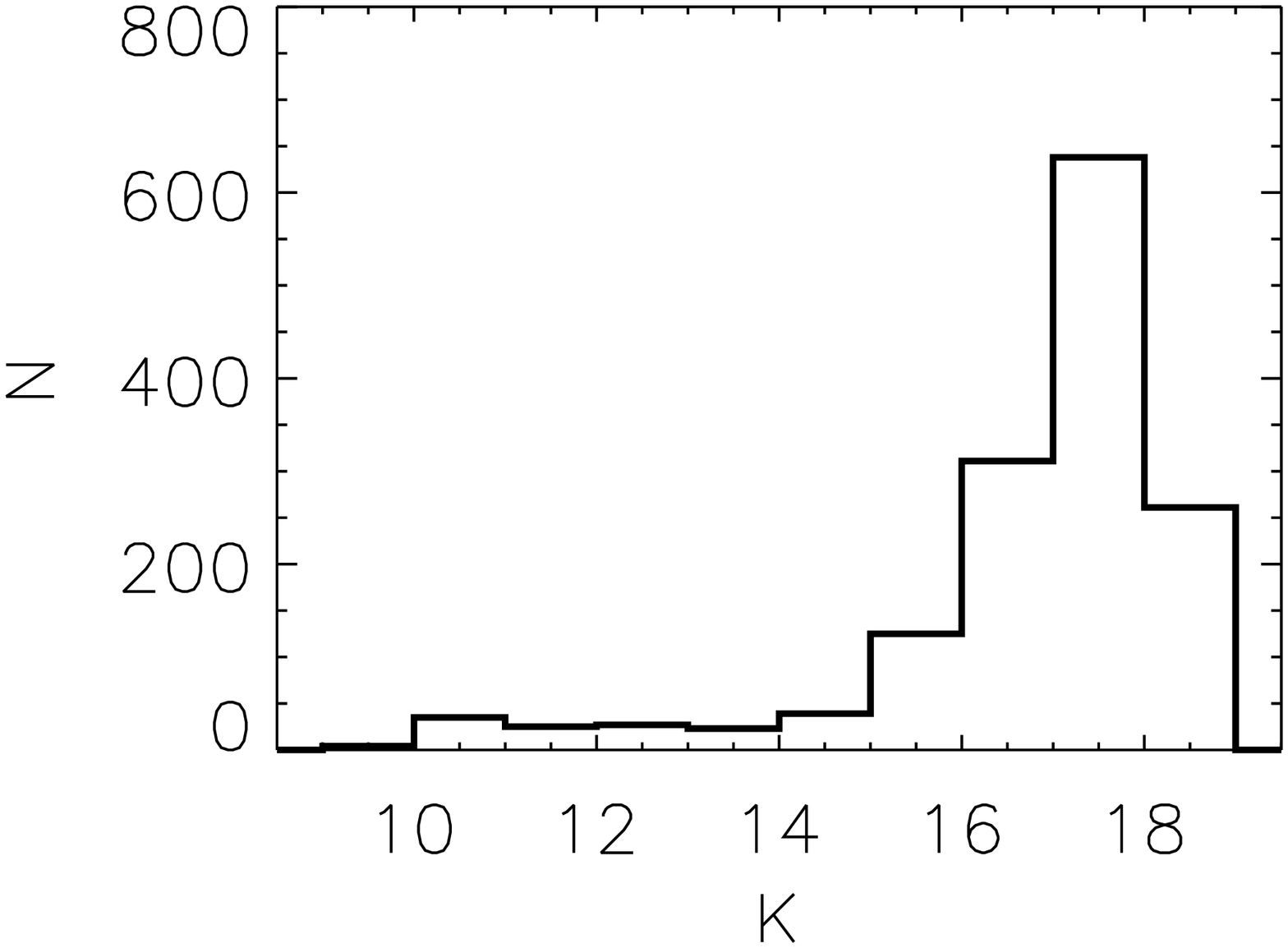}}~
\subfigure[]{\includegraphics[scale=0.2,angle=90]{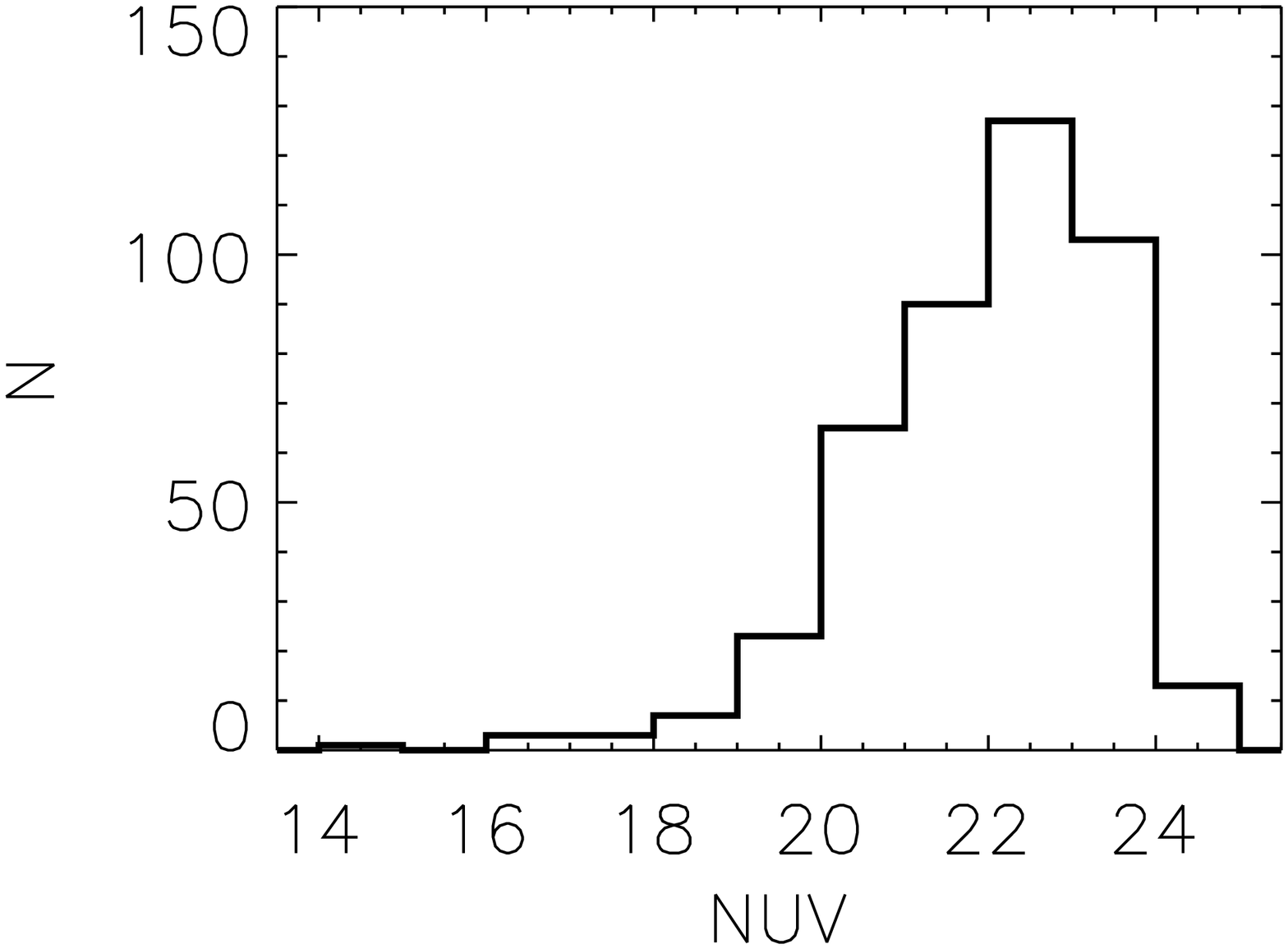}}
\subfigure[]{\includegraphics[scale=0.2,angle=90]{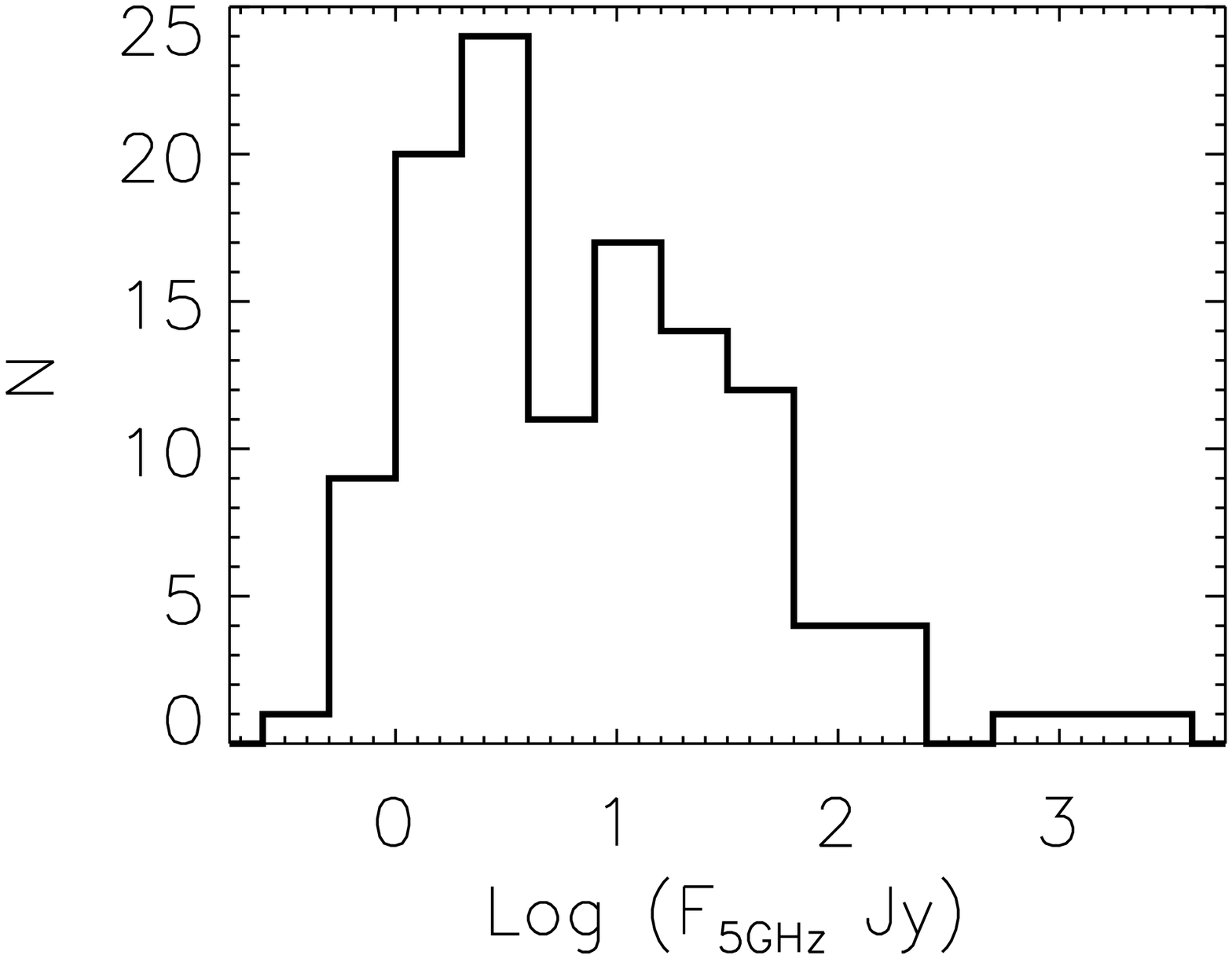}}
\caption[]{\label{mag_distr} Magnitude/flux density distributions of X-ray counterparts from (a) SDSS, (b) {\it WISE}, (c) UKIDSS, (d) {\it GALEX} and (e) FIRST.}
\end{figure}

In the on-line catalogues \cite[available at CDS and searchable with VizieR,][]{vizier}, we list the X-ray sources, fluxes and matches to the ancillary multi-wavelength catalogs, including the non-aperture matched photometry. Duplicate observations of the same X-ray object between the {\it Chandra} and {\it XMM-Newton} source lists are marked in the on-line tables. Objects not included in the Log$N$-Log$S$ relations, i.e., targets of observations and for {\it Chandra} objects, all sources identified in observations targeting galaxy clusters, are also noted. If the X-ray flux is not detected at a significant level in any individual band ($<4.5\sigma$ for {\it Chandra} and $det\_ml <$15 for {\it XMM-Newton}), the flux is listed as null in the on-line catalogues. A high level summary of the number of sources matched to each optical, near-infrared and ultraviolet catalog is reported in Table \ref{cp_summary}, with the magnitude/flux density distributions for these counterparts shown in Figure \ref{mag_distr}. Appendix B details the columns for the on-line versions of the catalogs.

\subsection{Sloan Digital Sky Survey}
Due to the high density of sources in SDSS, as well as sub-arcsecond astrometry precision, we matched the X-ray sources separately to the $u$, $g$, $r$, $i$ and $z$ bands, using single-epoch photometry from Data Release 9 \citep[][DR9]{dr9}. A uniform 0.$^{\prime\prime}$1 error was assumed for all SDSS positions \citep{csc_sdss}. Comparing the reliability distributions for each band with the distributions for randomly shifted X-ray positions, we chose $R_{\rm crit}$=0.5 for both the {\it Chandra} and {\it XMM-Newton} source lists.

After vetting each individual band source list to include only objects exceeding $R_{\rm crit}$, we combined these source lists into a matched SDSS/{\it Chandra} catalog and an SDSS/{\it XMM-Newton} catalog. We visually inspected the cases where multiple SDSS objects (from separate band matchings) were paired to one X-ray source and selected the most likely counterpart by selecting the one with the greatest number of matches and/or the brightest object. We also imposed quality control cuts to assure the broad-band SEDs and derived photometric redshifts we will generate in a future paper (after careful aperture matching) are robust. We therefore require the SDSS objects to not be saturated\footnote{(NOT SATUR) OR (SATUR AND (NOT SATUR\_CENTER))} or blended\footnote{(NOT BLENDED) OR (NOT NODEBLEND)} and to have the photometry well measured\footnote{(NOT BRIGHT) AND (NOT DEBLEND\_TOO\_MANY\_PEAKS) AND (NOT PEAKCENTER) AND (NOT NOTCHECKED) AND (NOT NOPROFILE)}. After this vetting, every remaining SDSS match was visually inspected to remove objects contaminated by optical artifacts from e.g., diffraction spikes, or proximity to a close object that was not caught in the pipeline flagging.

We identified 748 and 1444 SDSS counterparts to {\it Chandra} and {\it XMM-Newton} sources, corresponding to 65\% and 61\% of the sample respectively, that exceeded $R_{\rm crit}$. However, 72 and 161 of these were rejected due to failing the quality control checks and visual inspection described above (marked as `yes' in the `SDSS\_rej' flag in the on-line catalogues, leaving 676 and 1283 reliable matches to {\it Chandra}  and {\it XMM-Newton} sources, or 59\% and 54\% of the X-ray sources. In a follow-up paper in which we will generate the broad-band SEDs, we will use co-added data of the 50-60 epochs of Stripe 82 scans to search for counterparts for the remaining $\sim$35\% of the X-ray sources \citep[][for studies of $z>5$ QSOs using co-added SDSS Stripe 82 data, see]{jiang,mcgreer}.

\subsubsection{Spectroscopy}
We searched spectroscopic databases to find redshifts corresponding to our matched X-ray/SDSS catalogs, using SDSS DR9, 2SLAQ \citep{2slaq}, WiggleZ \citep{wigglez} and DEEP2 \citep{spec_deep2}. This yielded spectroscopic redshifts for 306 {\it Chandra} sources ($\sim$27\% of the sample): 286 from SDSS DR9; 10 from 2SLAQ; 3 from WiggleZ and 7 from DEEP2. For the {\it XMM-Newton} sources, 497 optical counterparts had spectroscopic redshifts ($\sim$21\% of the sample): 468 from SDSS DR9, 20 from 2SLAQ, 4 from WiggleZ and 5 from DEEP2. We manually checked the spectra for the 25 SDSS sources where warning flags were set or for any object with $z > 5$: three spectra were re-fit to give more reliable redshifts, 11 were discarded due to poor spectra that could not be reliably fitted and we confirmed the redshifts for the remaining 11 objects.  In Table \ref{cp_summary}, the number of reported redshifts do not include the 11 that we discarded. Twenty-eight {\it XMM-Newton} sources had spectroscopic redshifts but unreliable photometry; we retain the redshift, but not the photometric, information for these objects. In the online catalogues, we indicate the database from which the spectroscopic redshifts were found, with $z$-source of 0, 1, 2, 3 and 4 referring to SDSS, 2SLAQ, WiggleZ, DEEP2 and SDSS spectra refitted/verified by us, respectively.

\subsection{{\it WISE}}
For a source to be included in the {\it WISE} All Sky Source Catalog \citep{wright,wise_cat}, a SNR $>$ 5 detection was required for one of the four photometric bands, W1, W2, W3 or W4, corresponding to wavelengths 3.4, 4.6, 12, and 22 $\mu$m, with resolution 6.$^{\prime\prime}$1, 6.$^{\prime\prime}$4, 6.$^{\prime\prime}$5 and 12.$^{\prime\prime}$0. The X-ray sources were matched to the W1 band since this band has the greatest number of non-null values, including both detections and upper limits. In the full Stripe 82 area, no {\it WISE} sources had null W1 detections, so we do not miss any potential {\it WISE} counterparts by matching to only the W1 band. The matching was performed on all W1 values, regardless of whether the magnitude corresponded to a detection or an upper limit (i.e., where the W1 SNR is below 2).  The R.A. and Dec errors were added in quadrature to provide an estimate of the {\it WISE} astrometric error.

If any bands suffered from saturation,\footnote{We consider the band to be affected by saturation if the fraction of saturated pixels exceeded 0.05, i.e., we could not rule out saturation at the 2$\sigma$ level.} spurious detections associated with artifacts (i.e., diffraction spikes, persistence from a short-term latent image, scattered halo light from a nearby bright source, or optical ghost image from nearby bright source), contamination from artifacts, or moon level contamination\footnote{We consider {\it moon\_lev} $\geq$5 as contaminated, where {\it moon\_lev} is the number of frames affected by scattered moonlight normalized by the total number of frames in the exposure multiplied by 10, and spans from $0 \leq$ {\it moon\_lev} $\leq 9$.}, we consider the magnitude in that band unreliable. If every band did not pass these quality control tests, then the source is not included in our final tally since we will not use the {\it WISE} data for generating the SEDs. For extended sources (where {\it ext\_flag} $>$0), the magnitudes measured from the profile-fitting photometry (i.e., {\it wnmpro}, where $n$ goes from 1-4) are unreliable. For these objects, we therefore focus on the magnitudes and quality flags associated with the elliptical apertures, {\it wngmag}, where $n$ goes from 1-4. Again, if all bands have null elliptical magnitudes and/or non-zero quality control flags, the source is not included in our catalog. The extended sources have the {\it WISE}\_ext flag set to `yes' in the online catalogues.

When matching the {\it Chandra} catalog to {\it WISE}, we imposed an $R_{\rm crit}$ of 0.75 and found 595 counterparts that passed the photometry quality control checks, or 52\% of the {\it Chandra} sample. Eight of these were extended. Photometry of 30 sources was compromised, 20 of which were extended. Our $R_{\rm crit}$ threshold for the {\it XMM-Newton} source list was 0.9, with 1324 counterparts identified with acceptable photometry (56\% of the sample), of which 8 were extended. Sixty-five sources did not pass the quality control checks, of which 40 were extended. The X-ray sources with {\it WISE} counterparts removed for not passing the quality control checks are marked as `yes' in the {\it WISE}\_rej field in the on-line catalogues.

\subsection{UKIDSS}
We searched for the UKIDSS Large Area Survey (LAS) Data Release 8 \citep{lawrence, casali, hewett, warren} for NIR counterparts to the Stripe 82 X-ray sources; details regarding maintenance of the UKIDSS science archive are described by \citet{hambly}. We used the LAS $YJHK$ Source table, which contains only fields that have coverage in every filter and merges the data from multiple detections of the same object. Only primary objects were selected\footnote{The `priOrSec' flag is set to zero if there are no duplicate observations of the same source or to the best `frameSetId' for duplicated observations. The SQL syntax to isolate primary observations is then `(priOrSec = 0 OR priORSec=frameSetId)'.} so that we worked with a clean input list with no duplicate NIR sources. We {\it a priori} removed objects flagged as noise, i.e., those sources with {\it mergedClass} set to zero and {\it PNoise} $\leq$ 0.05; that is, we only retained objects that are consistent with real detections (not noise) at greater than the 2$\sigma$ level for our candidate source list and background histograms. The UKIDSS positional uncertainties are set to NULL in the catalog. \citet{dye} quote that the internal accuracy can be $\sim$100 mas in each coordinate and the external accuracy is $\sim$80 mas in each coordinate. Adding the 180 mas uncertainty in quadrature for each coordinate gives a positional error of $\sim$0.$^{\prime\prime}$25 which we apply uniformly to all UKIDSS sources.

The X-ray source catalogs were matched separately to each UKIDSS band: $Y$ (0.97-1.07 $\mu$m), $J$ (1.17-1.33 $\mu$m), $H$ (1.49-1.78 $\mu$m) and $K$ (2.03-2.37 $\mu$m). The output matches were culled to include only sources exceeding $R_{\rm crit}$ and these individual band lists were then combined. Based on our test of shifting the X-ray positions by random amounts, we chose the values $R_{\rm crit}$=0.85, 0.75, 0.8 and 0.75 for the $Y$, $J$, $H$ and $K$ bands, respectively, for the {\it Chandra} matches; we used $R_{\rm crit}$=0.6, 0.6, 0.7 and 0.5 for the $Y$, $J$, $H$ and $K$ bands, respectively, in the {\it XMM-Newton} source matching.

When merging the individual UKIDSS matches with the X-ray source lists, more than one UKIDSS counterpart was matched to 1 {\it Chandra} object and to 45 {\it XMM-Newton} objects. We inspected these cases by eye and generally chose the brightest potential candidate in the most number of bands as the preferred match. In cases where the brightnesses were similar, we favored the candidate with the greatest number of matches among the UKIDSS bands. We note that a handful of these multiple potential candidates were duplicate observations of a bright star or a bright star and associated diffraction spike. We found 543 UKIDSS counterparts to the 1146 {\it Chandra} sources (47\%) and 1266 UKIDSS counterparts to the 2358 {\it XMM-Newton} objects (54\%). None of these IR sources was affected by saturation, i.e., there were no instances where `MergedClass' was set to -9 and `PSaturated' (probability of saturation) was 0 for all objects.

\subsection{{\it GALEX}}
The {\it GALEX} catalog comprises sources detected over several surveys, including Deep, Medium and All Sky Imaging Surveys (DIS, MIS and AIS, respectively) as well as a Guest Investigator program. For a trade-off between depth and coverage, and to cleanly remove duplicate observations of the same source, we extracted objects from the MIS survey only. Since the survey has overlapping tiles \citep[see][for observation details]{morrissey}, multiple observations of the same source can appear in the catalog. To chose the best candidate list, we queried the MIS database from Galex Release 7 for primary sources, i.e., those that are inside the pre-defined position (`SkyGrid') cell within the field \citep[see][]{budavari}. We further require that each primary is within 5$^{\prime}$ of the field center. Following the prescription of \cite{bianchi}, we considered objects within 2.5$^{\prime\prime}$ as possible duplicates: if they are part of the same observation, i.e., had the same `photoextractid,' they are considered unique sources but if they are from different observations, the data corresponding to the longest exposure was used. We note that in many cases, sources with the same `photextractid' but different `objids' (which identifies unique sources) were actually unmerged FUV and NUV detections of the same source, where one observation had either a FUV non-detection while there was a  NUV detection or vice versa. However, since we matched the X-ray source lists separately to the NUV and FUV catalogs, such duplicates do not affect the results of our analysis. 

The {\it Chandra} and {\it XMM-Newton} source lists were matched to this cleaned {\it GALEX} catalog using $R_{\rm crit}$ = 0.5 for each band. We used the individual source positional errors reported in the {\it GALEX} database, rather than applying a systematic positional error to all sources. Matching the NUV and FUV detections separately, rather than focusing on the {\it GALEX} sources with detections in both bands, has the advantage that we locate ultraviolet counterparts that are detected in one band and not the other. We then merged the results of the individual band matching, locating {\it GALEX} counterparts for 164 {\it Chandra} and 249 {\it XMM-Newton} objects, corresponding to 14\% and 11\% of each parent sample, respectively.

\subsection{FIRST}
Due the to low space density of both radio and X-ray sources, we matched our X-ray source lists to the FIRST \citep{first,first_cat1} catalog using a simple nearest neighbor approach rather than MLE: the closest radio object within a search radius of 5$^{\prime\prime}$ for {\it Chandra} sources and within 7$^{\prime\prime}$ for {\it XMM-Newton} objects was chosen as the true counterpart. We used the FIRST catalog released in 2012, which contains all sources detected between 1993 and 2011, with a detection limit of 0.75 mJy over part of Stripe 82 ($319.6^{\circ} < $ R.A. $<$ 49.5$^{\circ}$, $-1^{\circ} <$ Dec $< 1^{\circ}$), and 1 mJy detection limit for the rest of the region \citep{first_cat}. We identified radio counterparts for 42 {\it Chandra} sources (4\% of the sample) and 82 {\it XMM-Newton} objects (3\% of the sample). From shifting the X-ray positions by random amounts, we expect spurious associations for 1 {\it Chandra} source within 5$^{\prime\prime}$ and 4 {\it XMM-Newton} objects within 7.$^{\prime\prime}$ Two {\it Chandra} sources had 2 potential radio counterparts within $r_{\rm search}$, but these X-ray sources were within the search radius of each other, so these duplicate potential matches are expected. Within the 7$^{\prime\prime}$ {\it XMM-Newton} search radius, 2 potential counterparts were found for 4 X-ray sources. In all of these cases, the nearest neighbor was also the brightest radio object.

\begin{table}
\caption{\label{cp_summary}Number of X-ray Sources Detected in Ancillary Databases}
\begin{tabular}{lrrr}
\hline
Catalog & {\it Chandra} & {\it XMM-Newton} & Total$^{1}$ \\
\hline

X-ray     & 1146 & 2358 & 3362 \\
SDSS      &  676 & 1283 & 1892 \\
{\it WISE}      &  595 & 1324 & 1855 \\
UKIDSS    &  543 & 1266 & 1754 \\
{\it GALEX}     &  164 &  301 & 447  \\
FIRST     &   42 &   82 & 119  \\
Spec-$z$s &  306 &  497 & 759  \\

\hline
\multicolumn{4}{l}{$^{1}$Duplicate sources between the {\it Chandra} and}\\
\multicolumn{4}{l}{{\it XMM-Newton} catalogs removed.}
\end{tabular}
\end{table}

\section{Discussion}
Here we use the results of the catalog matching to discuss general characteristics of the X-ray sources in Stripe 82, highlighting the science areas our survey is uniquely poised to investigate. 

\subsection{Probing the High X-ray Luminosity Regime of Black Hole Growth}
We calculated full band X-ray luminosities for the sources with spectroscopic redshifts. After removing duplicate matches between the {\it Chandra} and {\it XMM-Newton} source catalogs and isolating the objects with luminosities exceeding 10$^{42}$ erg s$^{-1}$, the X-ray luminosity above which there are few or no starburst-dominated X-ray sources \citep[e.g.,][]{persic,bh}, we confirm that 645 of the 759 Stripe 82 X-ray sources with optical spectra are AGN; the remaining sources have X-ray luminosities consistent with star-forming galaxies or low-luminosity AGN or are stars. Below, we compare the X-ray luminosity distribution with other X-ray surveys and with model predictions and comment on the interesting sources we have discovered.

\subsubsection{Comparison with Other X-ray Surveys}
\begin{figure}
\subfigure[]{\includegraphics[scale=0.4,angle=90]{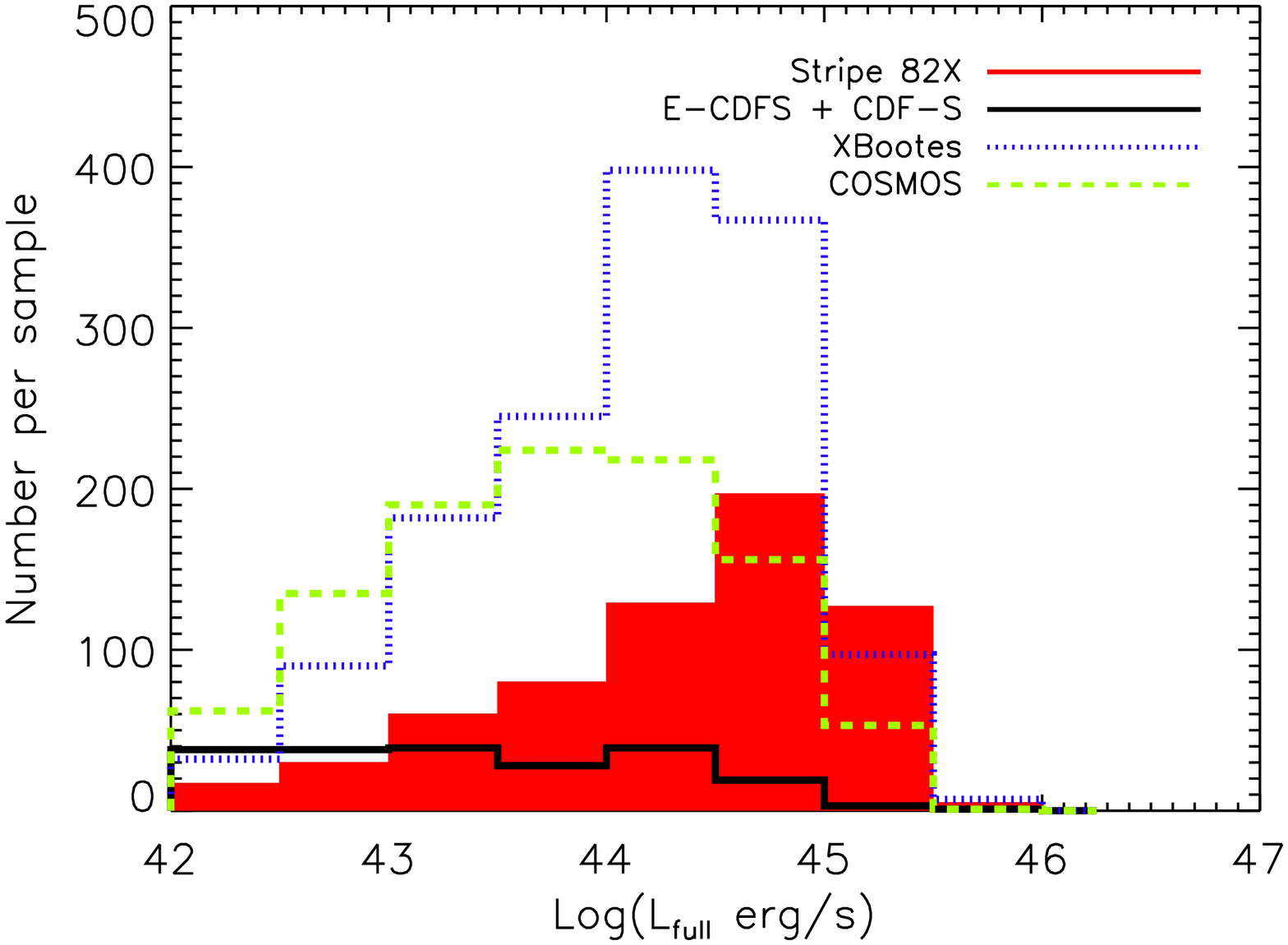}}
\subfigure[]{\includegraphics[scale=0.4,angle=90]{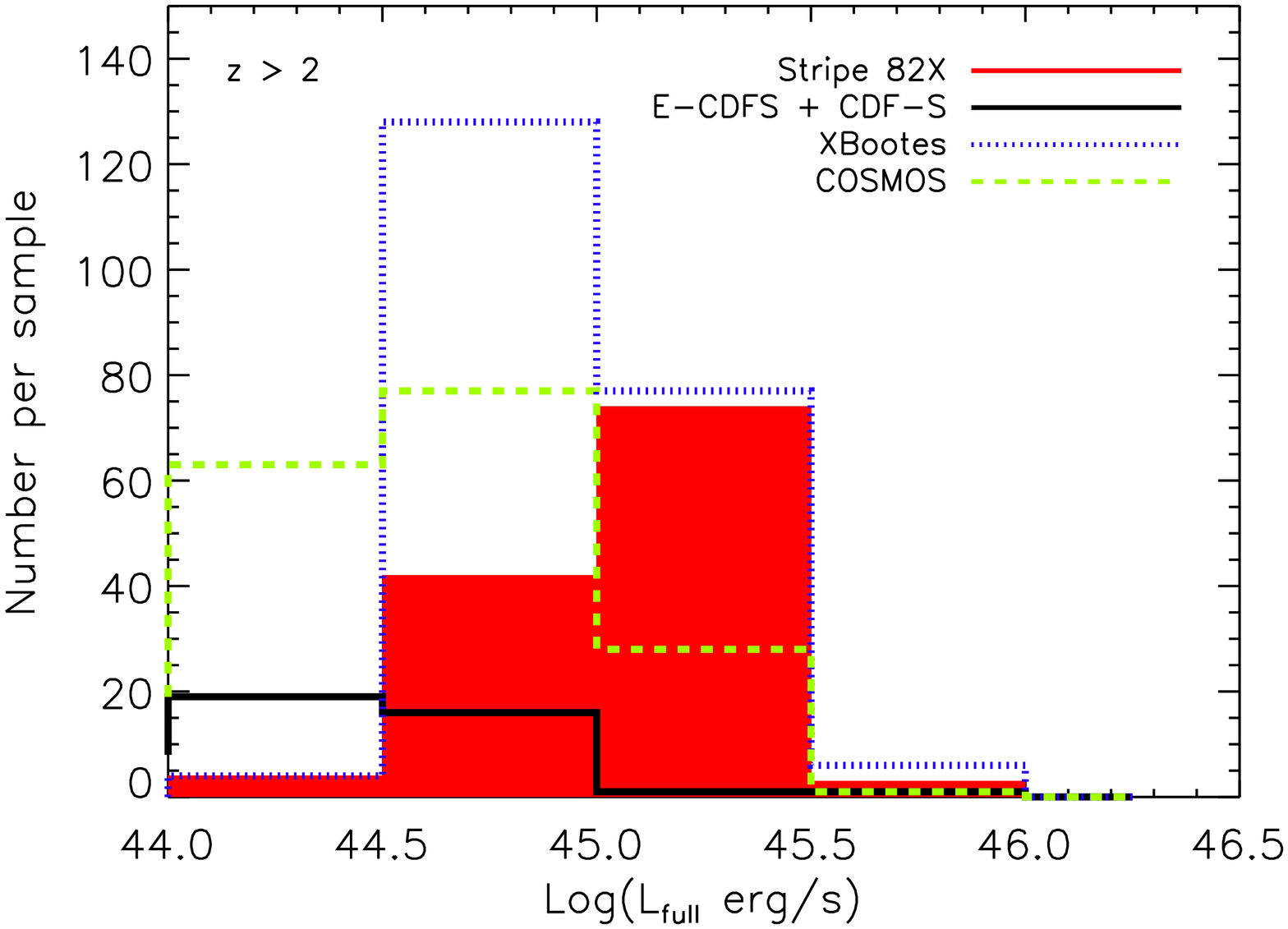}}
\caption[]{\label{lum_distr} X-ray luminosity distribution for Stripe 82X sources (red filled) compared to other X-ray selected samples for (a) all sources with spectroscopic redshifts and (b) objects with spectroscopic redshifts greater than 2. Wider area surveys are necessary to sample appreciable numbers of high-redshift and high-luminosity AGN; small area surveys \citep[e.g.,E-CDFS + CDF-S, black,][]{giavalisco,treister_04,cardamone2} lack sufficient volume to detect rare objects. Though Stripe 82X has a lower level of spectroscopic completeness than the other surveys, and {\it XMM}- and {\it Chandra}-COSMOS \citep[green dashed,][]{cap09,brusa3,civano_mle} and XBo\"otes \citep[blue dot-dash,][]{kenter,kochanek} have benefited from dedicated multi-year follow-up campaigns, we immediately identify more L$_x > 10^{45}$ erg s$^{-1}$ AGN at all redshifts and comparable numbers at $z>2$.}
\end{figure}

The comparison X-ray surveys plotted in Figure \ref{lum_distr} span from deep, small area \citep[the GOODS and MUSYC survey of E-CDFS and CDF-S, $\sim$0.3 deg$^2$,][]{giavalisco,treister_04, cardamone2}, to moderate area and moderate depth \citep[{\it XMM} and {\it Chandra} COSMOS, $\sim$2.1 deg$^2$,][]{cap09,brusa3,civano_mle}, to wide area and shallow depth \citep[XBo\"ootes,$\sim$9 deg$^2$,][]{kenter, kochanek}. Again, we focus on only the X-ray sources with spectroscopic redshifts, with a completeness of $\sim$28\%, $\sim$45\% and $\sim$44\% for E-CDFS + CDF-S, COSMOS and XBo\"otes, respectively. For reference, the spectroscopic completeness of Stripe 82X, prior to any dedicated follow-up observations, is currently $\sim$23\%, though the smaller X-ray surveys peer deeper, garnering many more faint optical counterparts where spectroscopic follow-up opportunities are limited. We report observed, full-band luminosities, which is 0.5-10 keV for the {\it XMM-Newton} surveys (obtained for {\it XMM}-COSMOS by summing the individual soft and hard band fluxes while \citet{civano_mle} provides full band fluxes for {\it Chandra}-COSMOS objects), 0.5-7 keV for the Stripe 82 {\it Chandra} sources and XBo\"otes, and 0.5-8 keV for E-CDFS + CDF-S.

As Figure \ref{lum_distr} illustrates, survey area determines the AGN population sampled. Small area surveys (e.g., E-CDFS + CDF-S) identify faint objects but leave the high luminosity objects sparsely sampled. Moving to wider areas expands the parameter space to higher luminosities since these objects are rare and more volume must be probed in order to locate them. This becomes very apparent at $z > 2$ (Figure \ref{lum_distr} b). 

Wider area surveys, such as COSMOS and XBo\"otes have higher levels of spectroscopic completeness than Stripe 82X due to dedicated multi-year spectroscopic campaigns. However, prior to any follow-up, we have identified 1.5 times more high luminosity AGN (L$_{0.5-10keV} > 3\times10^{44}$ erg s$^{-1}$) than {\it XMM}- and {\it Chandra}-COSMOS at all redshifts. Compared to XBo\"otes, Stripe 82X pilot has $\sim$30\% more L$_{0.5-10keV} > 10^{45}$ erg s$^{-1}$ AGN when considering all redshifts, and finds almost as many in the young universe. Though the current spectroscopic completeness of Stripe 82X pilot is comparatively lower, more area is covered, enabling identification of more high luminosity AGN.

However, comparing the source density of the brightest objects among these 2 wider area surveys with Stripe 82X pilot indicates that there are still more high luminosity AGN left to find. The space density of L$_{0.5-10keV} >10^{45}$ erg s$^{-1}$ AGN found in COSMOS is 26 deg$^{-2}$ (i.e., 54 AGN in 2.1 deg$^2$) and in XBo\"otes is 12 deg$^{-2}$ (104 in 9 deg$^2$), while currently Stripe 82X has a space density of 8 deg$^{-2}$ (132 AGN in 16.5 deg$^2$). We therefore anticipate that with additonal spectroscopic completeness, the source density of high luminosity AGN will subsequently increase. Of the L$_{0.5-10keV} >10^{45}$ erg s$^{-1}$ AGN already identified in Stripe 82 that have optical classifications, one is a narrow line AGN (optically classified as a `galaxy') while the remaining have broad lines.

\subsubsection{Comparison with Model Predictions}
\begin{figure}
\subfigure{\includegraphics[scale=0.4,angle=90]{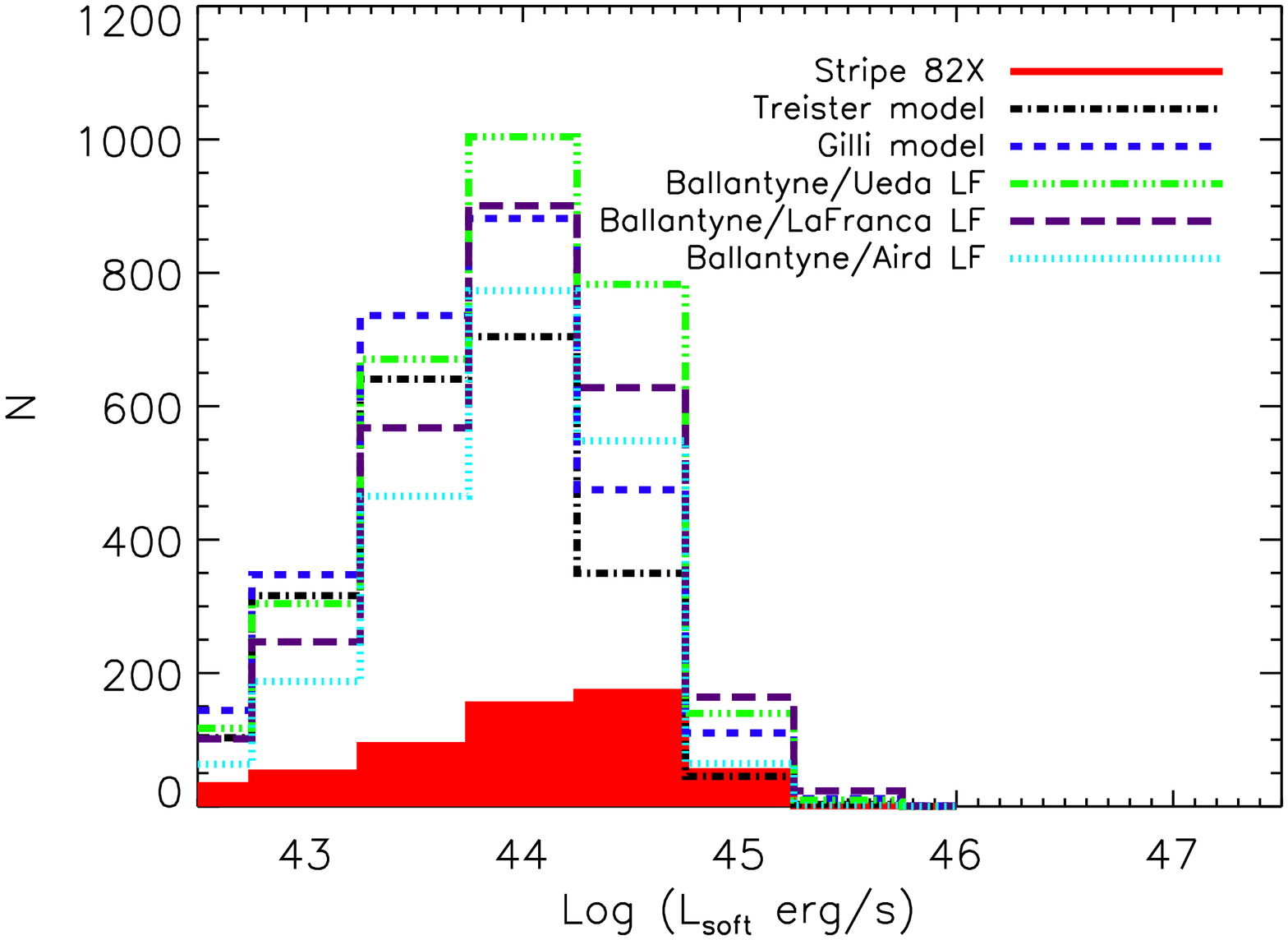}}
\subfigure{\includegraphics[scale=0.4,angle=90]{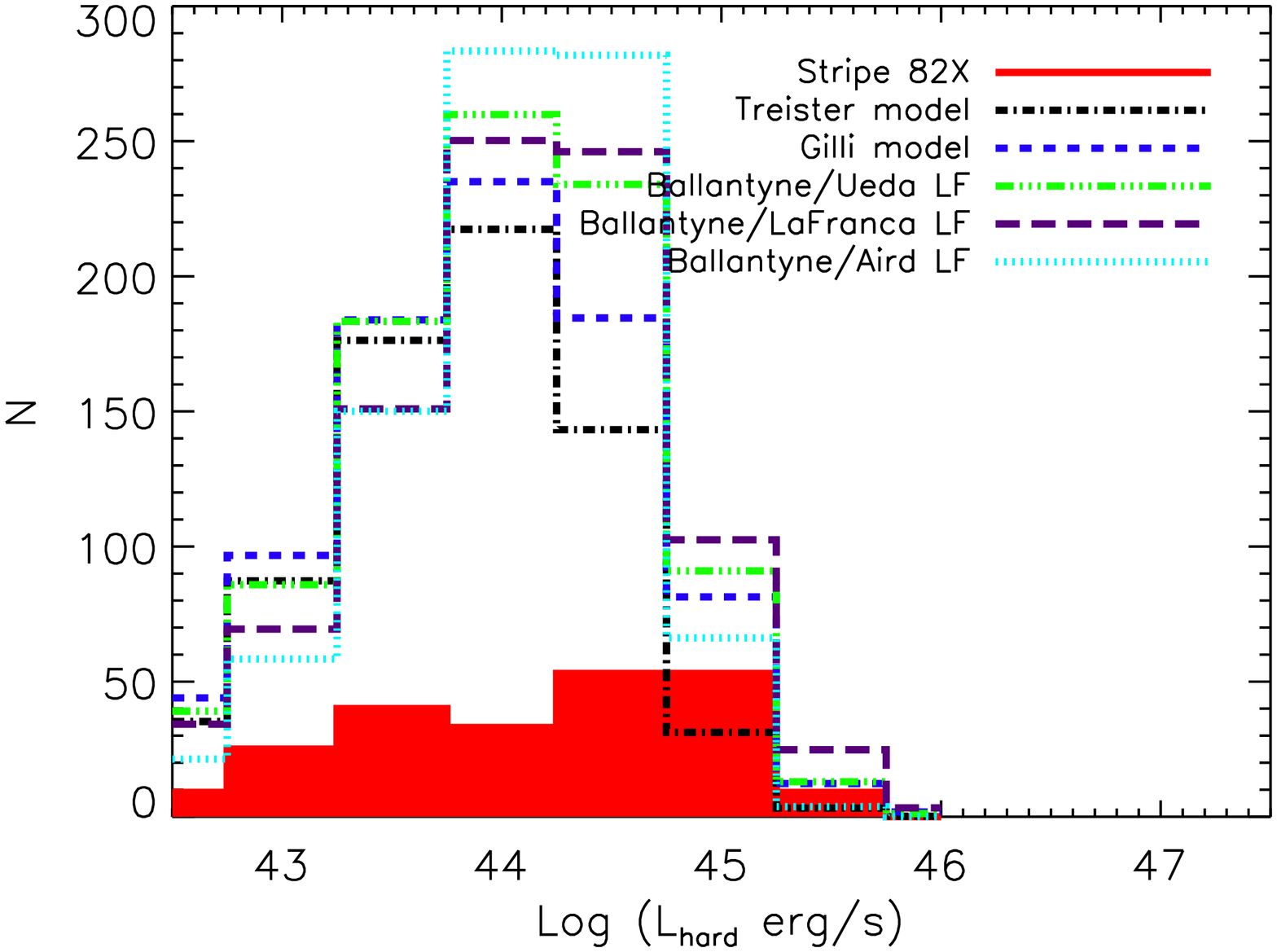}}
\subfigure{\includegraphics[scale=0.4,angle=90]{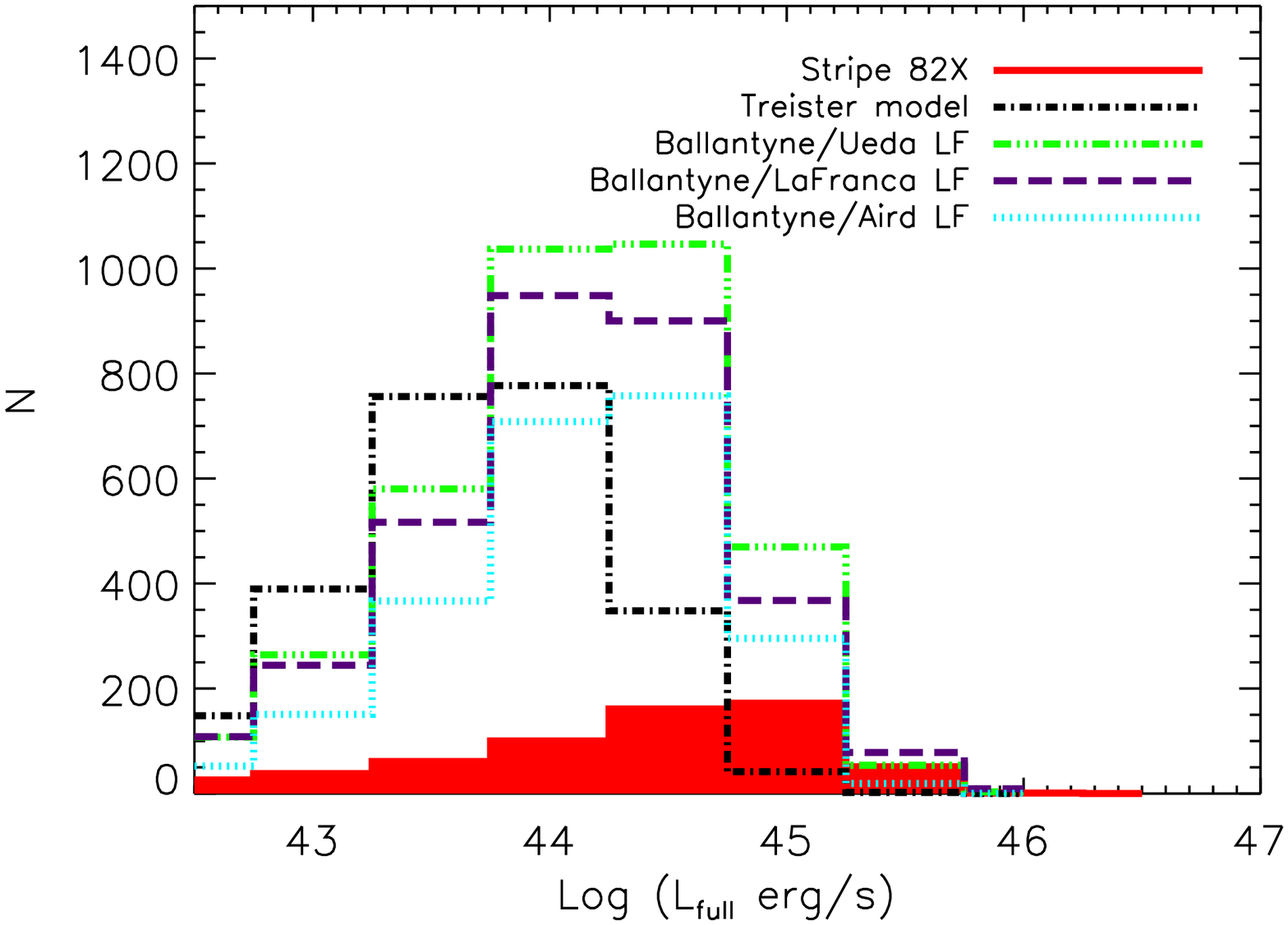}}
\caption[]{\label{pred_all} Luminosity distribution for Stripe 82X sources (red filled) for all sources with spectroscopic redshift compared to population synthesis models from \citet[][dash-dotted black line]{treister}, \citet[][dashed blue line]{Gilli} and \citet[][assuming different luminosity functions noted in the caption]{ballantyne} in the ({\it top}) soft, ({\it middle}) hard and ({\it bottom}) full energy bands;  models from \citet{Gilli} over the full X-ray band are not available. At high luminosities, we have already identified more AGN than predicted by the \citet{treister} model and almost as many as predicted by the \citet{Gilli} model and \citet{ballantyne} model, depending on the assumed luminosity function.}
\end{figure}

\begin{figure}
\subfigure{\includegraphics[scale=0.4,angle=90]{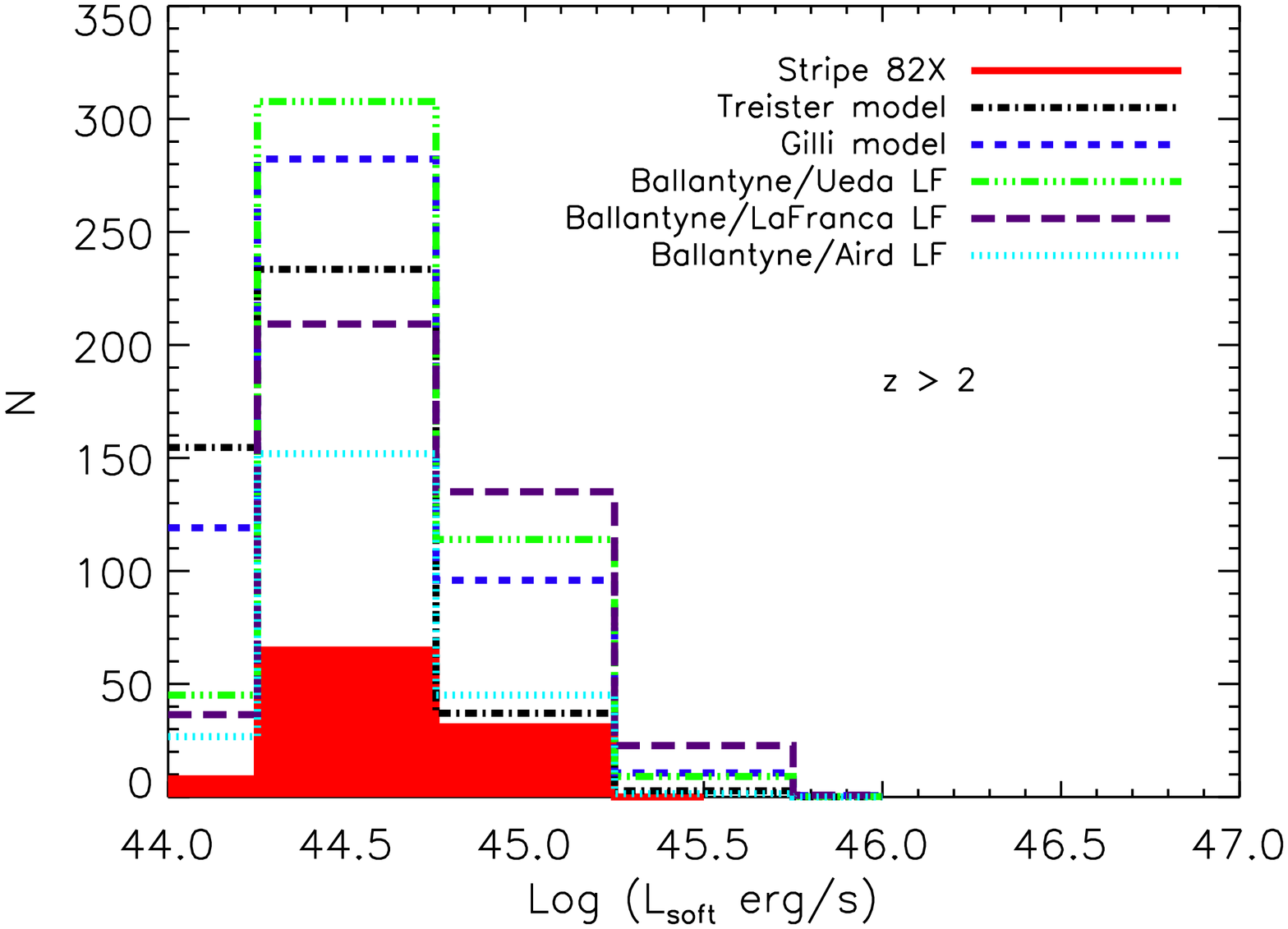}}
\subfigure{\includegraphics[scale=0.4,angle=90]{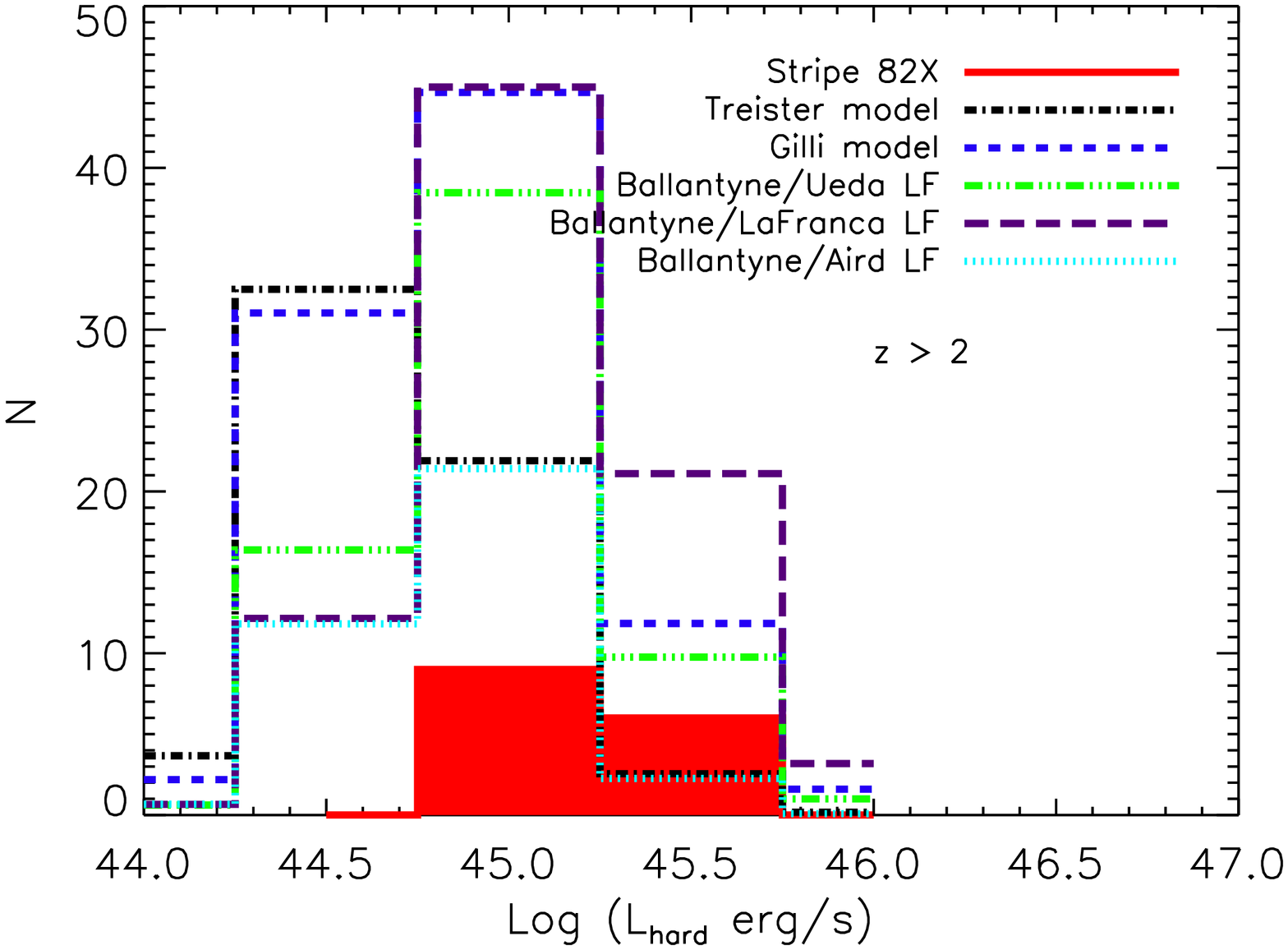}}
\subfigure{\includegraphics[scale=0.4,angle=90]{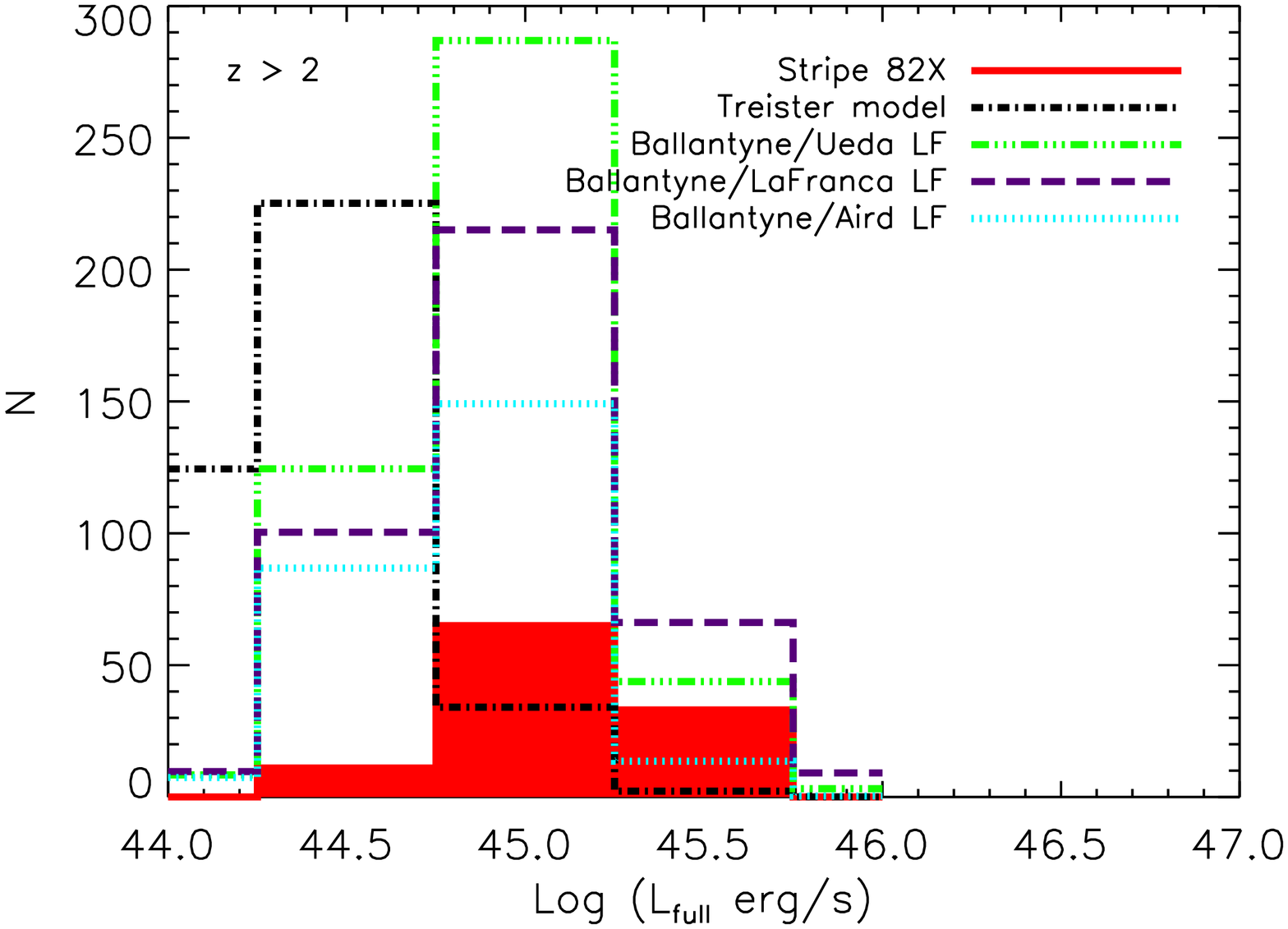}}
\caption[]{\label{pred_zgt2} Same as Figure \ref{pred_all} but for spectroscopic redshifts greater than 2. We identified more high luminosity AGN than predicted by \citet{treister} and \citet{ballantyne} with the \citet{aird} luminosity function as input in the hard and full bands, suggesting the Stripe 82X large area survey will provide important constraints to black hole growth in the high-luminosity, high-redshift regime.}
\end{figure}

With the larger dataset presented here, we expand on the work of \citet{me} and compare the luminosity distribution of X-ray AGN we immediately identify with X-ray background population synthesis predictions of \citet{treister}, \citet{Gilli}, and \citet{ballantyne}. We input the observed area-flux curves for {\it Chandra} and {\it XMM-Newton} into the \citet{treister} simulator,\footnote{Model predictions from the work of \citet{treister} for a range of input values are publicly available at http://agn.astroudec.cl/j\_agn/main.html} and convolved the predicted Log$N$-Log$S$ distributions from \citet{Gilli}\footnote{http://www.bo.astro.it/$\sim$gilli/counts.html where we use their assumed spectral model of $\Gamma$=1.9 to convert the hard band luminosity bins into soft band luminosity bins required by their code.} and \citet{ballantyne} with our observed area-flux curves; we note that since the \citet{Gilli} predictions only allow the hard band flux to be defined from 2-10 keV, we corrected the output fluxes in the {\it Chandra} band to our 2-7 keV range, using our assumed spectral model where $\Gamma$=1.7. As \citet{Gilli} do not provide model predictions in the full band, we compare our observed full-band numbers to the models from \citet{treister} and \citet{ballantyne}. The predicted luminosity bins represent intrinsic, rest-frame luminosities, while the Stripe 82X data are observed luminosities. We removed any {\it Chandra} pointings that overlapped {\it XMM-Newton} pointings since the latter has more effective area. Since we did detect a handful of {\it Chandra} sources in these removed pointings that were not identified by {\it XMM-Newton}, the histograms presented in Figures \ref{pred_all} and \ref{pred_zgt2} are a subset of the total data.

The models from \citet{Gilli} predict more AGN at all redshifts and more high luminosity ($>10^{45}$ erg s$^{-1}$) AGN at $z>2$ than the \citet{treister} models while the \citet{ballantyne} model predicts more L$>10^{44}$ erg s$^{-1}$ AGN than the \citet{treister} model. As \citet{ballantyne} produces predictions based on three different input luminosity functions \citep{ueda,lafranca,aird}, which show a significant range in expected AGN numbers within most luminosity bins, discrepancies among models can be attributed to differences in luminosity functions. \citet{Gilli} uses the XLF from \citet{hasinger} in the 0.5-2 keV band, estimating the contribution of moderately obscured AGN  ($10^{21}$ cm$^{-2} <$ N$_H < 10^{24}$ cm$^{-2}$) in the hard band by calculating the difference between the \citet{ueda} and \citet{lafranca} XLFs and the \citet{hasinger} XLF (after converting the latter to the hard band). The predictions from \citet{treister} are calibrated on the hard band XLF from \citet{ueda}.

Due to our limited spectroscopic completeness (see Section 4.1.1), we have identified fewer AGN than predicted given the constraints from our data. However, these `missing' objects are predominantly at low to moderate luminosities ($<10^{45}$ erg s$^{-1}$). When considering objects at all spectroscopic redshifts, we found more high luminosity AGN than predicted by the \citet{treister} model, most of those predicted by the \citet{Gilli} model and a significant fraction to most, depending on the luminosity function, of those predicted by the \citet{ballantyne} model. The same result applies to objects at $z>2$ in the hard and full bands (though we do find more high luminosity objects than the \citet{ballantyne} predictions based on the \citet{aird} models in these bands), while in the soft band, the \citet{treister} and \citet{ballantyne} models predict slightly more high luminosity objects than we have yet discovered. Currently, the discrepancies between our observations and the \citet{treister} model are within $\sim$2$\sigma$ assuming Poisson uncertainties. Given the lower space density of these objects compared to surveys with higher spectroscopic completeness (i.e., Section 5.1.1), it seems clear that more high luminosity AGN will be confirmed. Even a small increase in the high luminosity population would surpass the predictions of \citet{Gilli} and the more conservative numbers of \citet{ballantyne}. We also expect that our luminosity distribution is systematically lower than the predictions as the latter use intrinsic, rather than observed, luminosities as input. The systematic effect would shift the Stripe 82X sources into higher luminosity bins if corrected for absorption, making our comparison at high luminosities conservative. 

Finding a greater number of high X-ray luminosity AGN relative to the model predictions is consistent with what was reported earlier by \citet{me}, namely, that population synthesis models need to be refined to properly account for the high luminosity AGN regime. As this unexplored population is more numerous than predicted, quantifying its impact on AGN demography and evolution is critical for fully understanding black hole growth. Increased spectroscopic completeness will inform us as to the significance of the offset.

\subsection{SMBH Growth at High Redshift}
\begin{figure}
\includegraphics[scale=0.4,angle=90]{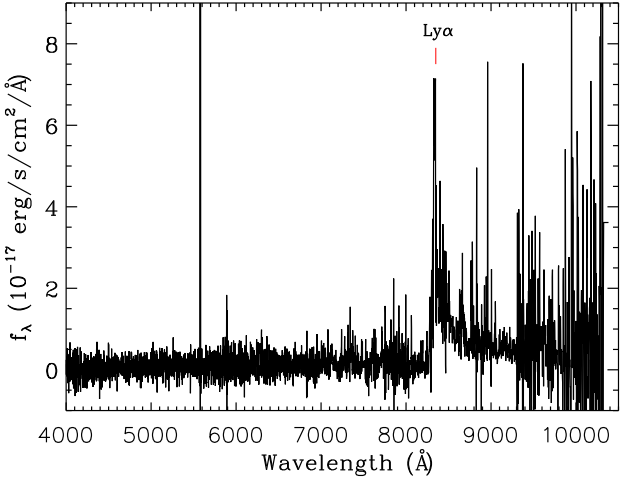}
\caption[]{\label{highz_spec} SDSS spectrum of the highest redshift quasar yet discovered in an X-ray survey at $z=5.86$ with an X-ray luminosity of 4.4$\times10^{45}$ erg s$^{-1}$. The Ly$\alpha$ transition is marked. This source was discovered in the archival {\it Chandra} data (MSID = 165442).}
\end{figure}

Though $z>5$ quasars identified by SDSS have been followed up with dedicated {\it Chandra} observations \citep[e.g.,][]{brandt, shemmer, vignali}, not many have been found in X-ray surveys. Thus far, only 6 have been confirmed spectroscopically: $z=5.19$ in CDF-N \citep{barger}, $z = 5.4$ from the {\it Chandra} Large Area Synoptic X-ray Survey \citep{steffen}, $z = 5.3$ and $z = 5.07$ from {\it Chandra}-COSMOS \citep{civano_11}, and 2 from ChaMP, with the most distant object having a redshift of 5.41 \citep{trichas}. None have been located in the 4 Ms of CDF-S, demonstrating that area trumps depth for locating these very high redshift sources. In this pilot survey of Stripe 82X, we have discovered only one object beyond a redshift of 5, but it is the most distant X-ray selected quasar from an X-ray survey to date, at $z = 5.86$ with L$_{0.5-10keV}$ = 4.4$\times10^{45}$ erg s$^{-1}$ ({\it Chandra} source, MSID = 165442). The SDSS spectrum of this source is shown in Figure \ref{highz_spec}, revealing broad Ly$\alpha$ (i.e., this source is classified as a broad line AGN). 

Such objects are expected to be quite rare. For instance, model predictions from \citet{treister} estimate that only 3 AGN at $z> 5$ with L$_{0.5-10keV}>10^{45}$ erg s$^{-1}$ exist in this survey area, given the observed full band area-flux curves. Similarly, using the soft band area-flux curves, the models from \citet{Gilli} predict 5 AGN at $z > 5$ with L$_{0.5-2keV}>10^{45}$ erg s$^{-1}$, but less than one when applying an exponential decline to the high-$z$ luminosity function. These types of objects will be below the flux limit of eRosita \citep{merloni,kolodzig}, making the Stripe 82X survey important for constraining black hole formation models.

\subsection{Obscured AGN Beyond the Local Universe}

\subsubsection{{\it WISE} AGN Candidates}
\begin{figure}
\includegraphics[scale=0.4,angle=90]{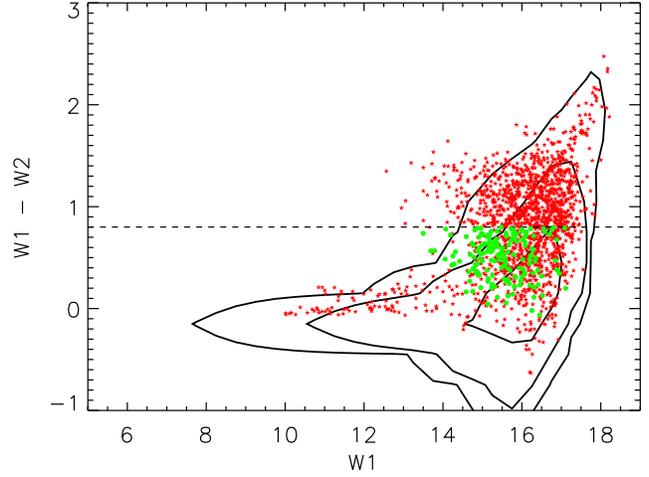}
\caption[]{\label{w1_w2} {\it WISE} color $W1 - W2$ as a function of $W1$ with the contours indicating the density (10$^3$, 10$^4$, 10$^5$, 10$^6$ objects per contour) of all {\it WISE} objects in 300 deg$^2$ if Stripe 82, with our X-ray objects overplotted as red stars and the $W1 - W2 \geq 0.8$ AGN candidate color cut \citep[e.g.,][]{stern,assef} marked by the dashed line. About half of our X-ray objects have redder colors, while 2/3 of the objects with spectra that are X-ray identified as AGN also exceed this boundary. The 166 spectroscopically identified X-ray AGN (L$_x > 10^{42}$ erg s$^{-1}$) with bluer colors are shown by the green circles.}
\end{figure}

In Figure \ref{w1_w2}, we plot the WISE $W1 - W2$ color as a function of $W1$ for the 1713 Stripe 82 X-ray sources with significant detections (SNR $\geq$ 2) in both bands on top of the contours for all {\it WISE} sources with significant $W1$ and $W2$ colors in the full 300 deg$^2$ Stripe 82 area. The color cut of $W1 - W2 \geq 0.8$ used to identify {\it WISE} AGN candidates \citep{stern,assef} is overplotted, with 904 of our X-ray/{\it WISE} objects falling within this region, or 53\% of the total. This contrasts with the results of \citet{stern} who find that in the COSMOS field, a majority of X-ray sources with {\it WISE} counterparts have blue colors, i.e., $W1 - W2 < 0.8$; for instance, only 91 out the 244 {\it XMM-Newton}/WISE sources in COSMOS (38\%) have {\it WISE} AGN candidate colors. A higher fraction of the Stripe 82 X-ray sources have infrared colors consistent with obscured AGN.

Five hundred nine of the 1713 Stripe 82 X-ray sources with significant $W1$ and $W2$ detections have spectroscopic redshifts and X-ray luminosities indicative of AGN  activity \citep[L$_x > 10^{42}$ erg s$^{-1}$][]{persic,bh}. Of these, 165, or 32\%, have {\it WISE} color $W1 - W2 < 0.8$ (green circles in Figure \ref{w1_w2}). These results indicate that two-thirds of our spectroscopically confirmed AGN are obscured (red {\it WISE} colors) and that identifying AGN candidates based on a simple color cut can miss up to a third of bluer AGN that can be recognized via other selection mechanisms, e.g., the optical and X-ray. As pointed out by \citet{stern}, this result reinforces the complementarity of MIR and X-ray selection in providing comprehensive views of SMBH growth.

\subsubsection{Optically Normal Galaxies}
At $z > 0.5$, diagnostic line ratio diagrams used to discriminate between Type 2 AGN and star-forming galaxies \citep[e.g.,][]{bpt, kewley, kauff} become challenging in optical surveys as H$\alpha$ is shifted out of the rest-frame bandpass. Candidates for obscured AGN would then have to be identified via alternative optical diagnostics, such as ratios of narrow emission lines vs. stellar mass \citep[MEx,][]{juneau} or vs. rest frame $g - z$ color \citep[TBT,][]{trouille}. Follow-up of type 2 AGN candidates with ground-based NIR spectroscopy to observe the traditional BPT line diagnostics is also possible, but as the metallicity of host galaxies evolves with redshift, the applicability of line ratios calibrated for galaxies at $z < 0.5$ to higher redshift systems may not cleanly separate star-forming from active galaxies. Conversely, calculating an object's X-ray luminosity provides a more efficient identification mechanism. In our survey, we have identified 22 X-ray AGN at $z > 0.5$ with luminosities exceeding 10$^{43}$ erg s$^{-1}$ that were classified as galaxies in SDSS, 2SLAQ or DEEP2 based on their optical spectra. One of these objects is extremely bright as noted above, L$_x = 10^{45}$ erg s$^{-1}$, and is an example of the kind of highly luminous obscured AGN our survey is designed to uncover. Currently these sources only represent 3\% of our AGN sample, but we expect that more of these objects will be discovered during our spectroscopic follow-up campaign. 

\subsubsection{Optical Dropouts}
We identified 748 and 1444 optical counterparts to the {\it Chandra} and {\it XMM-Newton} sources, respectively, though 72 and 161 are discarded due to poor photometry. How many of the $\sim$400 {\it Chandra} and $\sim$900 {\it XMM-Newton} X-ray objects lacking SDSS counterparts ($r > 23$) do we find in the infrared? Most of these optical dropouts are either reddened by large amounts of dust or live at high redshift, so that the rest-frame optical light is shifted to redder wavelengths.

To answer this question, we look at two classes of optical drop-outs: the X-ray sources with optical counterparts below $R_{\rm crit}$, including objects where no SDSS counterparts are found within the search radius, and the subset of X-ray sources without any optical counterpart within $r_{\rm search}$. In the former case, a true counterpart can be misclassified as a random association, especially if it is faint. The latter number then gives us a lower limit on the number of infrared bright optical dropout X-ray sources. Comparison of the flux limits for SDSS, {\it WISE} and UKIDSS to the type 1 quasar SED (i.e., broad line AGN) from \citet{elvis_sed} demonstrate that SDSS is deeper than the {\it WISE} or UKIDSS observations, making the detection of IR sources that are SDSS drop-outs a significant finding. We summarize these results in Table \ref{opt_drop}, detailing the number of optical dropouts found in the IR generally and the numbers identified in the {\it WISE} and UKIDSS catalogs specifically. We note that the greater percentage of optical dropouts that have no counterpart within the search radius for the {\it Chandra} catalog compared to {\it XMM-Newton} can be understood by the larger search radius used for the latter catalog.

\begin{figure}
\includegraphics[scale=0.50,angle=90]{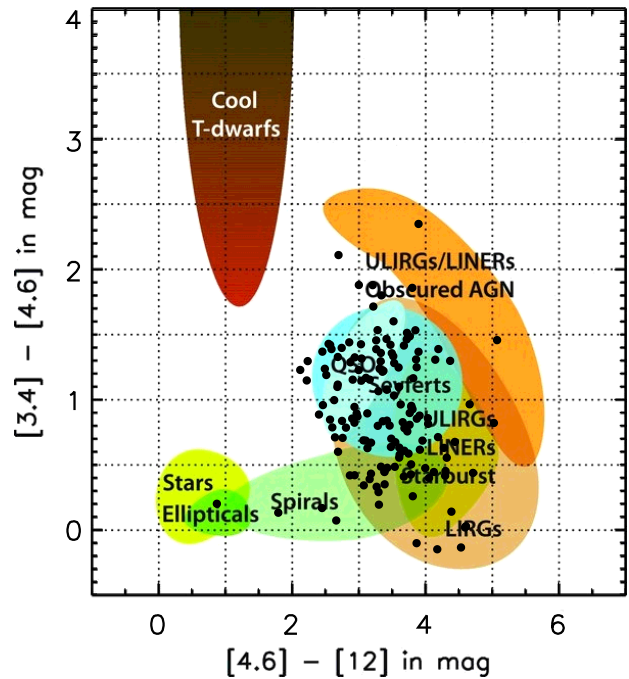}
\caption[]{\label{wise_color} {\it WISE} color-color diagram for X-ray sources (filled circles) detected significantly in the W1, W2 and W3 bands that have no optical counterpart within the search radius or where optical sources are found within $r_{\rm search}$ but are below $R_{\rm crit}$ and are therefore not likely candidates for the true optical counterpart to the X-ray source. The colored loci represent the classes of objects with these {\it WISE} colors, defined by \citet{wright}. Most of the optical dropouts are consistent with active galaxies.}
\end{figure}

Over 30\% of the optical droputs are detected in the infrared, making them candidates for the elusive population of obscured high luminosity AGN at high redshift. We plot the {\it WISE} colors of the 151 dropouts ($\sim$12\% of optical dropouts) that have significant W1, W2 and W3 detections (i.e., SNR $>$ 2 in each band) in Figure \ref{wise_color} for the optical dropout X-ray sources. The {\it WISE} colors are overlaid on the diagram from \citet{wright}, where the colored loci represent different classes of astronomical objects. A majority of the optical dropouts detected in X-rays have {\it WISE} colors consistent with active galaxies, with nearly half having infrared colors akin to quasars. These are prime candidates for high-luminosity Type 2 AGN or highly reddened quasars and will be followed up by us with NIRSPEC on Keck and ISAAC on ESO's VLT. For the remaining 840 optical dropouts without infrared associations (25\% of the X-ray sample), deeper optical and infrared imaging is necessary is identify the multi-wavelength counterparts to the X-ray sources. 

\begin{table}
\caption{\label{opt_drop}Number of Optical Dropouts Detected in X-rays}
\begin{tabular}{lrrr}
\hline
Catalog & {\it Chandra} & {\it XMM-Newton} & Total$^{1}$ \\
\hline
\multicolumn{4}{l}{No SDSS counterpart within $r_{\rm search}$ or above $R_{\rm crit}$}\\
\hline

X-ray & 398 & 914 & 1312 \\
IR$^{2}$    & 112 & 371 & 472 \\
{\it WISE}  &  95 & 313 & 401 \\
UKIDSS      &  43 & 149 & 189 \\

\hline
\multicolumn{4}{l}{No SDSS counterpart within $r_{\rm search}$}\\
\hline

X-ray & 317 & 486 &  781 \\
IR$^{2}$    &  88 & 161 & 240 \\
{\it WISE}  &  73 & 124 & 192 \\
UKIDSS      &  37 &  82 & 116 \\
\hline
\multicolumn{4}{l}{$^{1}$After removing duplicate sources between the}\\
\multicolumn{4}{l}{{\it Chandra} and {\it XMM-Newton} catalogs.}\\
\multicolumn{4}{l}{$^2$Detected in {\it WISE} or UKIDSS.}
\end{tabular}
\end{table}

\section{Conclusions}
We have reduced and analyzed the $\sim$10.5 deg$^2$ of {\it XMM-Newton} data overlapping SDSS Stripe 82, including $\sim$4.6 deg$^2$ of proprietary data awarded to us in AO 10. From these observations, we detected 2358 unique X-ray sources at high significance, with 2073, 607 and 2079 in the soft (0.5-2 keV), hard (2-10 keV), and full (0.5-10 keV) bands, respectively. The Log$N$-Log$S$ relations show general agreement with previous surveys in these bands, given the effect that choice of spectral model affects the normalization in the full band. 
Using a maximum likelihood estimator algorithm \citep{mle,brusa1,brusa2,cardamone,luo,brusa3,civano_mle}, we identified multi-wavelength counterparts to Stripe 82 X-ray sources, finding:
\begin{itemize}

\item 1892 optical matches from SDSS, of which 759 have spectroscopic redshifts; 1855 {\it WISE} counterparts; 1754 UKIDSS matches; 447 ultraviolet counterparts from {\it GALEX}; and 119 radio sources from FIRST (using nearest neighbor matching rather than MLE due to low source densities).

\item Focusing on the subset of sources with spectroscopic redshifts, Stripe 82X harbors more high luminosity (L$_x \geq 10^{45}$ erg s$^{-1}$) AGN than E-CDFS and CDFS \citep[$\sim$0.3 deg$^2$,][]{cardamone2}, {\it XMM}- and {\it Chandra}-COSMOS \citep[$\sim$2.1 deg$^2$,][]{cap09,brusa3,civano_mle} and even the larger XBo\"otes survey \citep[$\sim$9 deg$^2$,][]{kenter,kochanek}. Though these other surveys benefited from years of spectroscopic follow-up, Stripe 82X covers a wider area and thereby already uncovers more rare objects (high luminosity AGN at all redshifts and in the early universe, at $z >$ 2). These numbers will increase with the spectroscopic follow-up we are currently undertaking.

\item We have compared the luminosity distribution of X-ray sources with spectroscopic redshifts with the population synthesis model predictions from \citet{treister}, \citet{Gilli} and \citet{ballantyne} taking into account the observational constraints of our observed area-flux curves in the soft, hard and full X-ray bands. As we showed in \citet{me} using a subset of these data and the full-band \citep{treister} model predictions given our area-flux curves, we discovered more high luminosity ($>10^{45}$ erg s$^{-1}$) AGN than predicted by \citet{treister}. Though the \citet{Gilli} and \citet{ballantyne} models predict more AGN, we have found most of those predicted at high luminosity (depending on the luminosity function for the \citet{ballantyne} model), and this number will continue to increase with our spectroscopic follow-up.  Refinement to models is clearly indicated to better account for this important regime of black hole growth. As these rare, high luminosity AGN are more numerous than previously predicted, understanding their census, evolution and connection to the host galaxy becomes an important piece in completing the puzzle of cosmic black hole growth.

\item We have found the most distant, spectroscopically confirmed X-ray selected quasar in an X-ray survey to date, at $z = 5.86$.

\item About a third of the X-ray sources that are optical dropouts are identified in the infrared, making them candidates for reddened quasars and/or high luminosity Type 2 AGN at high redshift. Most of those with significant detections in the W1, W2 and W3 {\it WISE} bands have colors consistent with active galaxies, with more than half of them having quasar colors. We have a Keck-NIRSPEC campaign and were awarded ESO VLT ISAAC DDT time to follow-up these objects.

\end{itemize}

The Stripe 82X survey provides an important pathfinder mission to eRosita, scheduled to be launched in 2014, which will survey the entire sky in 0.5-10 keV X-rays, though with a poorer resolution than {\it Chandra} and {\it XMM-Newton} ($\sim$25$^{\prime\prime}$) and with an 0.5-2 keV flux limit that is 5 times higher than our proprietary {\it XMM-Newton} mosaicked observations \citep{merloni}. eRosita expects to uncover millions of AGN, of which a few tens will be $z>$6 QSOs. An efficient method will then need to be devised to isolate the very high redshift population: results of Stripe 82X, with the wealth of multi-wavelength data, will help to inform robust identification techniques applicable to eRosita.

In other luminosity ranges, X-ray selection has uncovered a different, if overlapping, population of AGN compared to optical selection. While the jury is still out on how this impacts black hole growth at high luminosity, it is clear that the answer requires large samples selected at X-ray energies, so that the optical- and X-ray samples can be compared.

\section*{Acknowledgments}
We thank the referee for a careful reading of this manuscript and useful comments and suggestions. We also thank M. Brusa for helpful discussions.

This publication makes use of data products from the Wide-field Infrared Survey Explorer, which is a joint project of the University of California, Los Angeles, and the Jet Propulsion Laboratory/California Institute of Technology, funded by the National Aeronautics and Space Administration.

Funding for SDSS-III has been provided by the Alfred P. Sloan Foundation, the Participating Institutions, the National Science Foundation, and the U.S. Department of Energy Office of Science. The SDSS-III web site is http://www.sdss3.org/.

SDSS-III is managed by the Astrophysical Research Consortium for the Participating Institutions of the SDSS-III Collaboration including the University of Arizona, the Brazilian Participation Group, Brookhaven National Laboratory, University of Cambridge, Carnegie Mellon University, University of Florida, the French Participation Group, the German Participation Group, Harvard University, the Instituto de Astrofisica de Canarias, the Michigan State/Notre Dame/JINA Participation Group, Johns Hopkins University, Lawrence Berkeley National Laboratory, Max Planck Institute for Astrophysics, Max Planck Institute for Extraterrestrial Physics, New Mexico State University, New York University, Ohio State University, Pennsylvania State University, University of Portsmouth, Princeton University, the Spanish Participation Group, University of Tokyo, University of Utah, Vanderbilt University, University of Virginia, University of Washington, and Yale University. Funding for Yale participation in SDSS-III was provided by Yale University.

\appendix

\section[]{Reliability Thresholds for Counterpart Selection}

As mentioned in the main text, our goal is to optimize selection of multi-wavelength counterparts to the Stripe 82 X-ray sources by maximizing the number of true associations while minimizing contamination from chance coincidences. We inspected the distribution of source `reliabilities' calculated via MLE and picked a critical threshold ($R_{\rm crit}$) above which we expect a vast majority of the ancillary objects represent true counterparts. By shifting the X-ray positions by random amounts and running the MLE code, the distribution of resulting reliabilities provides an empirical estimate of the contamination in our matched catalogs. 

In Figures A1 - A7, we compare the reliability distribution of each wavelength band to which we matched (solid black histogram) with the reliability distribution after shifting the X-ray positions (by $\sim$21$^{\prime\prime}$ to $\sim$35$^{\prime\prime}$) overplotted (blue histogram). The dotted line indicates the specific $R_{\rm crit}$ value we used for that band. In the captions, we note the number of spurious associations expected, i.e., ancillary counterparts matched to random positions on the sky, above $R_{\rm crit}$. We stress that contamination percentages that can be calculated from this test are not exact, but are instead meant to provide an empirical method for calibrating the reliabilities on a band-by-band basis. As in all multi-wavelength surveys, a handful of true counterparts may be missed, falling below $R_{\rm crit}$, while several random coincident matches may be promoted as real matches. However, our empirical tests indicate that this effect is at the few percent level at most.

\begin{centering}
\begin{figure}
\subfigure{\includegraphics[scale=0.2,angle=90]{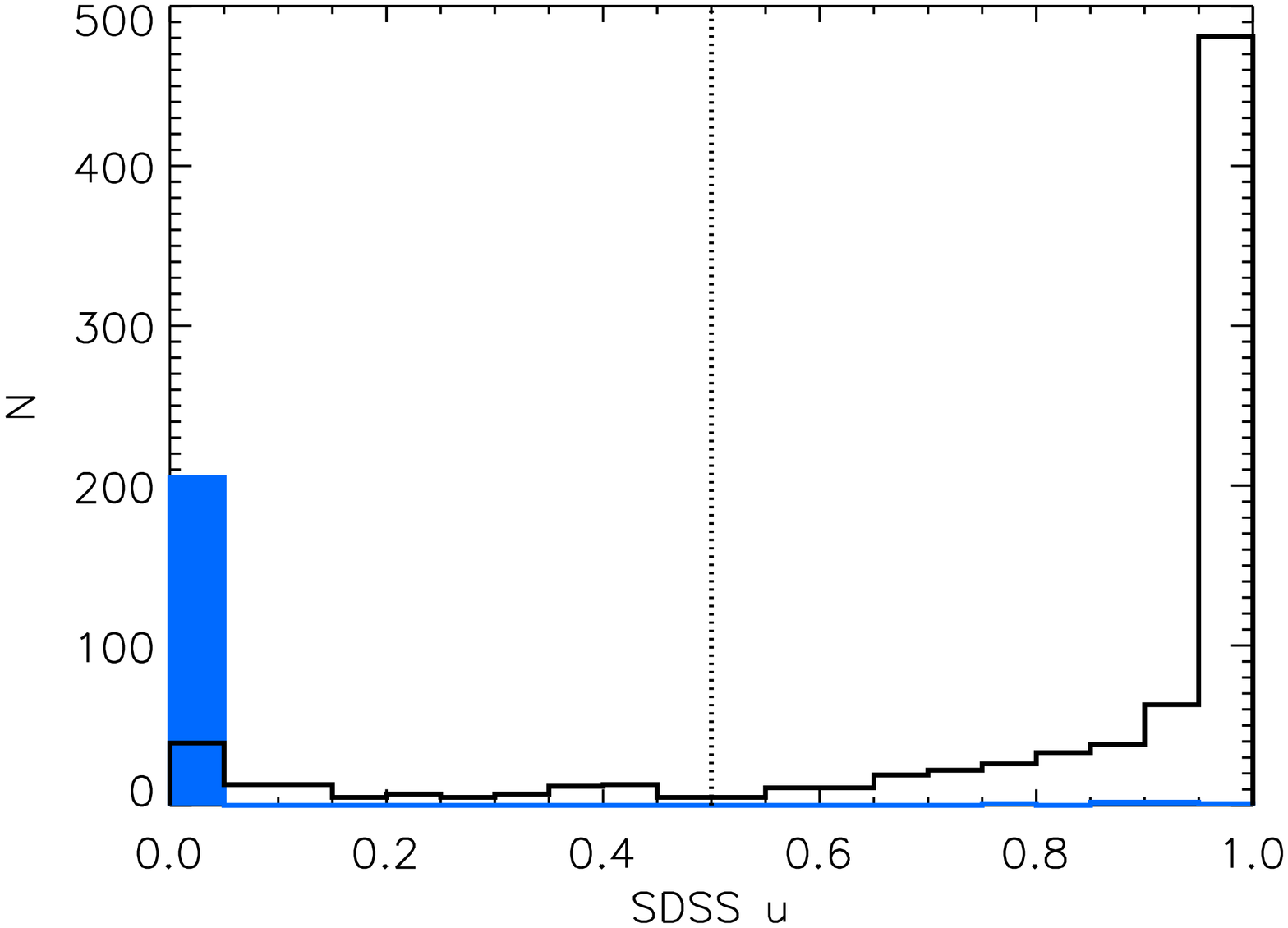}}~
\subfigure{\includegraphics[scale=0.2,angle=90]{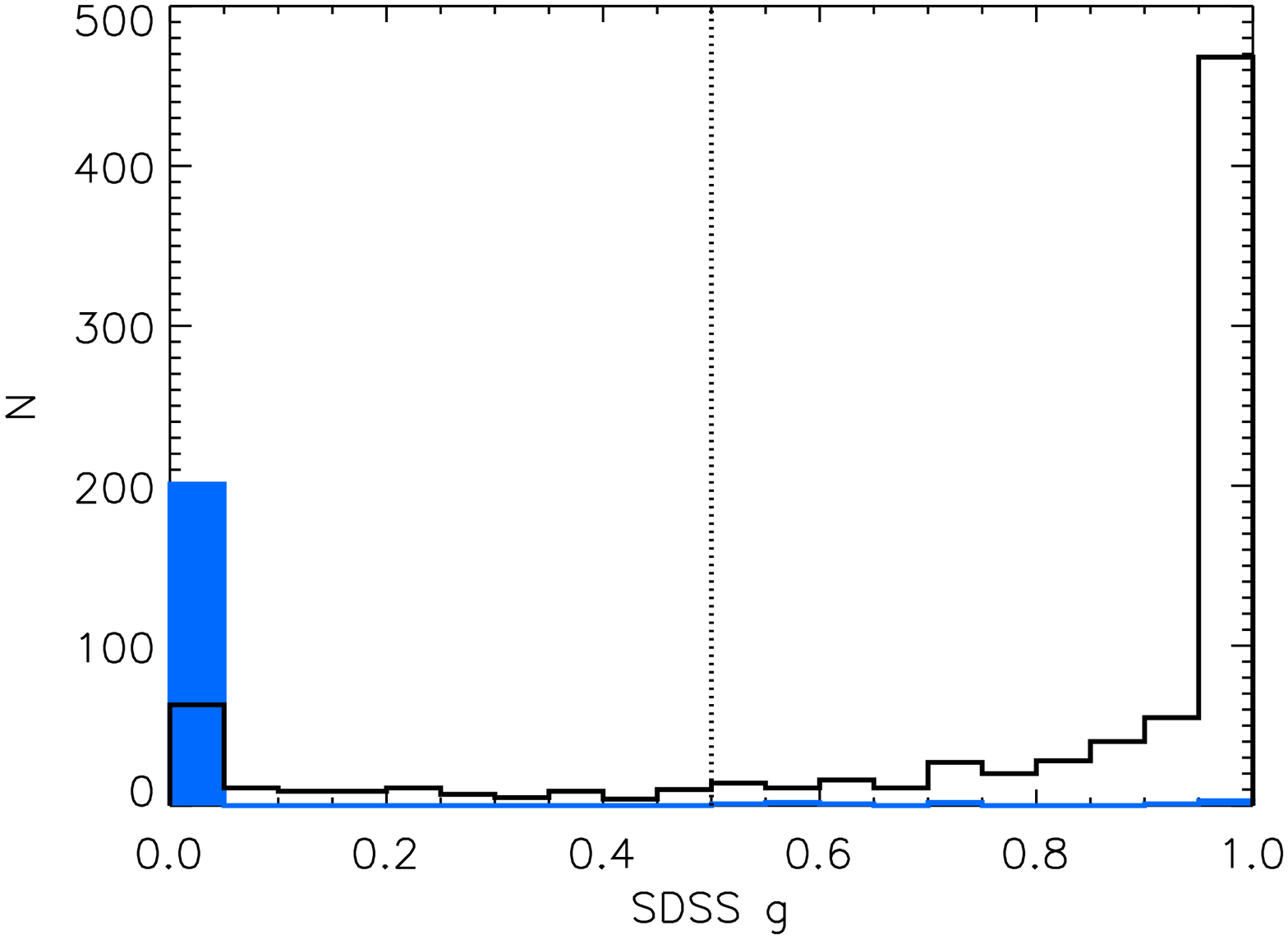}}
\subfigure{\includegraphics[scale=0.2,angle=90]{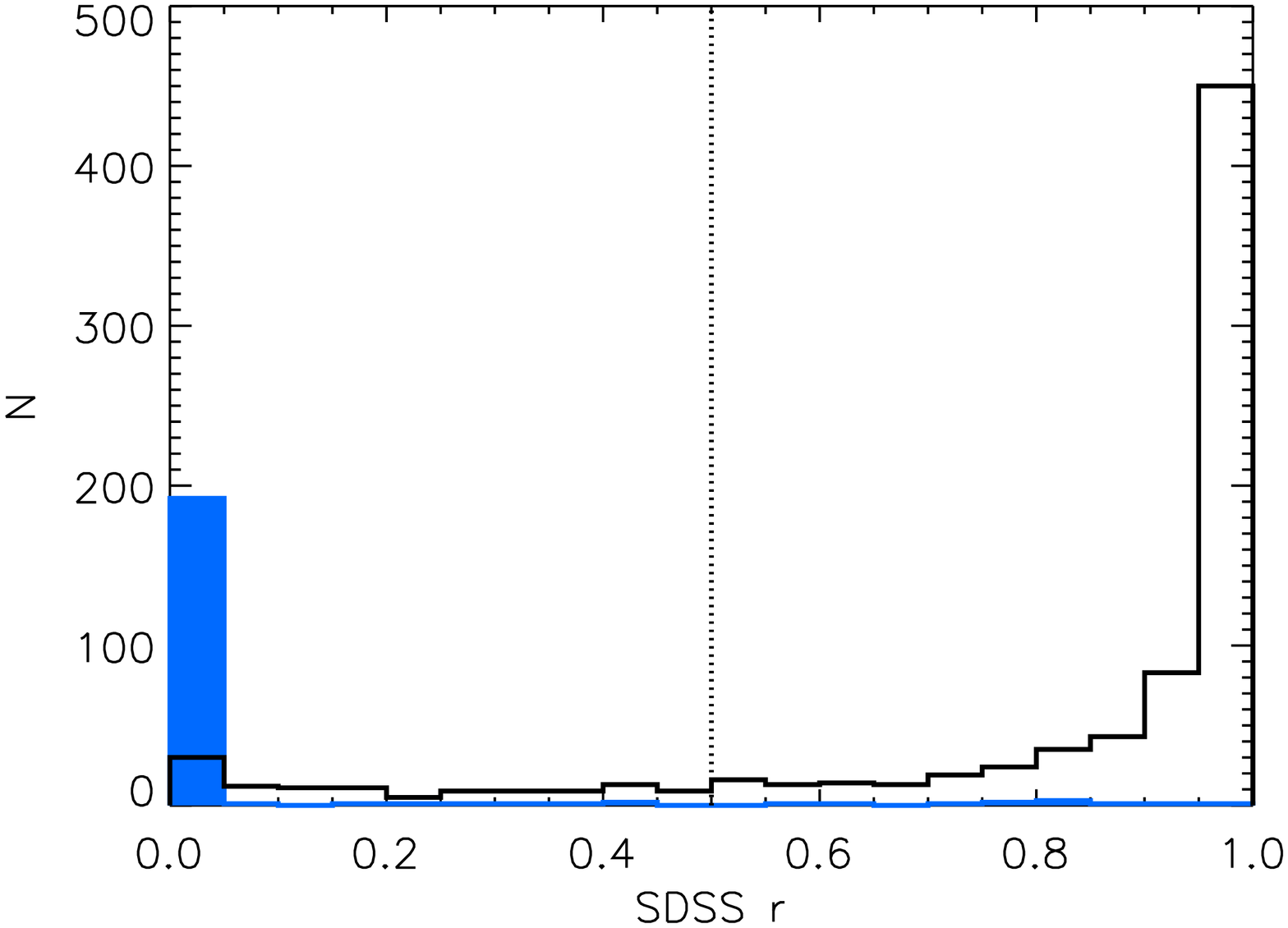}}~
\subfigure{\includegraphics[scale=0.2,angle=90]{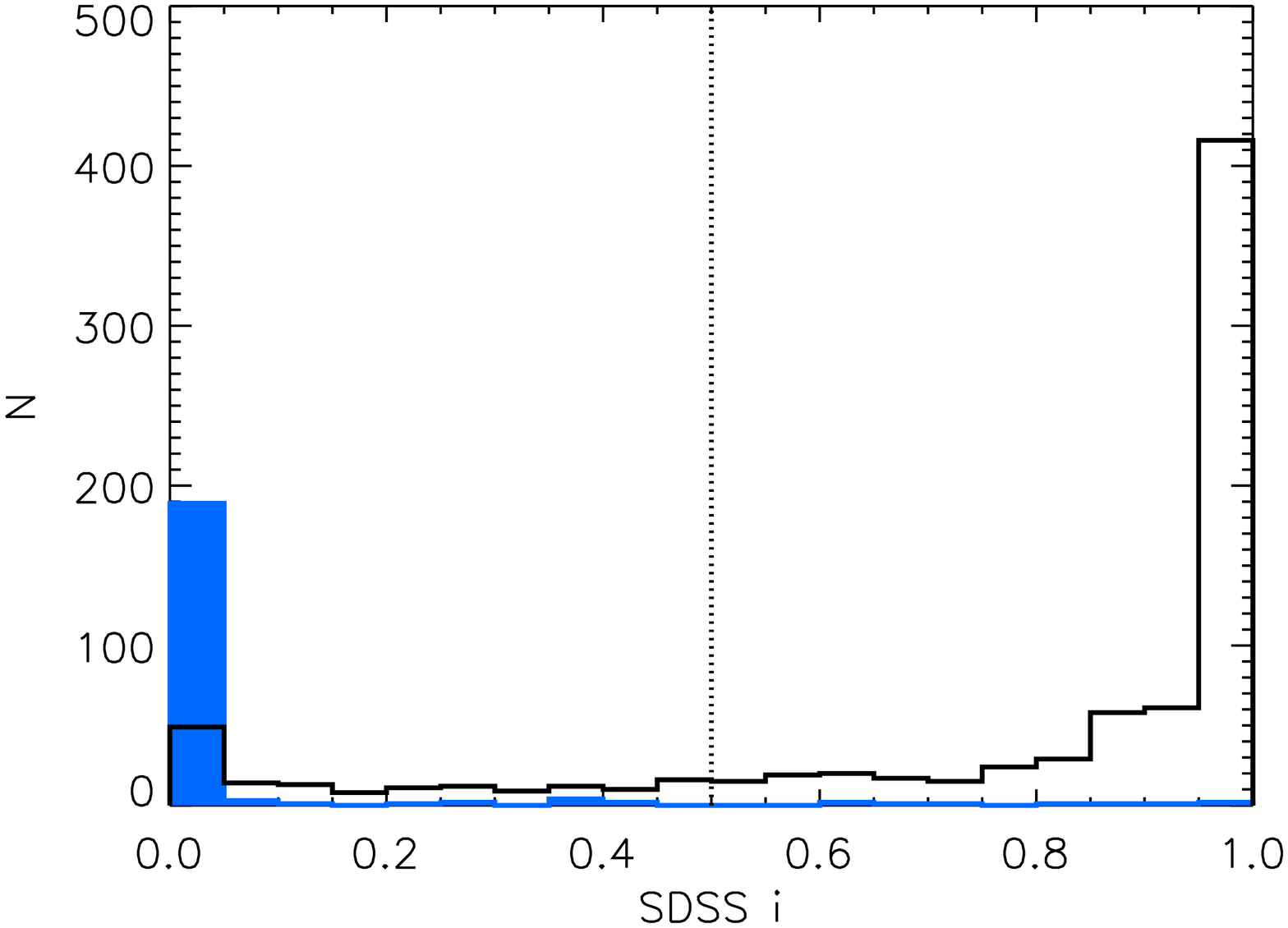}}
\subfigure{\includegraphics[scale=0.2,angle=90]{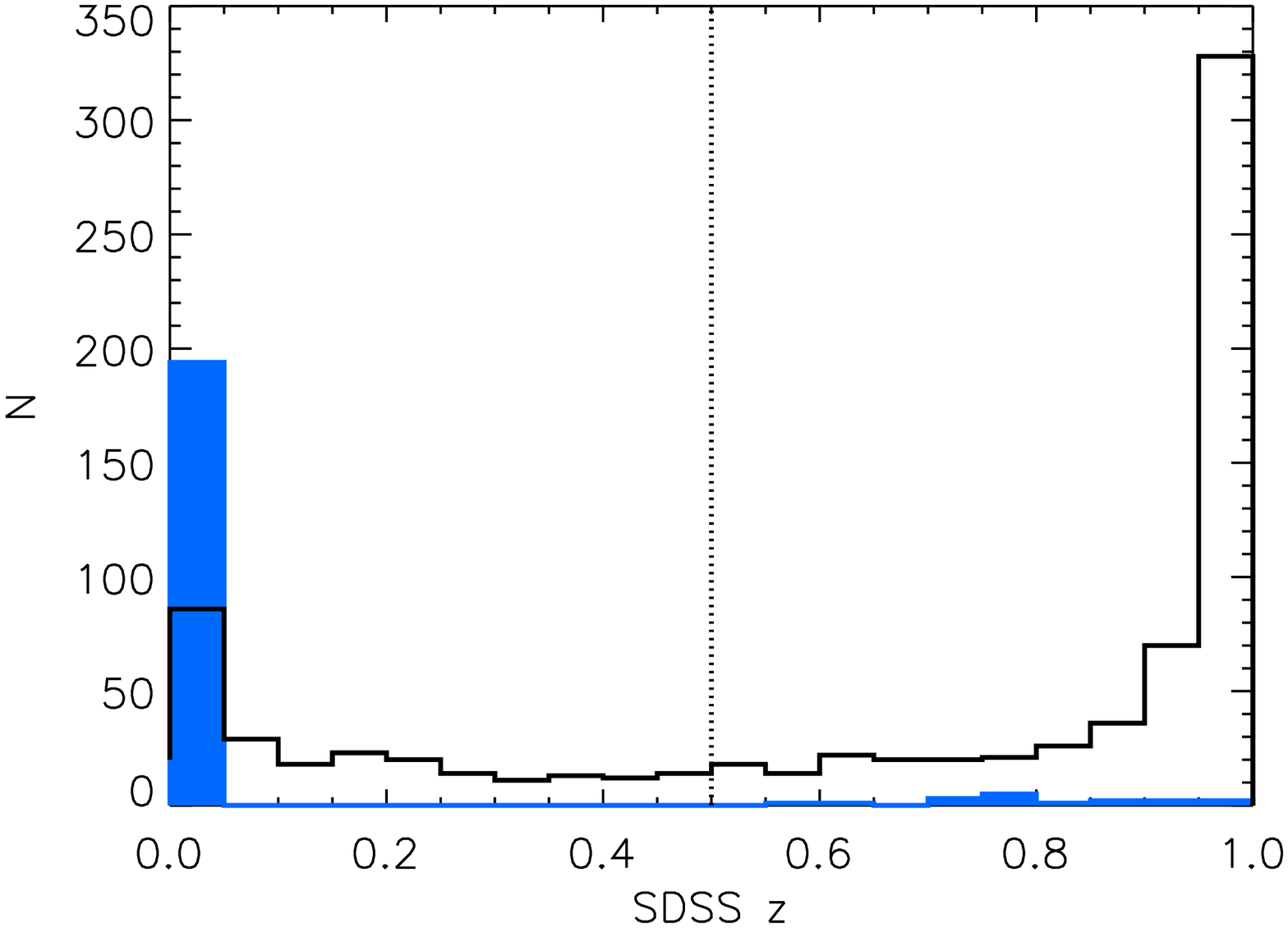}}
\caption[]{Reliability distributions for each SDSS band matched to {\it Chandra} sources. The number of spurious associations above $R_{\rm crit}$ is predicted to be 6, 10, 11, 9 and 17 in the $u$, $g$, $r$, $i$ and $z$ bands, respectively. The dot-dash line indicates the adopted reliability threshold for claiming a counterpart.}
\end{figure}
\end{centering}

\begin{centering}
\begin{figure}
\subfigure{\includegraphics[scale=0.2,angle=90]{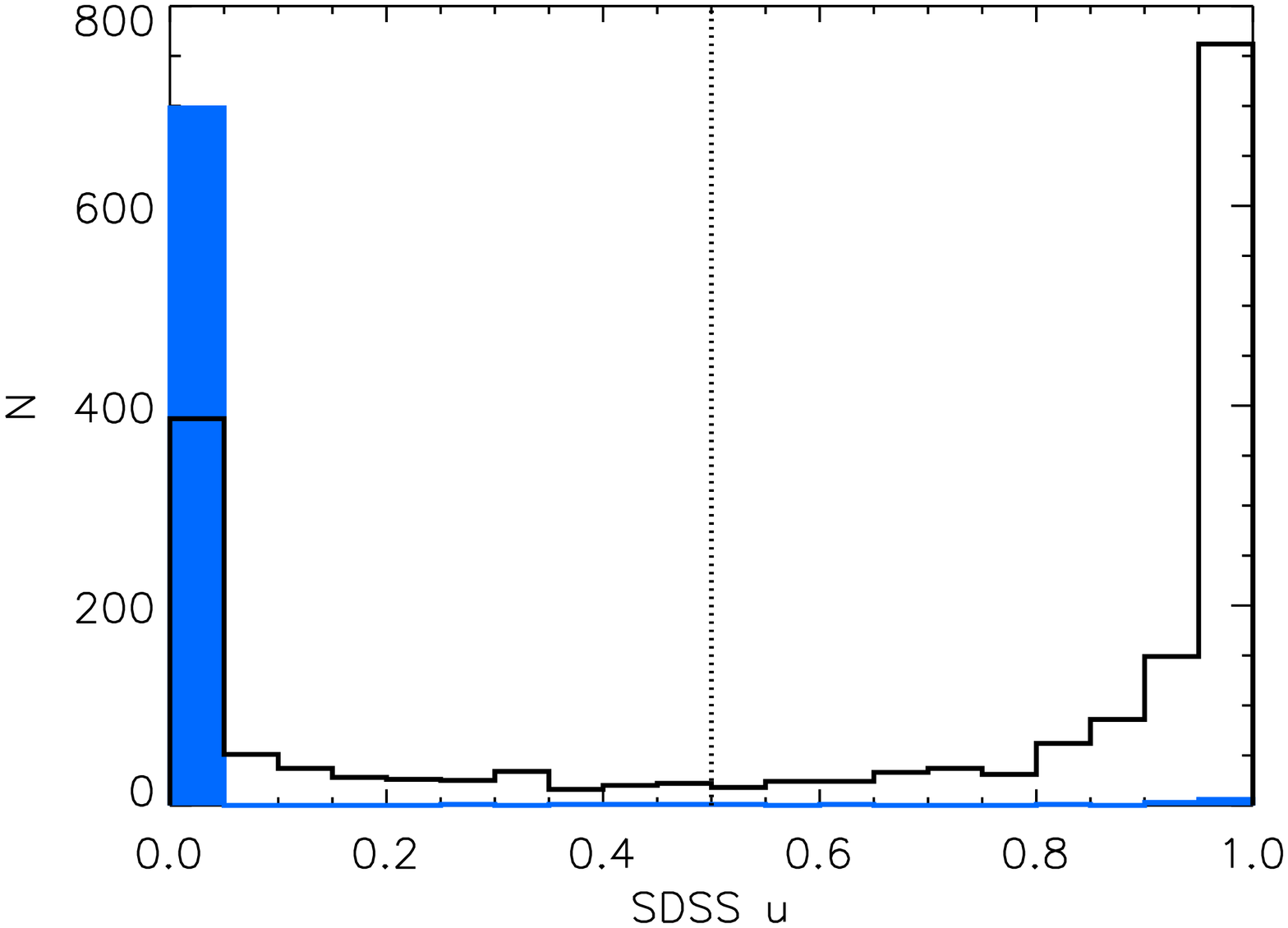}}~
\subfigure{\includegraphics[scale=0.2,angle=90]{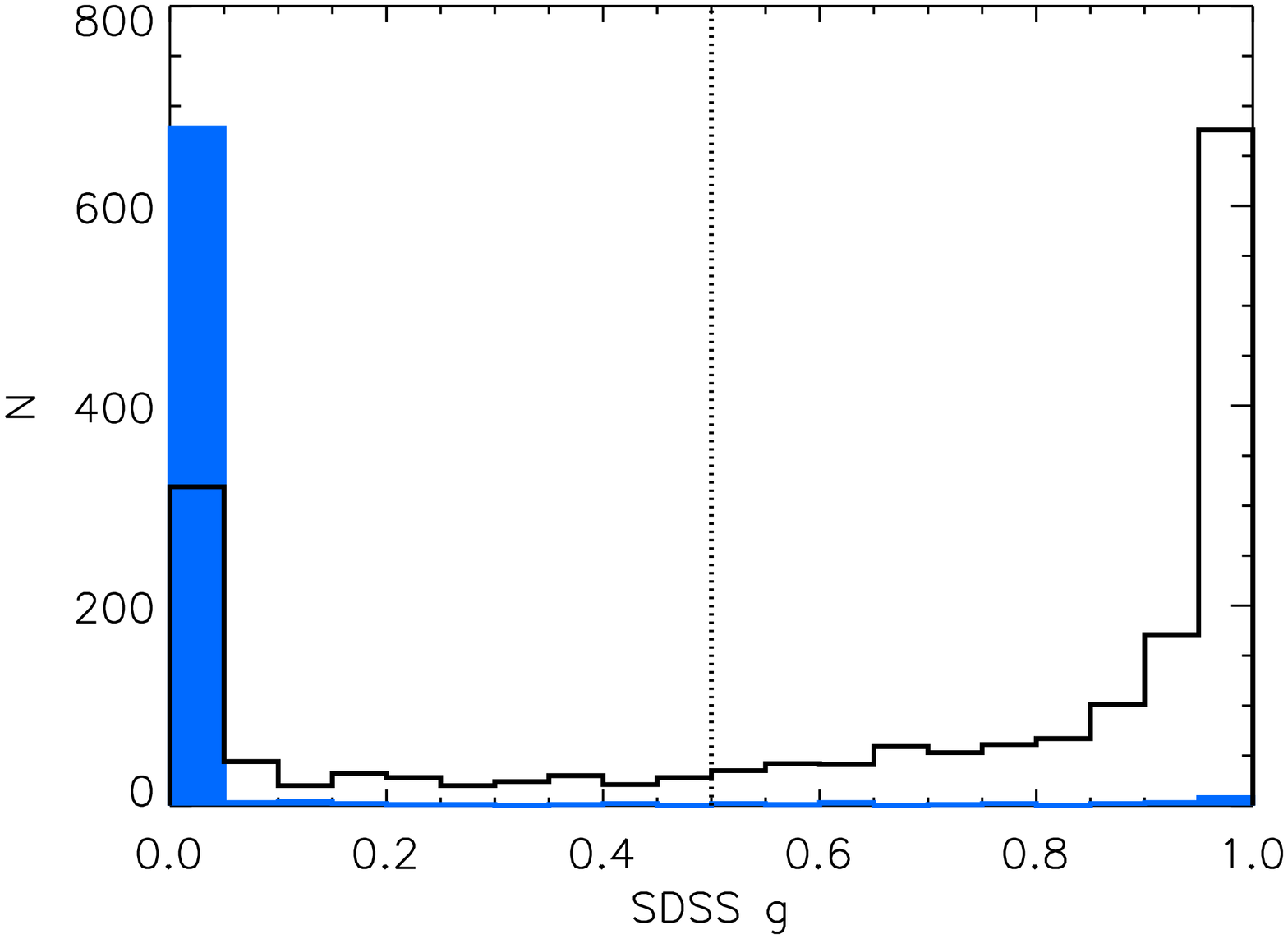}}
\subfigure{\includegraphics[scale=0.2,angle=90]{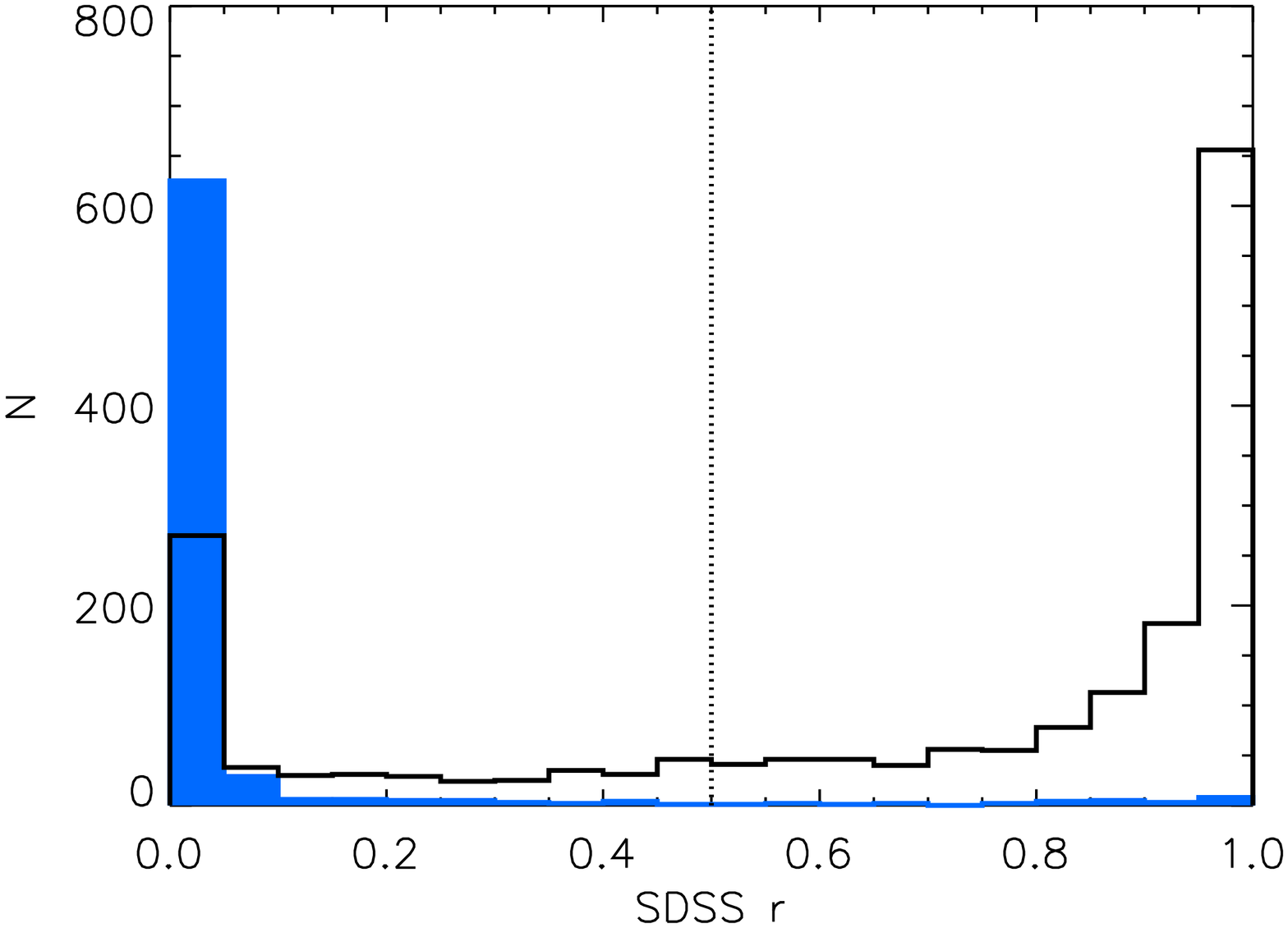}}~
\subfigure{\includegraphics[scale=0.2,angle=90]{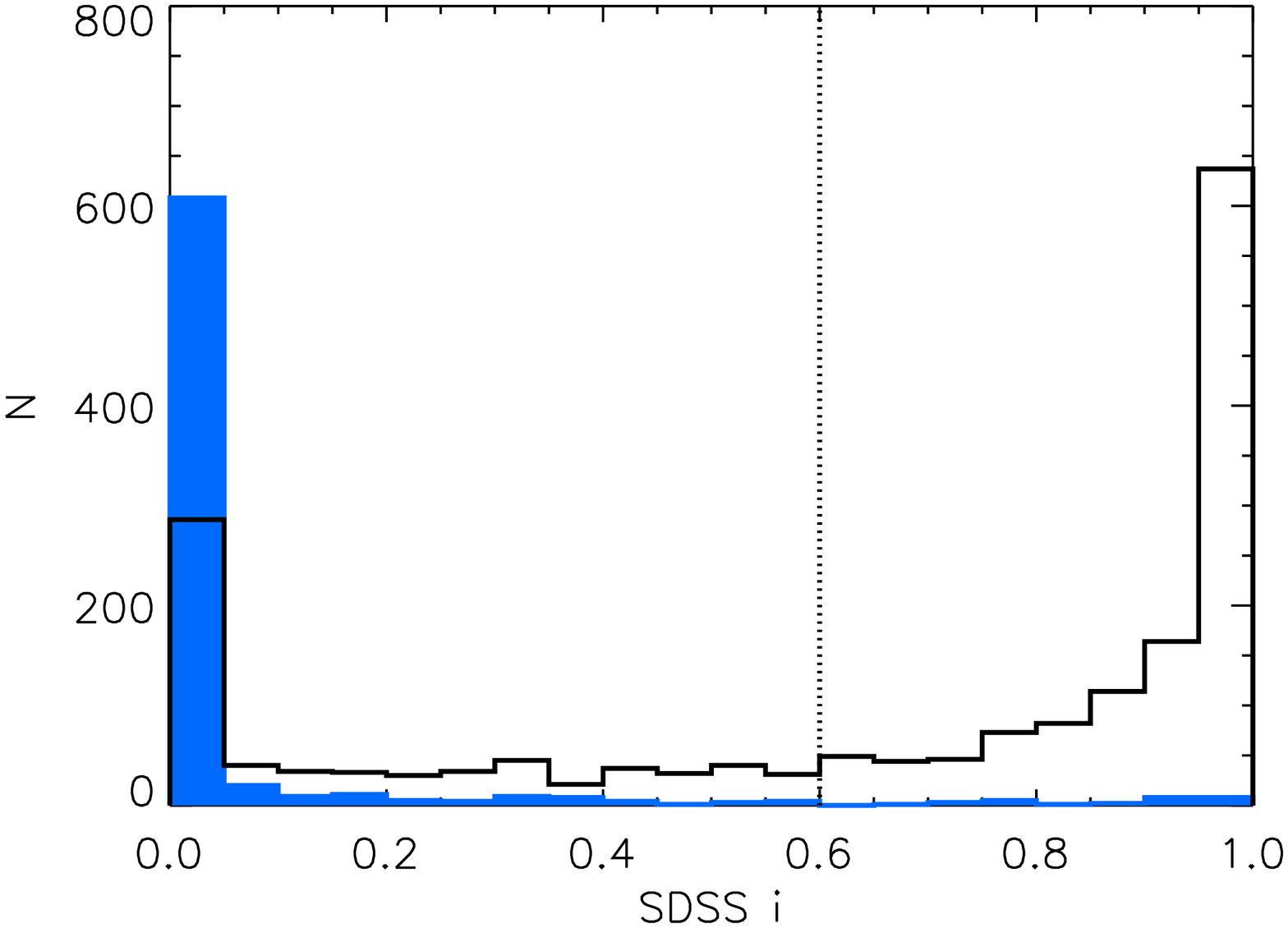}}
\subfigure{\includegraphics[scale=0.2,angle=90]{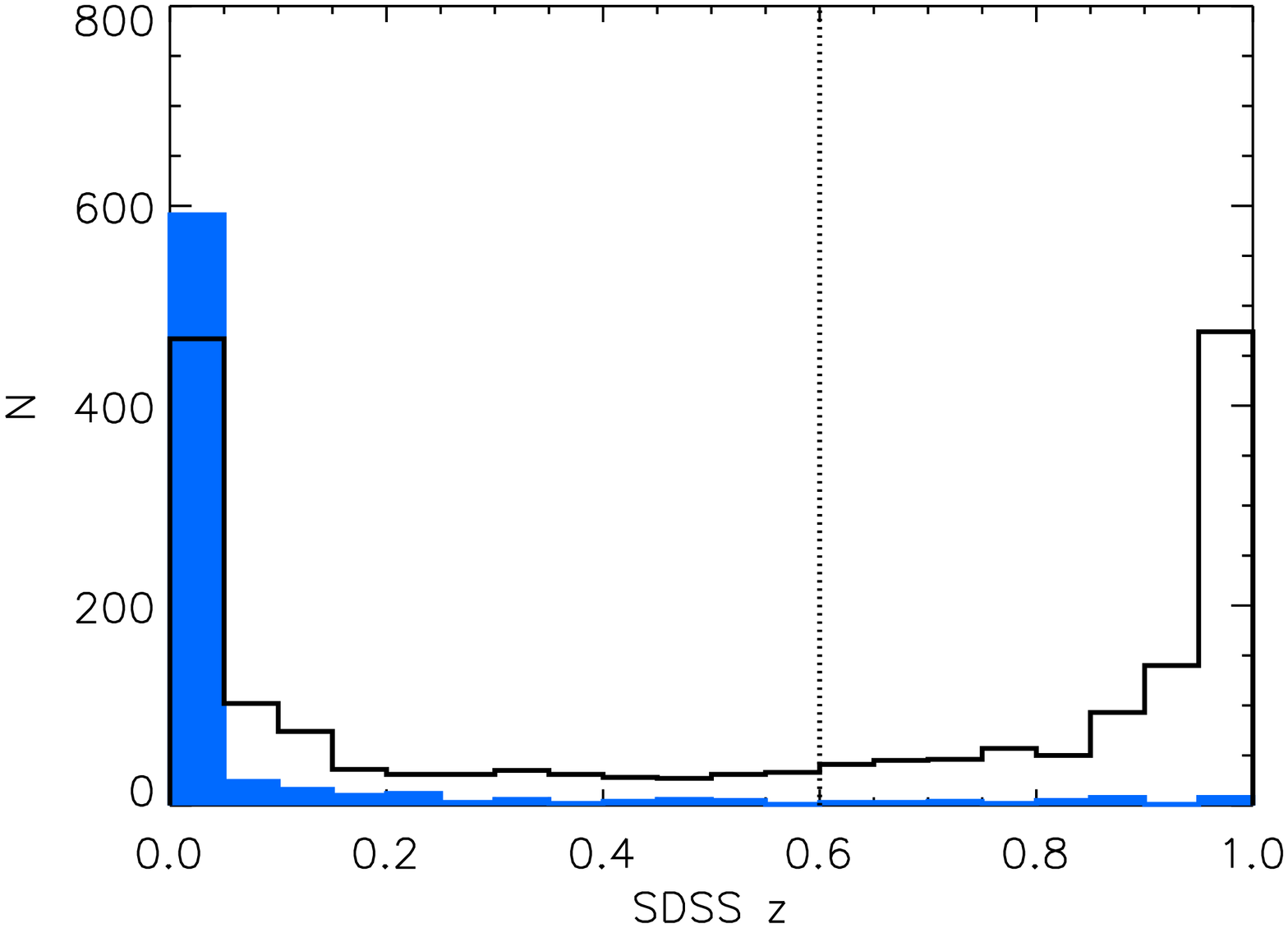}}
\caption[]{Reliability distributions for each SDSS band matched to {\it XMM-Newton} sources. The number of spurious associations above $R_{\rm crit}$ is predicted to be 12, 22, 28, 28 and 34 in the $u$, $g$, $r$, $i$ and $z$ bands, respectively. The dot-dash line indicates the adopted reliability threshold for claiming a counterpart.}
\end{figure}
\end{centering}

\begin{centering}
\begin{figure}
\subfigure{\includegraphics[scale=0.2,angle=90]{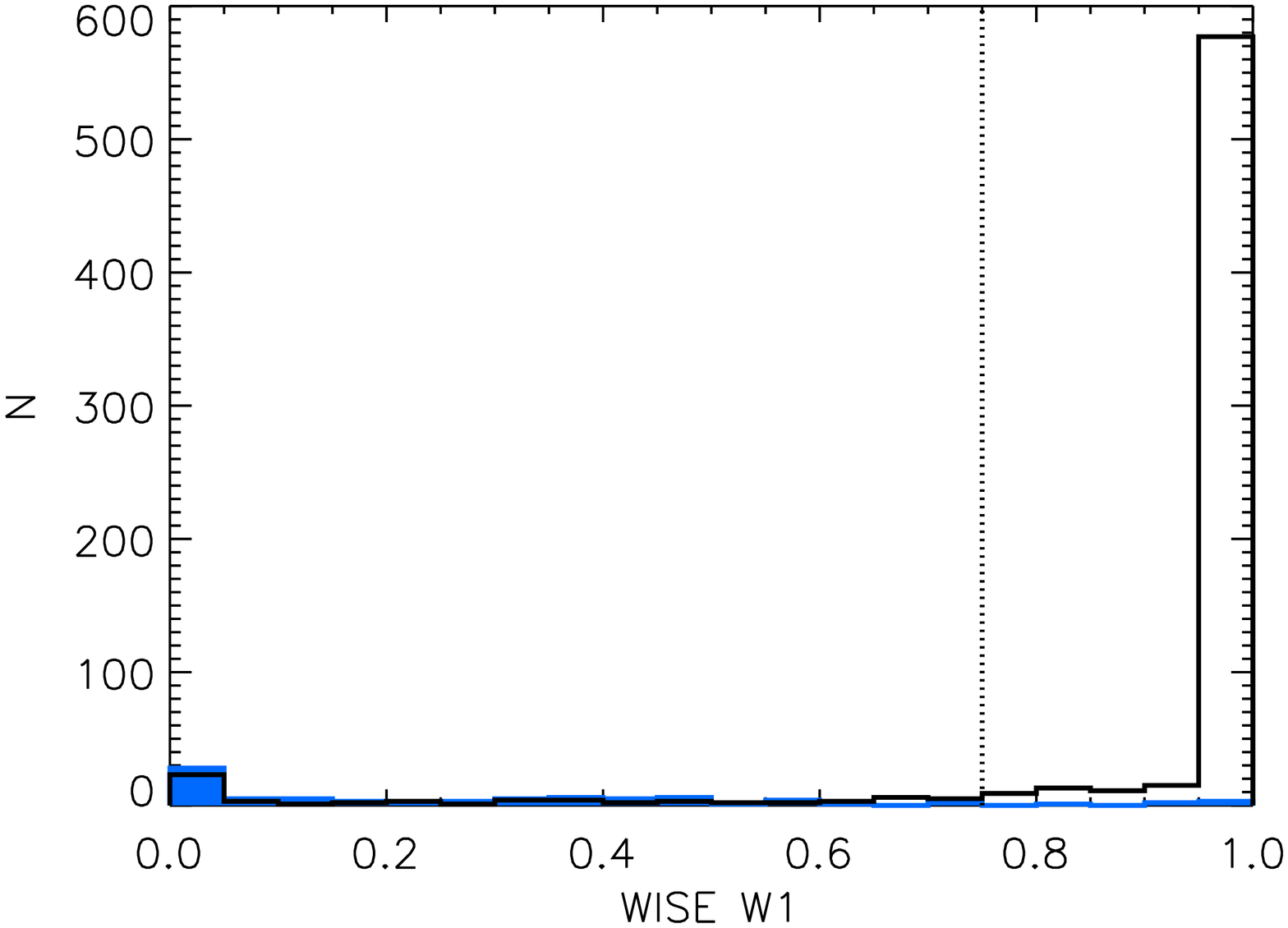}}~
\subfigure{\includegraphics[scale=0.2,angle=90]{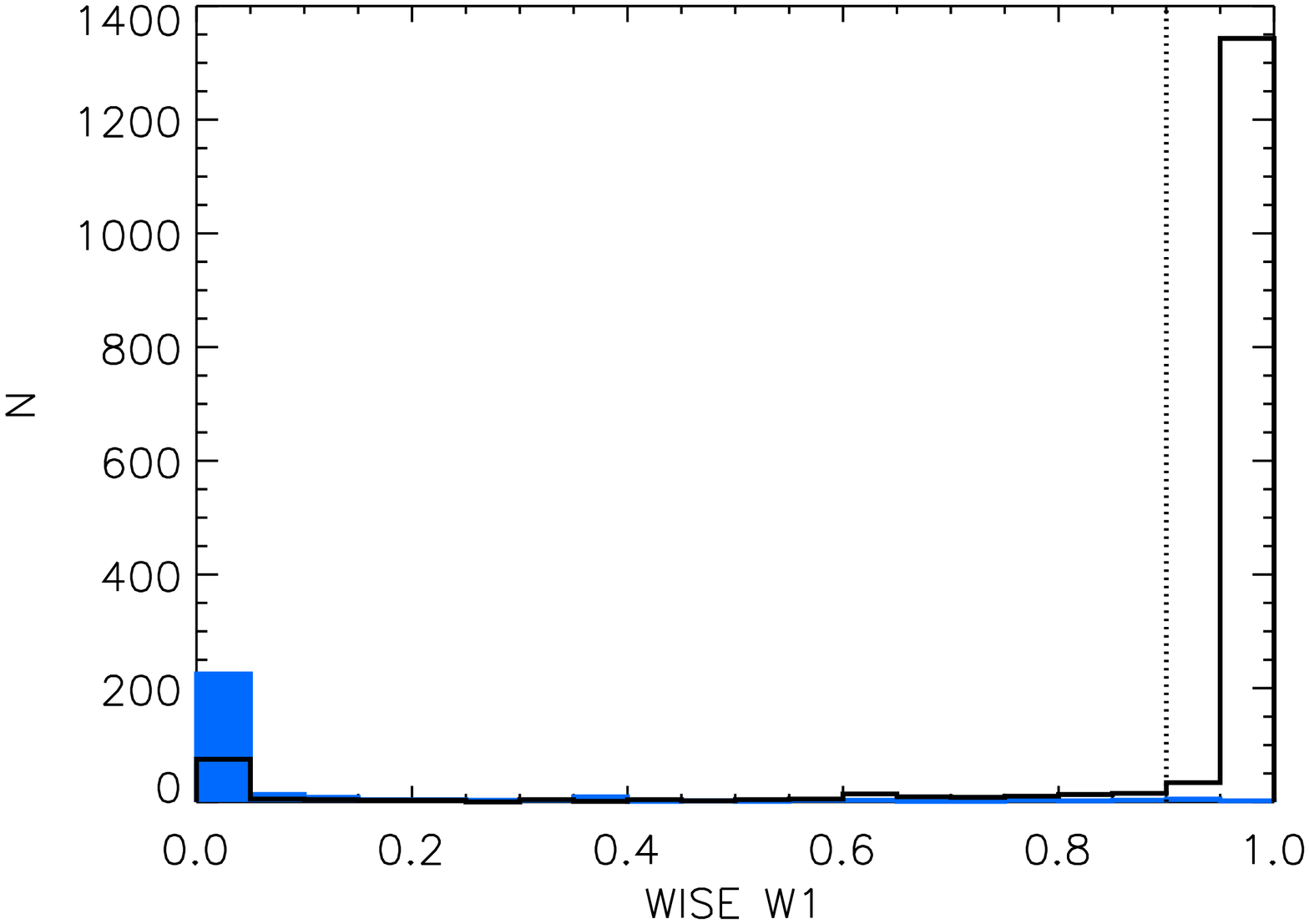}}
\caption[]{Reliability distributions for {\it WISE} band W1 matched to ({\it left}) {\it Chandra} sources and to ({\it right}) {\it XMM-Newton} objects. The number of spurious associations above $R_{\rm crit}$ is predicted to be 6 for the {\it Chandra} source list and 7 for the {\it XMM-Newton} catalog. The dot-dash line indicates the adopted reliability threshold for claiming a counterpart. }
\end{figure}
\end{centering}

\begin{centering}
\begin{figure}
\subfigure{\includegraphics[scale=0.2,angle=90]{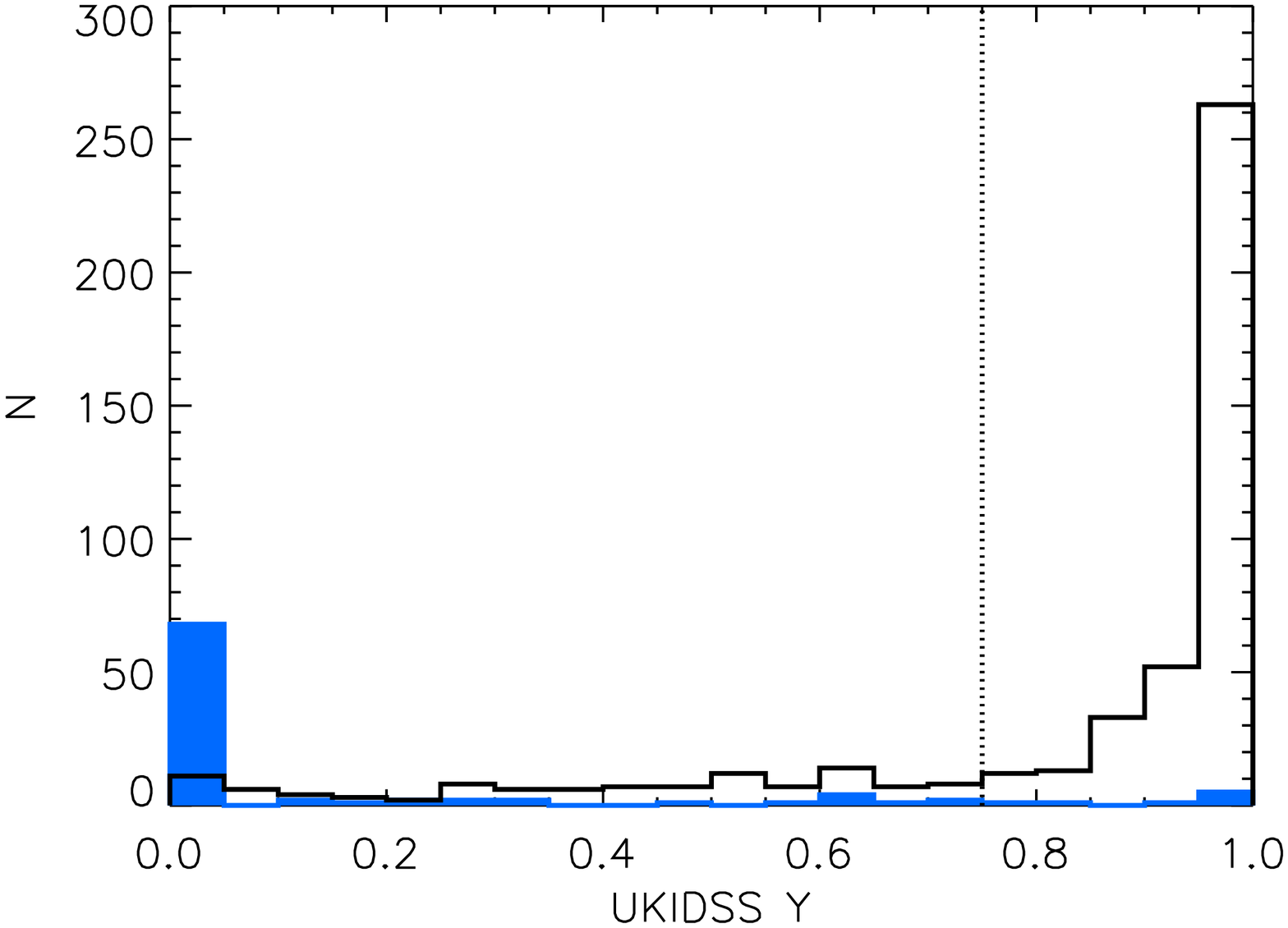}}~
\subfigure{\includegraphics[scale=0.2,angle=90]{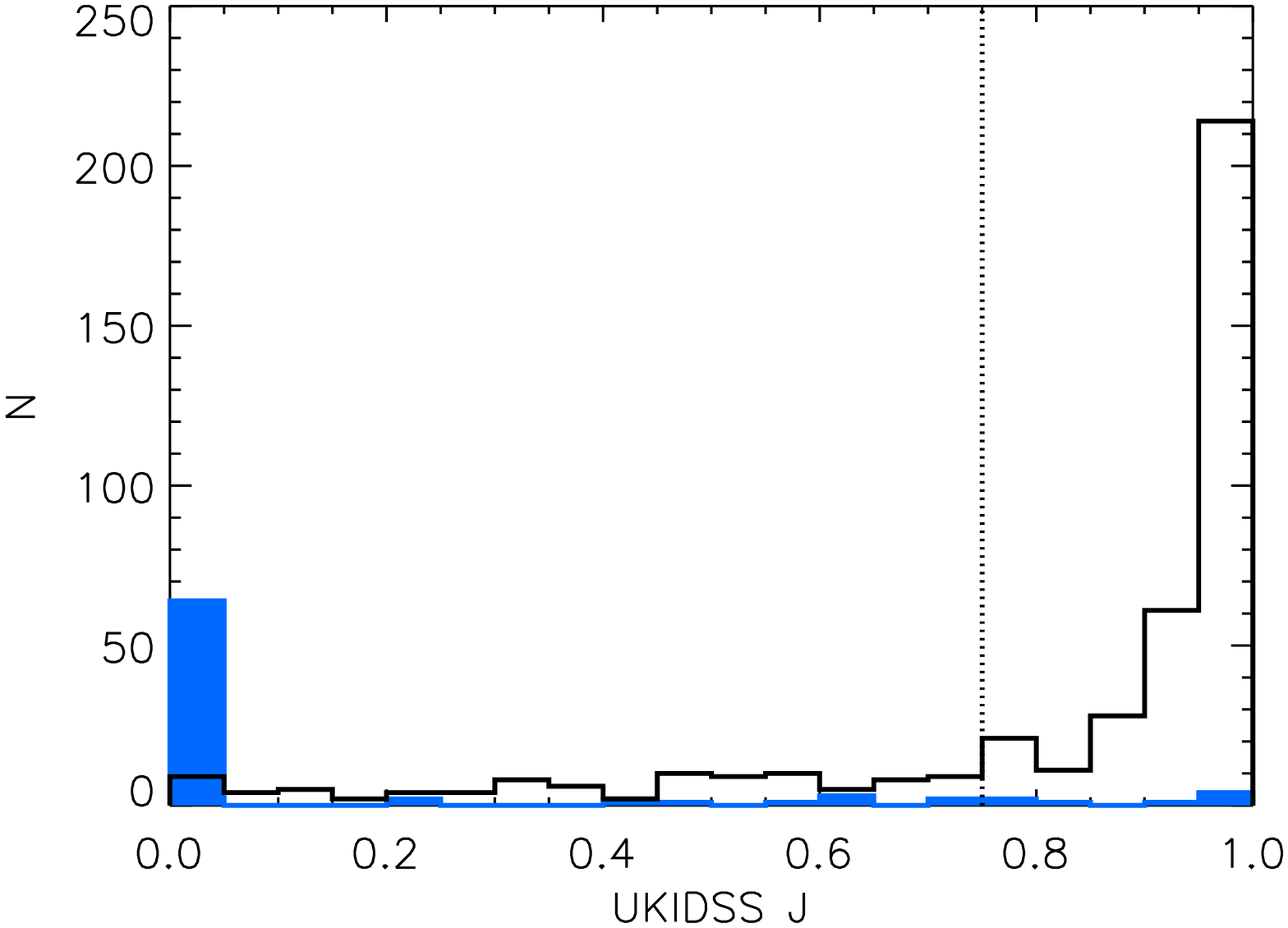}}
\subfigure{\includegraphics[scale=0.2,angle=90]{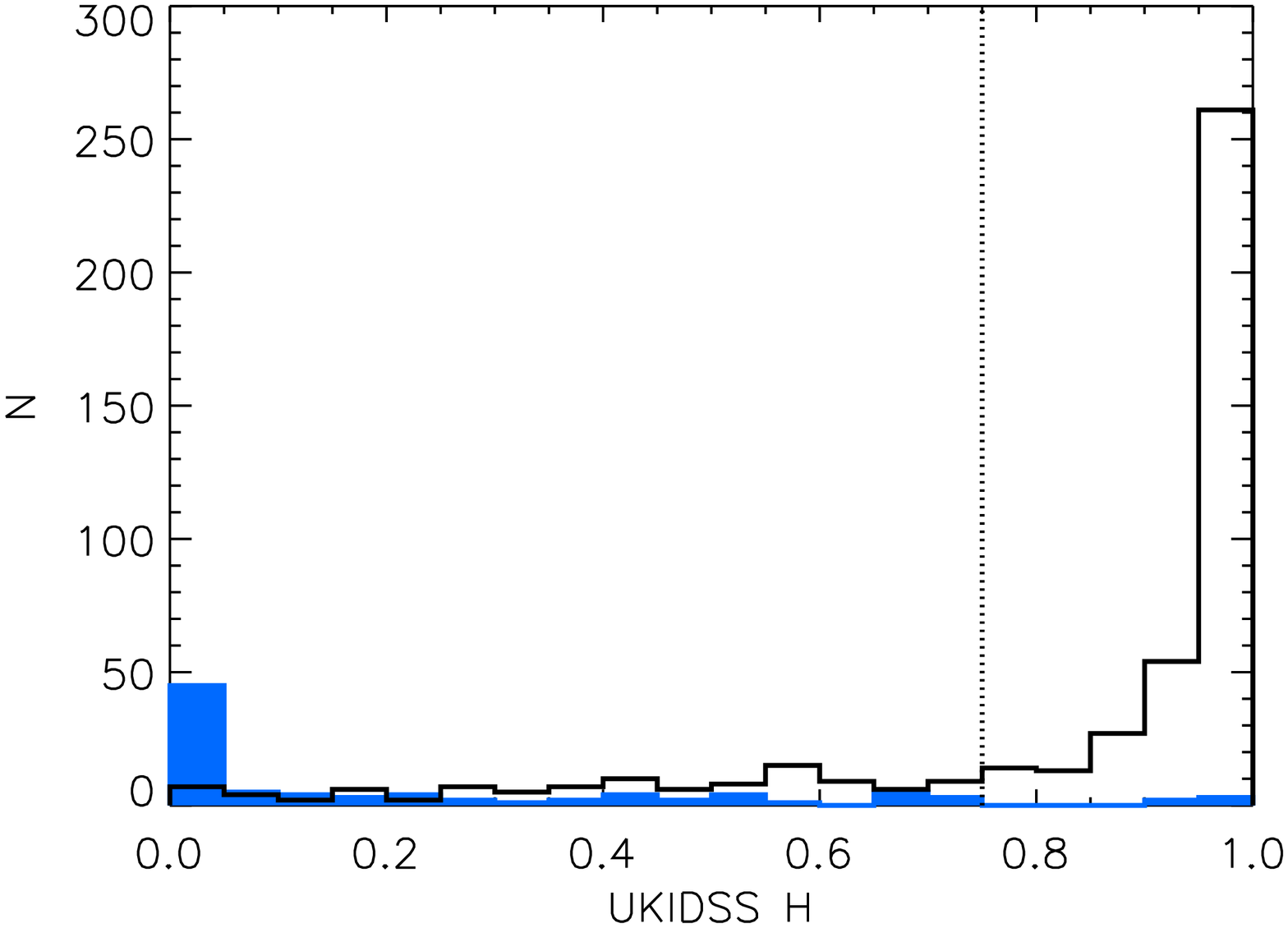}}~
\subfigure{\includegraphics[scale=0.2,angle=90]{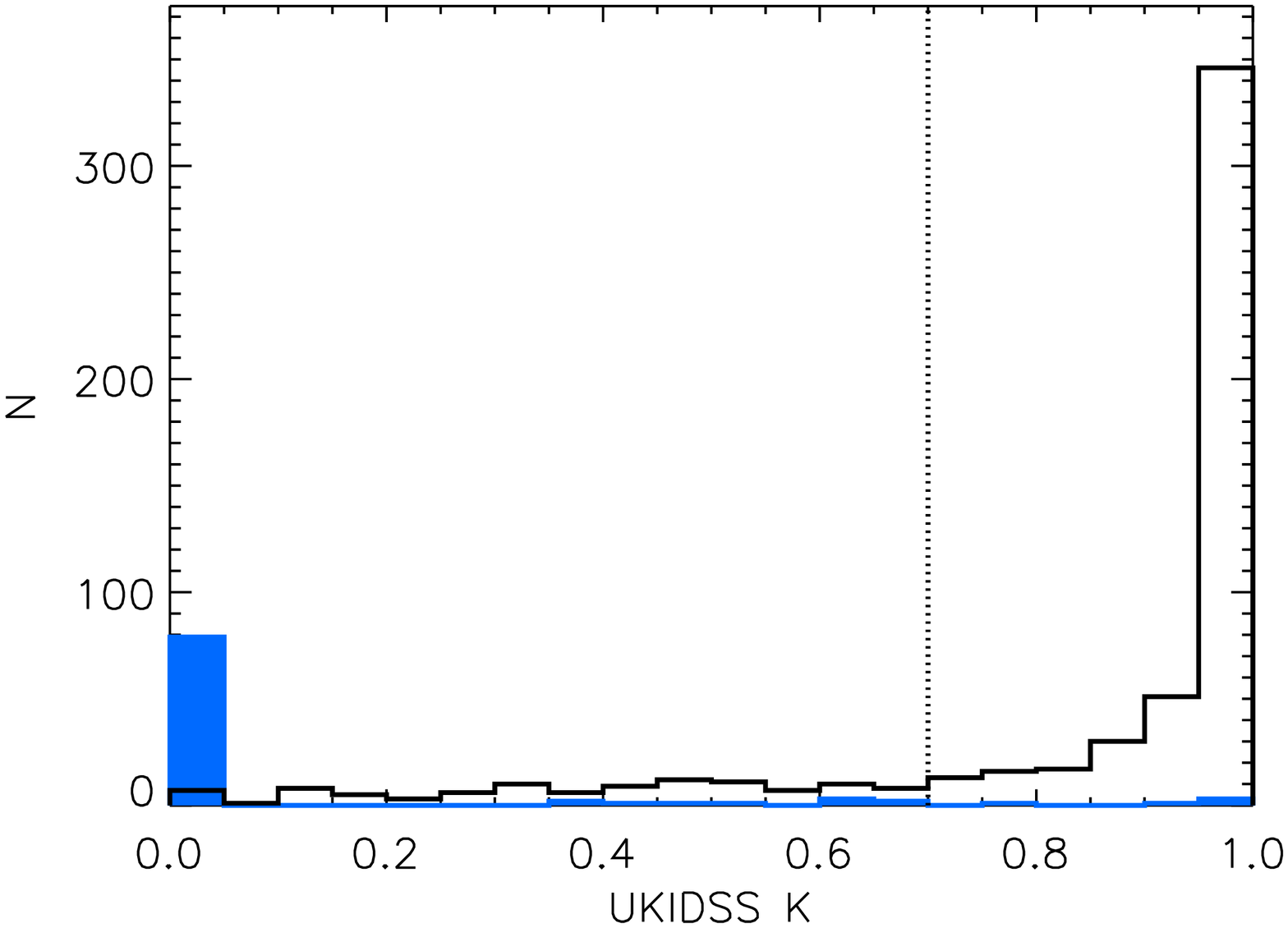}}
\caption[]{Reliability distributions for each UKIDSS band matched to {\it Chandra} sources. The number of spurious associations above $R_{\rm crit}$ is predicted at 8, 8, 5 and 5 in the $Y$, $J$, $H$ and $K$ bands, respectively. The dot-dash line indicates the adopted reliability threshold for claiming a counterpart.}
\end{figure}
\end{centering}

\begin{centering}
\begin{figure}
\subfigure{\includegraphics[scale=0.2,angle=90]{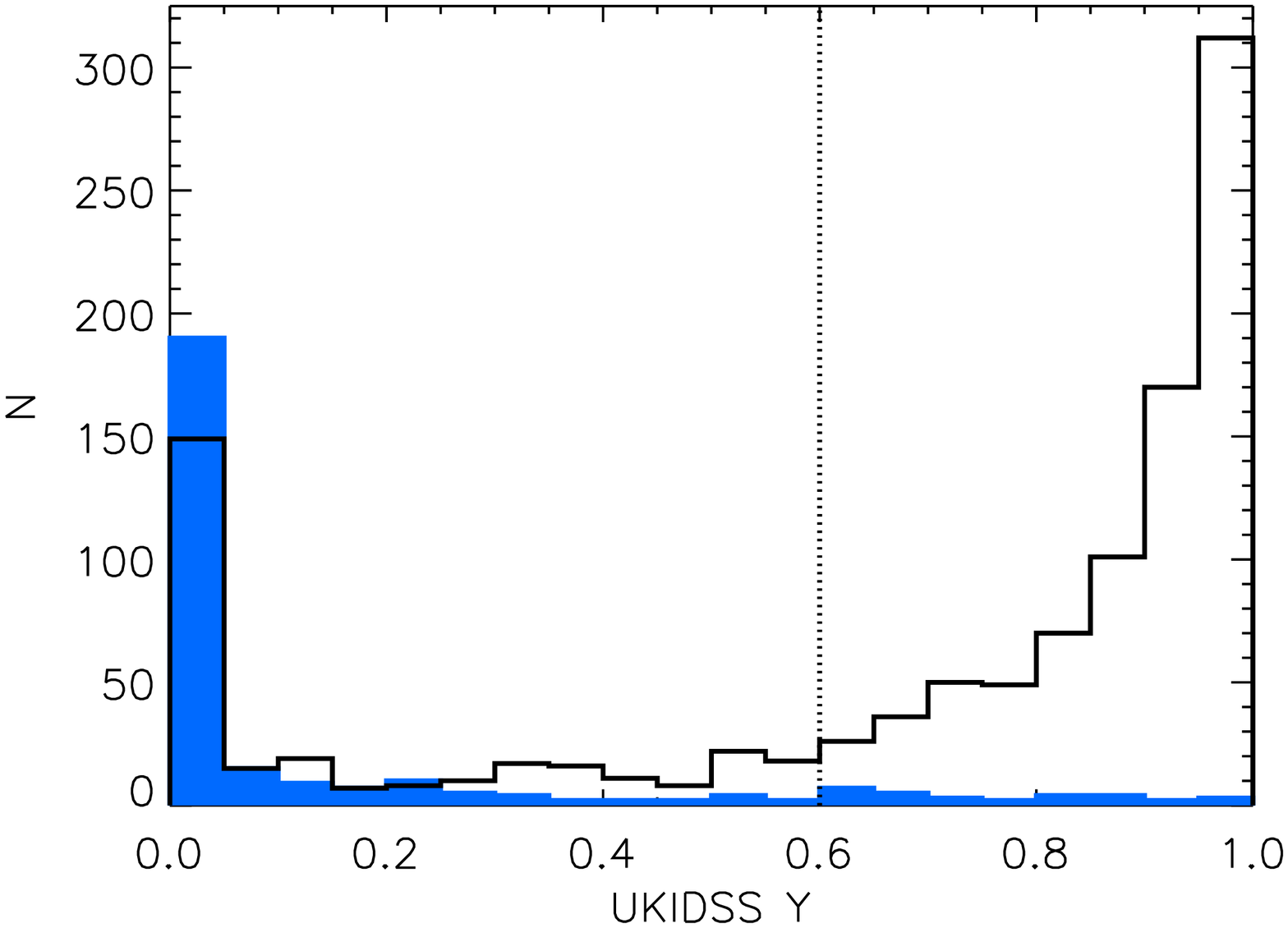}}~
\subfigure{\includegraphics[scale=0.2,angle=90]{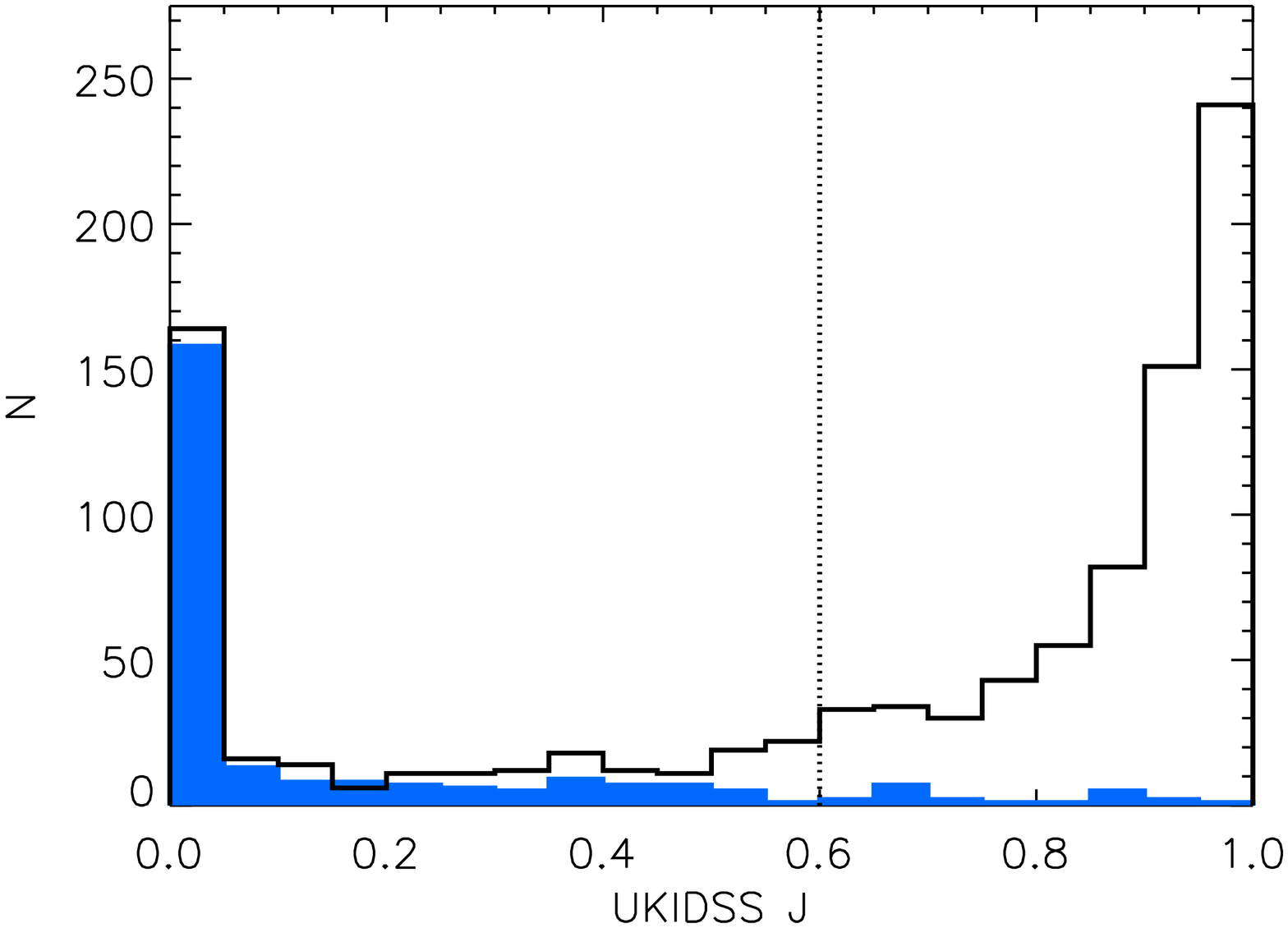}}
\subfigure{\includegraphics[scale=0.2,angle=90]{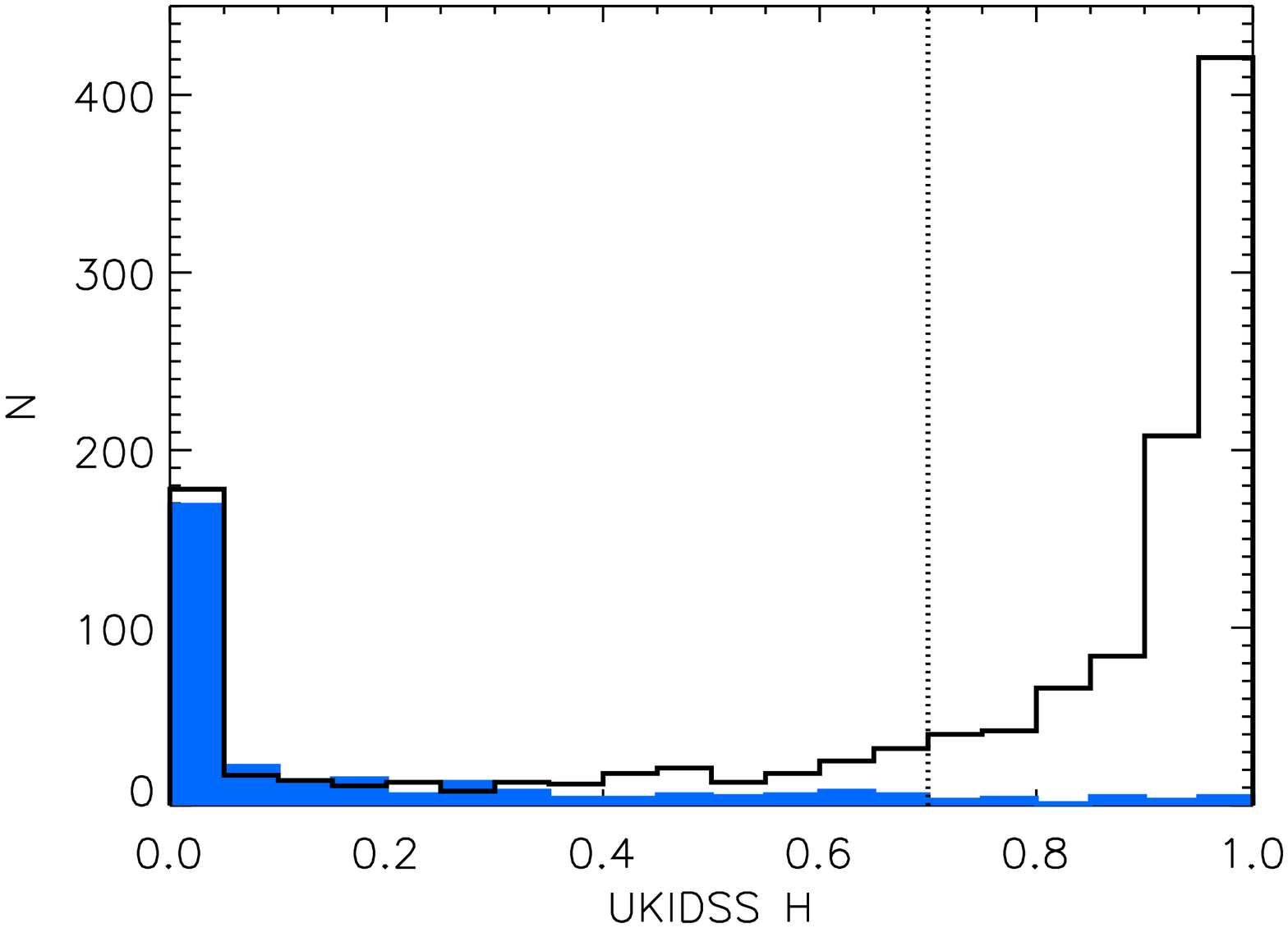}}~
\subfigure{\includegraphics[scale=0.2,angle=90]{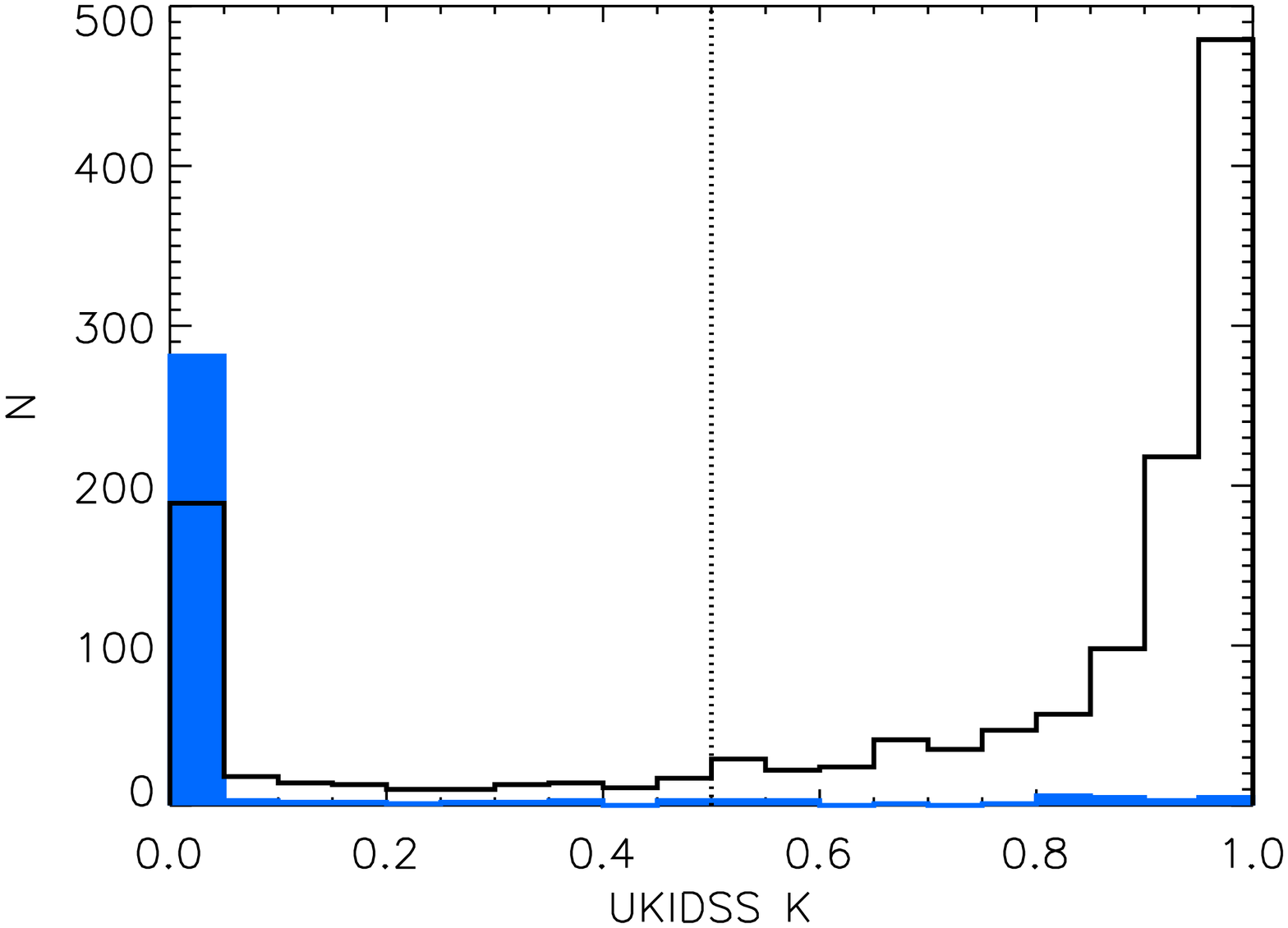}}
\caption[]{Reliability distributions for each UKIDSS band matched to {\it XMM-Newton} sources. The number of spurious associations above $R_{\rm crit}$ is predicted to be 21, 30, 21 and 27 in the $Y$, $J$, $H$ and $K$ bands, respectively. The dot-dash line indicates the adopted reliability threshold for claiming a counterpart.}
\end{figure}
\end{centering}

\begin{centering}
\begin{figure}
\subfigure{\includegraphics[scale=0.2,angle=90]{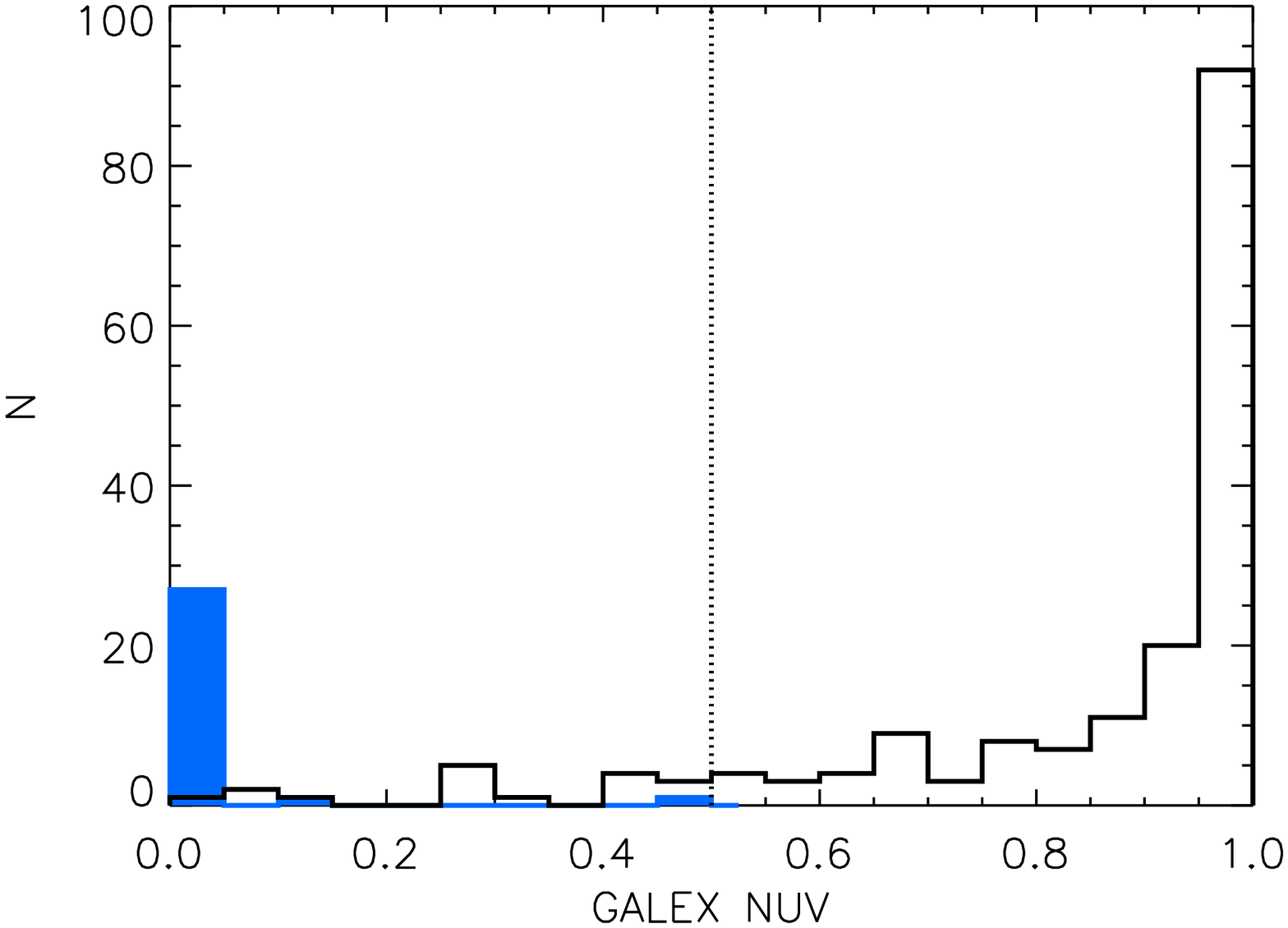}}~
\subfigure{\includegraphics[scale=0.2,angle=90]{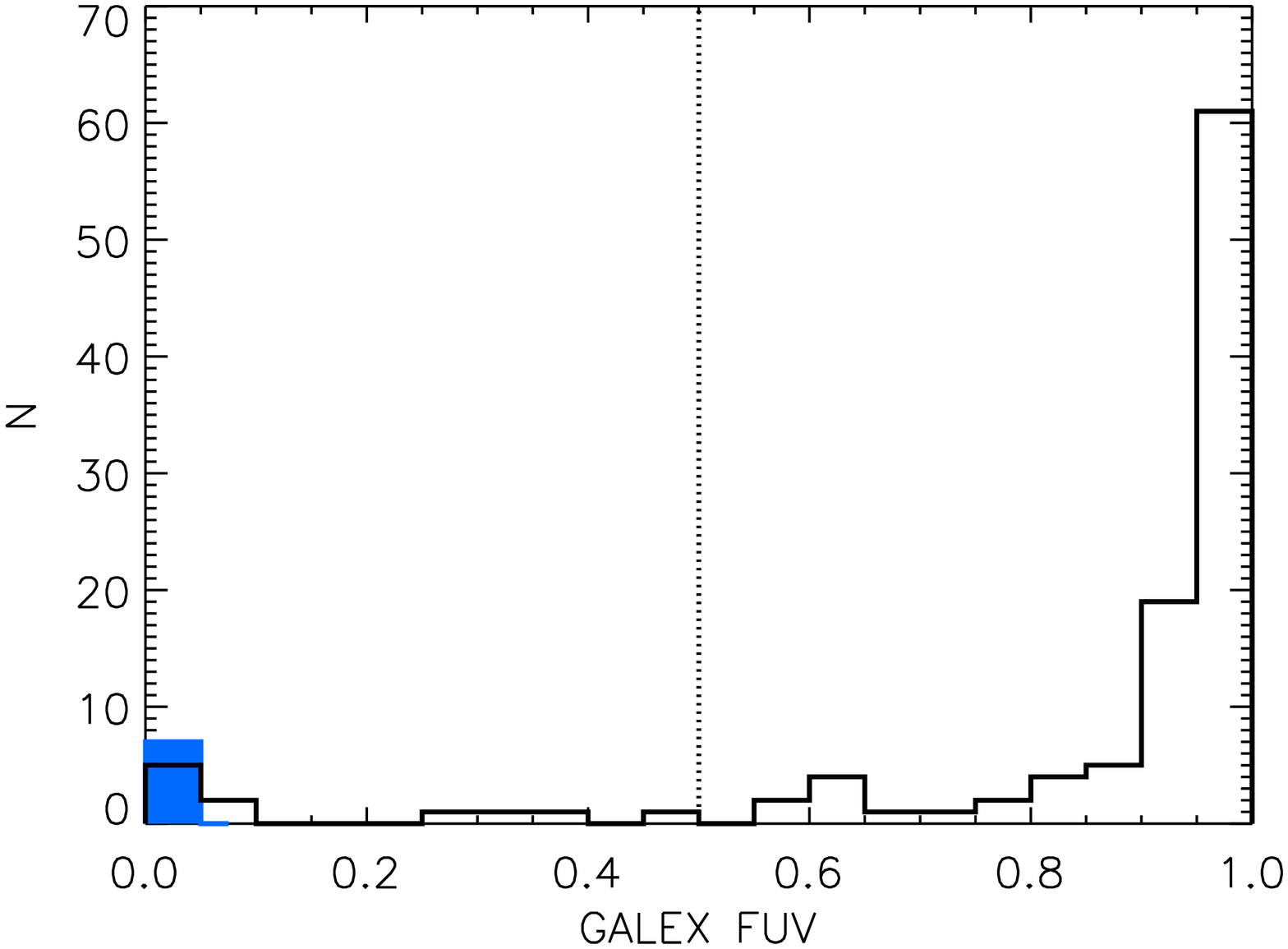}}
\caption[]{Reliability distributions for each {\it GALEX} band matched to {\it Chandra} sources. When shifting the X-ray positions by random amounts and re-running the MLE code, we find no spurious associations above $R_{\rm crit}$ in either band. The dot-dash line indicates the adopted reliability threshold for claiming a counterpart.}
\end{figure}
\end{centering}

\begin{centering}
\begin{figure}
\subfigure{\includegraphics[scale=0.2,angle=90]{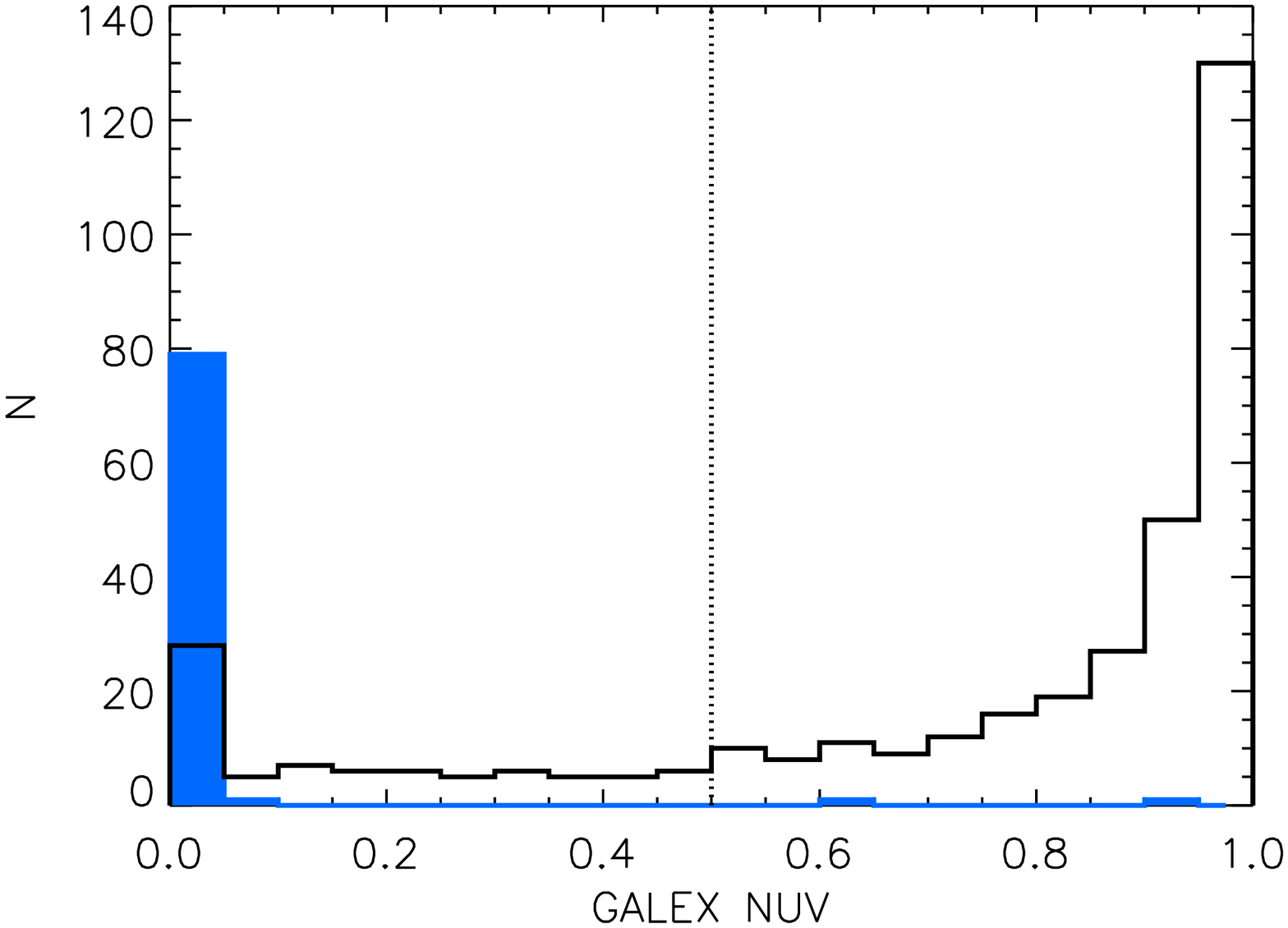}}~
\subfigure{\includegraphics[scale=0.2,angle=90]{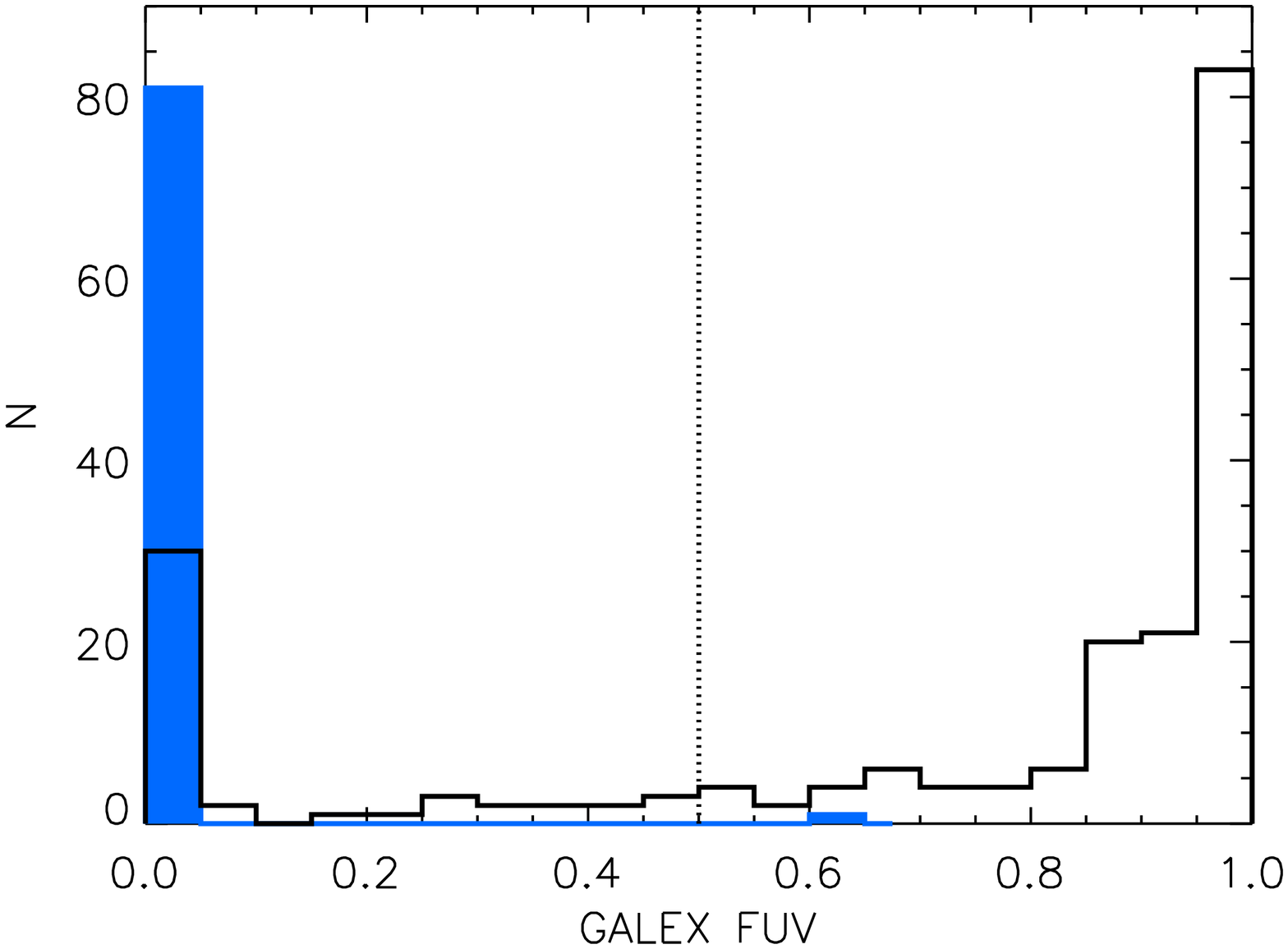}}
\caption[]{Reliability distributions for each {\it GALEX} band matched to {\it XMM-Newton} sources. We find two spurious counterparts above $R_{\rm crit}$ in the NUV band and one in the FUV band. The dot-dash line indicates the adopted reliability threshold for claiming a counterpart.}
\end{figure}
\end{centering}

\clearpage
\section[]{Column Descriptions for On-line Versions of the Catalogs}
Non-significant X-ray fluxes have zero values in the on-line catalogs. When reporting the ancillary multi-wavelength data, numeric values of -999 and null strings indicate that a reliable counterpart was not identified for that X-ray source.

\subsection{{\it Chandra}}

\renewcommand{\theenumi}{\arabic{enumi}}
\begin{enumerate}
\item {\bf MSID}: {\it Chandra} Source Catalog identification number \citep{csc}
\item {\bf ObsID}: {\it Chandra} observation identification number
\item {\bf RA}: {\it Chandra} RA (J2000)
\item {\bf Dec} {\it Chandra} Dec (J2000)
\item {\bf RADec\_err}: {\it Chandra} positional error (arcsec)
\item {\bf Dist\_nn}: Distance to nearest {\it Chandra} source (arcsec)
\item {\bf Soft\_Flux}: 0.5-2 keV flux (10$^{-14}$ erg cm$^{-2}$ s$^{-1}$). Set to 0 if flux is not significant at $>$4.5$\sigma$ level.
\item {\bf Soft\_flux\_error\_high}: higher bound on 0.5-2 keV flux (10$^{-14}$ erg cm$^{-2}$ s$^{-1}$). If flux is 0, this is the flux upper limit.
\item {\bf Soft\_flux\_error\_low}: lower bound on 0.5-2 keV flux (10$^{-14}$ erg cm$^{-2}$ s$^{-1}$)
\item {\bf Hard\_flux}: 2-7 keV flux (10$^{-14}$ erg cm$^{-2}$ s$^{-1}$). Set to 0 if flux is not significant at $>$4.5$\sigma$ level.
\item {\bf Hard\_flux\_error\_high}: higher bound on 2-7 keV flux (10$^{-14}$ erg cm$^{-2}$ s$^{-1}$). If flux is 0, this is the flux upper limit.
\item {\bf Hard\_flux\_error\_lo}: lower bound on 2-7 keV flux (10$^{-14}$ erg cm$^{-2}$ s$^{-1}$)
\item {\bf Full\_flux}: 0.5-7 keV flux (10$^{-14}$ erg cm$^{-2}$ s$^{-1}$). Set to 0 if flux is not significant at $>$4.5$\sigma$ level.
\item {\bf Full\_flux\_error\_high}: higher bound on 0.5-7 keV flux (10$^{-14}$ erg cm$^{-2}$ s$^{-1}$). If flux is 0, this is the flux upper limit.
\item {\bf Full\_flux\_error\_lo}: lower bound on 0.5-7 keV flux (10$^{-14}$ erg cm$^{-2}$ s$^{-1}$)
\item {\bf Lum\_soft}: log 0.5-2 keV luminosity (erg s$^{-1}$)
\item {\bf Lum\_hard}: log 2-7 keV luminosity (erg s$^{-1}$)
\item {\bf Lum\_full}: log 0.5-7 keV luminosity (erg s$^{-1}$)
\item {\bf In\_XMM}: Set to `yes' if X-ray source is in the {\it XMM-Newton} Stripe 82 catalog
\item {\bf Removed\_LogN\_LogS}: Set to `yes' if X-ray source was not part of the Log$N$-Log$S$ relation published in \citep{me}
\item {\bf SDSS\_Rej}: Set to `yes' if SDSS counterpart is found but rejected due to poor photometry.
\item {\bf SDSS\_Objid}: SDSS object identification number
\item {\bf SDSS\_RA}: SDSS RA (J2000)
\item {\bf SDSS\_Dec}: SDSS Dec (J2000)
\item {\bf SDSS\_Rel}: MLE reliability of SDSS match to X-ray source
\item {\bf SDSS\_Dist}: Distance between X-ray and SDSS source (arcsec)
\item {\bf u\_mag}: SDSS u mag
\item {\bf u\_err}: SDSS u mag error
\item {\bf g\_mag}: SDSS g mag
\item {\bf g\_err}: SDSS g mag error
\item {\bf r\_mag}: SDSS r mag
\item {\bf r\_err}: SDSS r mag error
\item {\bf i\_mag}: SDSS i mag
\item {\bf i\_err}: SDSS i mag error
\item {\bf z\_mag}: SDSS z mag
\item {\bf z\_err}: SDSS z mag error
\item {\bf Specobjid}: SDSS spectroscopic object identification number
\item {\bf Class}: optical spectroscopic class (if available)
\item {\bf Redshift}: spectroscopic redshift
\item {\bf z\_src}: source of spectroscopic redshift; 0 - SDSS, 1 - 2SLAQ, 2 - WiggleZ, 3 - DEEP2, 4 - SDSS spectra re-fit/verified by us
\item {\bf WISE\_Name}: {\it WISE} name
\item {\bf WISE\_RA}: {\it WISE} RA (J2000)
\item {\bf WISE\_Dec}: {\it WISE} Dec (J2000)
\item {\bf WISE\_sigra}: {\it WISE} RA error (arcsec)
\item {\bf WISE\_sigdec}: {\it WISE} Dec error (arcsec)
\item {\bf WISE\_Rel}: MLE reliability of {\it WISE} match to X-ray source
\item {\bf WISE\_Dist}: Distance between X-ray and {\it WISE} source (arcsec)
\item {\bf W1}: {\it WISE} W1 mag. All {\it WISE} magnitudes are from profile-fitting photometry, unless the {\bf WISE\_ext} flag is set to `yes,' in which case the magnitudes are associated with elliptical apertures. 
\item {\bf W1sig}: {\it WISE} W1 error
\item {\bf W1SNR}: {\it WISE} W1 SNR. Any {\it WISE} magnitudes with SNR $<$2 are upper limits.
\item {\bf W2}: {\it WISE} W2 mag
\item {\bf W2sig}: {\it WISE} W2 error
\item {\bf W2SNR}: {\it WISE} W2 SNR
\item {\bf W3}: {\it WISE} W3 mag
\item {\bf W3sig}: {\it WISE} W3 error
\item {\bf W3SNR}: {\it WISE} W3 SNR
\item {\bf W4}: {\it WISE} W4 mag
\item {\bf W4sig}: {\it WISE} W4 error
\item {\bf W4SNR}: {\it WISE} W4 SNR
\item {\bf WISE\_ext}: Set to `yes' if {\it WISE} source is extended
\item {\bf WISE\_rej}: Set to `yes' if {\it WISE} counterpart is found but rejected for poor photometry
\item {\bf UKIDSS\_ID}: UKIDSS ID
\item {\bf UKIDSS\_RA}: UKIDSS RA (J2000)
\item {\bf UKIDSS\_Dec}: UKIDSS Dec (J2000)
\item {\bf UKIDSS\_Rel}: MLE reliability of UKIDSS match to X-ray source
\item {\bf UKIDSS\_Dist}: Distance between X-ray and UKIDSS source (arcsec)
\item {\bf Ymag}: UKIDSS Y mag
\item {\bf Ysig}: UKIDSS Y error
\item {\bf Hmag}: UKIDSS H mag
\item {\bf Hsig}: UKIDSS H error
\item {\bf Jmag}: UKIDSS J mag
\item {\bf Jsig}: UKIDSS J error
\item {\bf Kmag}: UKIDSS K mag
\item {\bf Ksig}: UKIDSS K error
\item {\bf UKIDSS\_Rej}: UKIDSS counterpart found but rejected for poor photometry
\item {\bf GALEX\_Objid}: {\it GALEX} object identification number
\item {\bf GALEX\_RA}: {\it GALEX} RA (J2000)
\item {\bf GALEX\_Dec}: {\it GALEX} Dec (J2000)
\item {\bf NUV\_poserr}: {\it GALEX} NUV positional error (arcsec)
\item {\bf FUV\_poserr}: {\it GALEX} FUV positional error (arcsec)
\item {\bf GALEX\_Rel}: MLE reliability of {\it GALEX} match to X-ray source
\item {\bf GALEX\_Dist}: Distance between X-ray and {\it GALEX source} (arcsec)
\item {\bf NUV\_mag}: {\it GALEX} NUV mag
\item {\bf NUV\_magerr}: {\it GALEX} NUV error
\item {\bf FUV\_mag}: {\it GALEX} FUV mag
\item {\bf FUV\_magerr}: {\it GALEX} FUV error
\item {\bf FIRST Name}: IAU Name of FIRST counterpart
\item {\bf FIRST\_RA}: FIRST RA (J2000)
\item {\bf FIRST\_Dec}: FIRST Dec (J2000)
\item {\bf FIRST\_Dist}: Distance between X-ray and FIRST source (arcsec)
\item {\bf FIRST\_Flux}: FIRST 5 GHz Flux Density (Jy)
\item {\bf FIRST\_err}: FIRST 5 GZ Flux Density error (Jy)

\end{enumerate}

\subsection{{\it XMM-Newton}}

\begin{enumerate}
\item {\bf Rec\_no}: Unique record number assigned to each {\it XMM-Newton} source
\item {\bf ObsID}: {\it XMM-Newton} observation identification number
\item {\bf RA}: {\it XMM-Newton} RA (J2000)
\item {\bf Dec}: {\it XMM-Newton} Dec(J2000)
\item {\bf RADec\_Err}: {\it XMM-Newton} positional error (arcsec)
\item {\bf Dist\_nn}: Distance to nearest {\it XMM-Newton} source (arcsec)
\item {\bf Soft\_flux}: 0.5-2 keV flux (10$^{-14}$ erg cm$^{-2}$ s$^{-1}$). Flux is 0 if $det\_ml < 15$ in the soft band.
\item {\bf Soft\_flux\_err}: error in 0.5-2 keV flux (10$^{-14}$ erg cm$^{-2}$ s$^{-1}$)
\item {\bf Hard\_flux}: 2-10 keV flux (10$^{-14}$ erg cm$^{-2}$ s$^{-1}$).  Flux is 0 if $det\_ml < 15$ in the hard band.
\item {\bf Hard\_flux\_err}:error in 2-10 keV flux (10$^{-14}$ erg cm$^{-2}$ s$^{-1}$)
\item {\bf Full\_flux}: 0.5-10 keV flux (10$^{-14}$ erg cm$^{-2}$ s$^{-1}$).  Flux is 0 if $det\_ml < 15$ in the full band.
\item {\bf Full\_flux\_err}:error in 0.5-10 keV flux (10$^{-14}$ erg cm$^{-2}$ s$^{-1}$)
\item {\bf Lum\_soft}: log 0.5-2k keV luminosity (erg s$^{-1}$)
\item {\bf Lum\_hard}: log 2-10 keV luminosity (erg s$^{-1}$)
\item {\bf Lum\_full}: log 0.5-10 keV luminosity (erg s$^{-1}$)
\item {\bf In\_Chandra}: Set to `yes' if source is found in the {\it Chandra} catalog.
\item {\bf Removed\_LogN\_LogS}: Set to `yes' if source is removed from Log$N$-Log$S$ calulation presented in the main text.
\item {\bf SDSS\_Rej}: Set to `yes' if SDSS counterpart is found but rejected due to poor photometry.
\item {\bf SDSS\_Objid}: SDSS object identification number
\item {\bf SDSS\_RA}: SDSS RA (J2000)
\item {\bf SDSS\_Dec}: SDSS Dec (J2000)
\item {\bf SDSS\_Rel}: MLE reliability of SDSS match to X-ray source
\item {\bf SDSS\_Dist}: Distance between X-ray and SDSS source (arcsec)
\item {\bf u\_mag}: SDSS u mag
\item {\bf u\_err}: SDSS u mag error
\item {\bf g\_mag}: SDSS g mag
\item {\bf g\_err}: SDSS g mag error
\item {\bf r\_mag}: SDSS r mag
\item {\bf r\_err}: SDSS r mag error
\item {\bf i\_mag}: SDSS i mag
\item {\bf i\_err}: SDSS i mag error
\item {\bf z\_mag}: SDSS z mag
\item {\bf z\_err}: SDSS z mag error
\item {\bf Specobjid}: SDSS spectroscopic object identification number
\item {\bf Class}: optical spectroscopic class (if available)
\item {\bf Redshift}: spectroscopic redshift
\item {\bf z\_src}: source of spectroscopic redshift; 0 - SDSS, 1 - 2SLAQ, 2 - WiggleZ, 3 - DEEP2, 4 - SDSS spectra re-fit/verified by us
\item {\bf WISE\_Name}: {\it WISE} name
\item {\bf WISE\_RA}: {\it WISE} RA (J2000)
\item {\bf WISE\_Dec}: {\it WISE} Dec (J2000)
\item {\bf WISE\_sigra}: {\it WISE} RA error (arcsec)
\item {\bf WISE\_sigdec}: {\it WISE} Dec error (arcsec)
\item {\bf WISE\_Rel}: MLE reliability of {\it WISE} match to X-ray source
\item {\bf WISE\_Dist}: Distance between X-ray and {\it WISE} source (arcsec)
\item {\bf W1}: {\it WISE} W1 mag. All {\it WISE} magnitudes are from profile-fitting photometry, unless the {\bf WISE\_ext} flag is set to `yes,' in which case the magnitudes are associated with elliptical apertures. 
\item {\bf W1sig}: {\it WISE} W1 error
\item {\bf W1SNR}: {\it WISE} W1 SNR. Any {\it WISE} magnitudes with SNR $<$2 are upper limits.
\item {\bf W2}: {\it WISE} W2 mag
\item {\bf W2sig}: {\it WISE} W2 error
\item {\bf W2SNR}: {\it WISE} W2 SNR
\item {\bf W3}: {\it WISE} W3 mag
\item {\bf W3sig}: {\it WISE} W3 error
\item {\bf W3SNR}: {\it WISE} W3 SNR
\item {\bf W4}: {\it WISE} W4 mag
\item {\bf W4sig}: {\it WISE} W4 error
\item {\bf W4SNR}: {\it WISE} W4 SNR
\item {\bf WISE\_ext}: Set to `yes' if {\it WISE} source is extended
\item {\bf WISE\_rej}: Set to `yes' if {\it WISE} counterpart is found but rejected for poor photometry
\item {\bf UKIDSS\_ID}: UKIDSS ID
\item {\bf UKIDSS\_RA}: UKIDSS RA (J2000)
\item {\bf UKIDSS\_Dec}: UKIDSS Dec (J2000)
\item {\bf UKIDSS\_Rel}: MLE reliability of UKIDSS match to X-ray source
\item {\bf UKIDSS\_Dist}: Distance between X-ray and UKIDSS source (arcsec)
\item {\bf Ymag}: UKIDSS Y mag
\item {\bf Ysig}: UKIDSS Y error
\item {\bf Hmag}: UKIDSS H mag
\item {\bf Hsig}: UKIDSS H error
\item {\bf Jmag}: UKIDSS J mag
\item {\bf Jsig}: UKIDSS J error
\item {\bf Kmag}: UKIDSS K mag
\item {\bf Ksig}: UKIDSS K error
\item {\bf UKIDSS\_Rej}: UKIDSS counterpart found but rejected for poor photometry
\item {\bf GALEX\_Objid}: {\it GALEX} object identification number
\item {\bf GALEX\_RA}: {\it GALEX} RA (J2000)
\item {\bf GALEX\_Dec}: {\it GALEX} Dec (J2000)
\item {\bf NUV\_poserr}: {\it GALEX} NUV positional error (arcsec)
\item {\bf FUV\_poserr}: {\it GALEX} FUV positional error (arcsec)
\item {\bf GALEX\_Rel}: MLE reliability of {\it GALEX} match to X-ray source
\item {\bf GALEX\_Dist}: Distance between X-ray and {\it GALEX source} (arcsec)
\item {\bf NUV\_mag}: {\it GALEX} NUV mag
\item {\bf NUV\_magerr}: {\it GALEX} NUV error
\item {\bf FUV\_mag}: {\it GALEX} FUV mag
\item {\bf FUV\_magerr}: {\it GALEX} FUV error
\item {\bf FIRST Name}: IAU Name of FIRST counterpart
\item {\bf FIRST\_RA}: FIRST RA (J2000)
\item {\bf FIRST\_Dec}: FIRST Dec (J2000)
\item {\bf FIRST\_Dist}: Distance between X-ray and FIRST source (arcsec)
\item {\bf FIRST\_Flux}: FIRST 5 GHz Flux Density (Jy)
\item {\bf FIRST\_err}: FIRST 5 GZ Flux Density error (Jy)

\end{enumerate}


\begin{thebibliography}{}


\bibitem[\protect\citeauthoryear{Ahn et al.}{2012}]{dr9} Ahn, C.~P., Alexandroff, R., Allende Prieto, C., et al.\ 2012, ApJS, 203, 21 

\bibitem[\protect\citeauthoryear{Aird et al.}{2010}]{aird} Aird, J., Nandra, K., Laird, E.~S., et al.\ 2010, MNRAS, 401, 2531 

\bibitem[\protect\citeauthoryear{Alexander et al.}{2003}]{cdfn} Alexander, D.~M., Bauer, F.~E., Brandt, W.~N., et al.\ 2003, AJ, 126, 539 

\bibitem[\protect\citeauthoryear{Assef et al.}{2012}]{assef} Assef, R.~J., Stern, D., Kochanek, C.~S., et al.\ 2013, ApJ, 772, 26

\bibitem[\protect\citeauthoryear{Baldwin et al.}{1981}]{bpt} Baldwin, J.~A., Phillips, M.~M., \& Terlevich, R.\ 1981, PASP, 93, 5 

\bibitem[\protect\citeauthoryear{Ballantyne et al.}{2011}]{ballantyne} Ballantyne, D.~R., Draper, A.~R., Madsen, K.~K., Rigby, J.~R., \& Treister, E.\ 2011, ApJ, 736, 56 

\bibitem[\protect\citeauthoryear{Barger et al.}{2003}]{barger} Barger, A.~J., Cowie, L.~L., Capak, P., et al.\ 2003, ApJL, 584, L61 

\bibitem[\protect\citeauthoryear{Becker et al.}{1995}]{first} Becker, R.~H., White, R.~L., \& Helfand, D.~J.\ 1995, ApJ, 450, 559 

\bibitem[\protect\citeauthoryear{Becker et al.}{2012}]{first_cat} Becker, R.~H., Helfand, D.~J., White, R.~L., Gregg, M.~D., \& Laurent-Muehleisen, S.~A.\ 2012, VizieR Online Data Catalog, 8090, 0 

\bibitem[\protect\citeauthoryear{Bianchi et al.}{2011}]{bianchi} Bianchi, L., Efremova, B., Herald, J., et al.\ 2011, MNRAS, 411, 2770 

\bibitem[\protect\citeauthoryear{Brandt \& Hasinger}{2005}]{bh} Brandt, W.~N., \& Hasinger, G.\ 2005, ARAA, 43, 827 

\bibitem[\protect\citeauthoryear{Brandt et al.}{2002}]{brandt} Brandt, W.~N., Schneider, D.~P., Fan, X., et al.\ 2002, ApJL, 569, L5 
\bibitem[\protect\citeauthoryear{Brunner et al.}{2008}]{brunner} Brunner, H., Cappelluti, N., Hasinger, G., et al.\ 2008, A\&A, 479, 283 


\bibitem[\protect\citeauthoryear{Brusa et al.}{2010}]{brusa3} Brusa, M., Civano, F., Comastri, A., et al.\ 2010, ApJ, 716, 348 

\bibitem[\protect\citeauthoryear{Brusa et al.}{2007}]{brusa2} Brusa, M., Zamorani, G., Comastri, A., et al.\ 2007, ApJS, 172, 353 

\bibitem[\protect\citeauthoryear{Brusa et al.}{2005}]{brusa1} Brusa, M., Comastri, A., Daddi, E., et al.\ 2005, A\&A, 432, 69 

\bibitem[\protect\citeauthoryear{Budav{\'a}ri et al.}{2009}]{budavari} Budav{\'a}ri, T., Heinis, S., Szalay, A.~S., et al.\ 2009, ApJ, 694, 1281 

\bibitem[\protect\citeauthoryear{Cappelluti et al.}{2009}]{cap09} Cappelluti, N., Brusa, M., Hasinger, G., et al.\ 2009, A\&A, 497, 635 

\bibitem[\protect\citeauthoryear{Cappelluti et al.}{2007}]{cap} Cappelluti, N., Hasinger, G., Brusa, M., et al.\ 2007, ApJS, 172, 

\bibitem[\protect\citeauthoryear{Cardamone et al.}{2010}]{cardamone2} Cardamone, C.~N., van Dokkum, P.~G., Urry, C.~M., et al.\ 2010, ApJS, 189, 270 

\bibitem[\protect\citeauthoryear{Cardamone et al.}{2008}]{cardamone} Cardamone, C.~N., Urry, C.~M., Damen, M., et al.\ 2008, ApJ, 680, 130 

\bibitem[\protect\citeauthoryear{Casali et al.}{2007}]{casali} Casali, M., Adamson, A., Alves de Oliveira, C., et al.\ 2007, A\&A, 467, 777 

\bibitem[\protect\citeauthoryear{Chiappetti et al.}{2013}]{lss2} Chiappetti, L., Clerc, N., Pacaud, F., et al.\ 2013, MNRAS, 429, 1652 

\bibitem[\protect\citeauthoryear{Civano et al.}{2012}]{civano_mle} Civano, F., Elvis, M., Brusa, M., et al.\ 2012, ApJS, 201, 30 

\bibitem[\protect\citeauthoryear{Civano et al.}{2011}]{civano_11} Civano, F., Brusa, M., Comastri, A., et al.\ 2011, ApJ, 741, 91 

\bibitem[\protect\citeauthoryear{Comastri et al.}{2011}]{comastri} Comastri, A., Ranalli, P., Iwasawa, K., et al.\ 2011, A\&A, 526, L9 

\bibitem[\protect\citeauthoryear{Croom et al.}{2009}]{2slaq} Croom, S.~M., Richards, G.~T., Shanks, T., et al.\ 2009, MNRAS, 392, 19 


\bibitem[\protect\citeauthoryear{Cutri et al.}{2012}]{wise_cat} Cutri, R.~M., Wright, E.~L., Conrow, T., et al.\ 2012, Explanatory Supplement to the WISE All-Sky Data Release Products, 1 

\bibitem[\protect\citeauthoryear{Davis et al.}{2007}]{aegis} Davis, M., Guhathakurta, P., Konidaris, N.~P., et al.\ 2007, ApJL, 660, L1

\bibitem[\protect\citeauthoryear{Donley et al.}{2012}]{donley} Donley, J.~L., Koekemoer, A.~M., Brusa, M., et al.\ 2012, ApJ, 748, 142 

\bibitem[\protect\citeauthoryear{Drinkwater et al.}{2010}]{wigglez} Drinkwater, M.~J., Jurek, R.~J., Blake, C., et al.\ 2010, MNRAS, 401, 1429 

\bibitem[\protect\citeauthoryear{Dye et al.}{2006}]{dye} Dye, S., Warren, S.~J., Hambly, N.~C., et al.\ 2006, MNRAS, 372, 1227 

\bibitem[\protect\citeauthoryear{Elvis et al.}{2009}]{C-Cosmos} Elvis, M., Civano, F., Vignali, C., et al.\ 2009, ApJS, 184, 158 

\bibitem[\protect\citeauthoryear{Elvis et al.}{1994}]{elvis_sed} Elvis, M., Wilkes, B.~J., McDowell, J.~C., et al.\ 1994, ApJS, 95, 1 

\bibitem[\protect\citeauthoryear{Evans et al.}{2010}]{csc} Evans, I.~N., Primini, F.~A., Glotfelty, K.~J., et al.\ 2010, ApJS, 189, 37 

\bibitem[\protect\citeauthoryear{Georgakakis et al.}{2007}]{aegis2} Georgakakis, A., Nandra, K., Laird, E.~S., et al.\ 2007, ApJL, 660, L15 

\bibitem[\protect\citeauthoryear{Giacconi et al.}{2001}]{giacconi} Giacconi, R., Rosati, P., Tozzi, P., et al.\ 2001, ApJ, 551, 624 

\bibitem[\protect\citeauthoryear{Giavalisco et al.}{2004}]{giavalisco} Giavalisco, M., Ferguson, H.~C., Koekemoer, A.~M., et al.\ 2004, ApJL, 600, L93 

\bibitem[\protect\citeauthoryear{Gilli et al.}{2007}]{Gilli} Gilli, R., Comastri, A., \& Hasinger, G.\ 2007, A\&A, 463, 79 

\bibitem[\protect\citeauthoryear{Goulding et al.}{2012}]{deep2} Goulding, A.~D., Forman, W.~R., Hickox, R.~C., et al.\ 2012, ApJS, 202, 6

\bibitem[\protect\citeauthoryear{Hambly et al.}{2008}]{hambly} Hambly, N.~C., Collins, R.~S., Cross, N.~J.~G., et al.\ 2008, MNRAS, 384, 637 

\bibitem[\protect\citeauthoryear{Hasinger et al.}{2005}]{hasinger} Hasinger, G., Miyaji, T., \& Schmidt, M.\ 2005, A\&A, 441, 417 

\bibitem[\protect\citeauthoryear{Hewett et al.}{2006}]{hewett} Hewett, P.~C., Warren, S.~J., Leggett, S.~K., \& Hodgkin, S.~T.\ 2006, MNRAS, 367, 454 
\bibitem[\protect\citeauthoryear{Jiang et al.}{2009}]{jiang} Jiang, L., Fan, X., Bian, F., et al.\ 2009, AJ, 138, 305 

\bibitem[\protect\citeauthoryear{Juneau et al.}{2011}]{juneau} Juneau, S., Dickinson, M., Alexander, D.~M., \& Salim, S.\ 2011, ApJ, 736, 104 

\bibitem[\protect\citeauthoryear{Kauffmann et al.}{2003}]{kauff} Kauffmann, G., Heckman, T.~M., Tremonti, C., et al.\ 2003, MNRAS, 346, 1055 

\bibitem[\protect\citeauthoryear{Kenter et al.}{2005}]{kenter} Kenter, A., Murray, S.~S., Forman, W.~R., et al.\ 2005, ApJS, 161, 9 

\bibitem[\protect\citeauthoryear{Kewley et al.}{2001}]{kewley} Kewley, L.~J., Dopita, M.~A., Sutherland, R.~S., Heisler, C.~A., \& Trevena, J.\ 2001, ApJ, 556, 121 

\bibitem[\protect\citeauthoryear{Kim et al.}{2007}]{champ} Kim, M., Wilkes, B.~J., Kim, D.-W., et al.\ 2007, ApJ, 659, 29 

\bibitem[\protect\citeauthoryear{Kochanek et al.}{2012}]{kochanek} Kochanek, C.~S., Eisenstein, D.~J., Cool, R.~J., et al.\ 2012, ApJS, 200, 8 

\bibitem[\protect\citeauthoryear{Kolodzig et al.}{2012}]{kolodzig} Kolodzig, A., Gilfanov, M., Sunyaev, R., Sazonov, S., \& Brusa, M.\ 2012, arXiv:1212.2151 


\bibitem[\protect\citeauthoryear{Lacy et al.}{2004}]{lacy} Lacy, M., Storrie-Lombardi, L.~J., Sajina, A., et al.\ 2004, ApJS, 154, 166 

\bibitem[\protect\citeauthoryear{La Franca et al.}{2005}]{lafranca} La Franca, F., Fiore, F., Comastri, A., et al.\ 2005, ApJ, 635, 864 

\bibitem[\protect\citeauthoryear{LaMassa et al.}{2013}]{me} LaMassa, S.~M., Urry, C.~M., Glikman, E., et al.\ 2013, MNRAS, 432, 1351 

\bibitem[\protect\citeauthoryear{Lawrence et al.}{2007}]{lawrence} Lawrence, A., Warren, S.~J., Almaini, O., et al.\ 2007, MNRAS, 379, 1599 

\bibitem[\protect\citeauthoryear{Lehmer et al.}{2005}]{lehmer} Lehmer, B.~D., Brandt, W.~N., Alexander, D.~M., et al.\ 2005, ApJS, 161, 21 

\bibitem[\protect\citeauthoryear{Loaring et al.}{2005}]{Loaring} Loaring, N.~S., Dwelly, T., Page, M.~J., et al.\ 2005, MNRAS, 362, 1371 

\bibitem[\protect\citeauthoryear{Luo et al.}{2010}]{luo} Luo, B., Brandt, W.~N., Xue, Y.~Q., et al.\ 2010, ApJS, 187, 560 

\bibitem[\protect\citeauthoryear{Mateos et al.}{2008}]{Mateos} Mateos, S., Warwick, R.~S., Carrera, F.~J., et al.\ 2008, A\&A, 492, 51 

\bibitem[\protect\citeauthoryear{McGreer et al.}{2013}]{mcgreer} McGreer, I.~D., Jiang, L., Fan, X., et al.\ 2013, ApJ, 768, 105 

\bibitem[\protect\citeauthoryear{Mendez et al.}{2013}]{mendez} Mendez, A.~J., Coil, A.~L., Aird, J., et al.\ 2013, ApJ, 770, 40 

\bibitem[\protect\citeauthoryear{Merloni et al.}{2012}]{merloni} Merloni, A., Predehl, P., Becker, W., et al.\ 2012, arXiv:1209.3114 

\bibitem[\protect\citeauthoryear{Morrissey et al.}{2007}]{morrissey} Morrissey, P., Conrow, T., Barlow, T.~A., et al.\ 2007, ApJS, 173, 682 

\bibitem[\protect\citeauthoryear{Murray et al.}{2005}]{murray} Murray, S.~S., Kenter, A., Forman, W.~R., et al.\ 2005, ApJS, 161, 1 

\bibitem[\protect\citeauthoryear{Newman et al.}{2012}]{spec_deep2} Newman, J.~A., Cooper, M.~C., Davis, M., et al.\ 2013, ApJS, 208, 5 

\bibitem[Ochsenbein et al.(2000)]{vizier} Ochsenbein, F., Bauer, P., \& Marcout, J.\ 2000, A\&AS, 143, 23

\bibitem[\protect\citeauthoryear{Persic et al.}{2004}]{persic} Persic, M., Rephaeli, Y., Braito, V., et al.\ 2004, A\&A, 419, 849 

\bibitem[\protect\citeauthoryear{Pierre et al.}{2004}]{lss1} Pierre, M., Valtchanov, I., Altieri, B., et al.\ 2004, JCAP, 9, 11 

\bibitem[\protect\citeauthoryear{Ranalli et al.}{2013}]{ranalli} Ranalli, P., Comastri, A., Vignali, C., et al.\ \ 2013, A\&A, 555, A42 

\bibitem[\protect\citeauthoryear{Rots \& Budav{\'a}ri}{2011}]{csc_sdss} Rots, A.~H., \& Budav{\'a}ri, T.\ 2011, ApJS, 192, 8 

\bibitem[\protect\citeauthoryear{Shemmer et al.}{2006}]{shemmer} Shemmer, O., Brandt, W.~N., Schneider, D.~P., et al.\ 2006, ApJ, 644, 86 

\bibitem[\protect\citeauthoryear{Steffen et al.}{2004}]{steffen} Steffen, A.~T., Barger, A.~J., Capak, P., et al.\ 2004, AJ, 128, 1483 

\bibitem[\protect\citeauthoryear{Stern et al.}{2012}]{stern} Stern, D., Assef, R.~J., Benford, D.~J., et al.\ 2012, ApJ, 753, 30 

\bibitem[\protect\citeauthoryear{Stern et al.}{2005}]{stern1} Stern, D., Eisenhardt, P., Gorjian, V., et al.\ 2005, ApJ, 631, 163 

\bibitem[\protect\citeauthoryear{Sutherland \& Saunders}{1992}]{mle} Sutherland, W., \& Saunders, W.\ 1992, MNRAS, 259, 413 

\bibitem[\protect\citeauthoryear{Treister et al.}{2009}]{treister} Treister, E., Urry, C.~M., \& Virani, S.\ 2009, ApJ, 696, 110 

\bibitem[\protect\citeauthoryear{Treister et al.}{2004}]{treister_04} Treister, E., Urry, C.~M., Chatzichristou, E., et al.\ 2004, ApJ, 616, 123 

\bibitem[\protect\citeauthoryear{Trichas et al.}{2012}]{trichas} Trichas, M., Green, P.~J., Silverman, J.~D., et al.\ 2012, ApJS, 200, 17 

\bibitem[\protect\citeauthoryear{Trouille et al.}{2011}]{trouille} Trouille, L., Barger, A.~J., \& Tremonti, C.\ 2011, ApJ, 742, 46 

\bibitem[\protect\citeauthoryear{Ueda et al.}{2003}]{ueda} Ueda, Y., Akiyama, M., Ohta, K., \& Miyaji, T.\ 2003, ApJ, 598, 886 

\bibitem[\protect\citeauthoryear{Vignali et al.}{2005}]{vignali} Vignali, C., Brandt, W.~N., Schneider, D.~P., \& Kaspi, S.\ 2005, AJ, 129, 2519 

\bibitem[\protect\citeauthoryear{Virani et al.}{2006}]{virani} Virani, S.~N., Treister, E., Urry, C.~M., \& Gawiser, E.\ 2006, AJ, 131, 2373 

\bibitem[\protect\citeauthoryear{Warren et al.}{2007}]{warren} Warren, S.~J., Cross, N.~J.~G., Dye, S., et al.\ 2007, arXiv:astro-ph/0703037 

\bibitem[\protect\citeauthoryear{Watson et al.}{2009}]{Watson} Watson, M.~G., Schr{\"o}der, A.~C., Fyfe, D., et al.\ 2009, A\&A, 493, 339 


\bibitem[\protect\citeauthoryear{White et al.}{1997}]{first_cat1} White, R.~L., Becker, R.~H., Helfand, D.~J., \& Gregg, M.~D.\ 1997, ApJ, 475, 479 

\bibitem[\protect\citeauthoryear{Wright et al.}{2010}]{wright} Wright, E.~L., Eisenhardt, P.~R.~M., Mainzer, A.~K., et al.\ 2010, AJ, 140, 1868 

\bibitem[\protect\citeauthoryear{Xue et al.}{2011}]{cdfs} Xue, Y.~Q., Luo, B., Brandt, W.~N., et al.\ 2011, ApJS, 195, 10 


\end{thebibliography}
\end{document}